\documentclass[aps,
	twocolumn,
	altaffilletter,
	nolongbibliography,
	numerical,
	flushbottom,
	secnumarabic,
	pra,
	superscriptaddress,
	floatfix,
	10pt,
        nobalancelastpage]{revtex4-2}

\usepackage[T1]{fontenc}
\usepackage{comment}
\usepackage{booktabs}
\usepackage{newtxtext,newtxmath}
\usepackage{graphicx}
\usepackage{braket}
\usepackage[dvipsnames]{xcolor}
\usepackage{hyperref}
\usepackage{multirow}
\hypersetup{%
	colorlinks=true, 
	breaklinks=true,
	urlcolor=blue, 
	linkcolor=blue, 
	citecolor=blue,
}

\newcommand{\commentref}[1]{#1}

\begin{document}

\title{The role of gaps in  digitized counterdiabatic QAOA for fully-connected spin models}%

\author{Mara Vizzuso}
\email{mara.vizzuso@unina.it}
\affiliation{Dipartimento di Fisica ``E. Pancini'', Universit\`a degli Studi di Napoli ``Federico II'', Complesso Universitario M. S. Angelo, via Cintia 21, 80126, Napoli, Italy}

\author{Gianluca Passarelli}
\affiliation{Dipartimento di Fisica ``E. Pancini'', Universit\`a degli Studi di Napoli ``Federico II'', Complesso Universitario M. S. Angelo, via Cintia 21, 80126, Napoli, Italy}

\author{Giovanni Cantele}
\affiliation{CNR-SPIN, c/o Complesso Universitario M. S. Angelo, via Cintia 21, 80126, Napoli, Italy}

\author{Procolo Lucignano}
\affiliation{Dipartimento di Fisica ``E. Pancini'', Universit\`a degli Studi di Napoli ``Federico II'', Complesso Universitario M. S. Angelo, via Cintia 21, 80126, Napoli, Italy}

\date{\today}%

\begin{abstract}
Recently, digitized-counterdiabatic (CD) corrections to the quantum approximate optimization algorithm (QAOA) have been proposed, yielding faster convergence within the desired accuracy than standard QAOA. In this manuscript, we apply this approach to a fully-connected spin model with random couplings. We show that the performances of the algorithm are related to the spectral properties of the instances analyzed. In particular, the larger the gap between the ground state and the first excited states, the better the convergence to the exact solution.

\end{abstract}
\maketitle

\section{\label{sec:Intro} Introduction}

In recent years, the field of quantum computing has experienced remarkable advancements, showing new ways for addressing computational challenges previously deemed impossible for classical computers. Among the others, Variational Quantum Algorithms (VQAs) have emerged as  particularly versatile tools~\cite{cerezo2021variational}. 
VQAs address optimization tasks by iteratively refining the parameters of a quantum circuit to minimize a desired cost function. This makes  VQAs adaptable and versatile to different applications, spanning quantum chemistry simulations~\cite{Colless2018,grimsley2019adaptive,kandala2017hardware}, machine learning~\cite{Yao2021,Zhang2023}, and combinatorial optimization~\cite{Anschuetz2019,karamlou2021analyzing}.

An example of VQAs is the Quantum Approximate Optimization Algorithm (QAOA)~\cite{farhi2014quantum}. QAOA serves as a variational technique tailored for tackling combinatorial optimization problems within gate-based quantum computing systems. In essence, combinatorial optimization consists in identifying, from a finite set of variables, the ones that  minimize a desired cost function. This finds practical applications across various domains such as streamlining supply chain costs, optimizing vehicle routes, allocating tasks efficiently, and more~\cite{venkatesh2024qubitefficient, sarmina2024parameter, montanezbarrera2024universal, tsvelikhovskiy2024equivariant, Boy2024}. In QAOA a combinatorial optimization problem is cast into the form of  searching the approximate ground state of a spin Hamiltonian~\cite{lucas:2014}, achieved through a specific variational ansatz. This ansatz is expressed via a gate circuit depending on free parameters to be  optimized using a classical computer minimization routine.

Many QAOA variants have proliferated over the years to further improve the performance of the original algorithm~\cite{Hadfield2019, Bartschi2020, Villalba-Diez2022, Golden2021, Fuchs2022, egger2021, Yoshioka2023, Wurtz2021, cadavid2023,ji2023improving,Yu2022,zhu2022,Bravyi2020,zou2023multiscale, Magann2022,BLEKOS20241,Misra-Spieldenner2023,zhu2022,Hadfield2019,Chandarana:2022,Chai:2022,Wurtz2022counterdiabaticity,chandarana2022digitizedcounterdiabatic}. A non-exhaustive list includes: QAOA$+$\cite{Chalupnik2022}, which improves the conventional QAOA adding an additional problem-independent layer with multiple parameters; adaptive-bias QAOA~\cite{Yu2022}, introducing local fields to QAOA operators; adaptive QAOA~\cite{zhu2022}, a QAOA variant iteratively selecting mixers based on a systematic gradient criterion; and recursive-QAOA~\cite{Bravyi2020}, aiming to decrease problem size by eliminating unnecessary qubits through a non-local scheme. %
A promising variant is the digitized counterdiabatic QAOA (QAOA-CD)~\cite{Chandarana:2022,Chai:2022,Wurtz2022counterdiabaticity}, which incorporates an additional driving Hamiltonian inspired by quantum shortcuts to adiabaticity~\cite{Chen2010,Guery-Odelin2019,vizzuso2023}. This additional Hamiltonian leads to faster convergence to the ground state energy, allowing for a reduction in the circuit depth. 

In this paper, we apply the variant of QAOA originally developed in Ref.~\cite{vizzuso2023} to study the randomly weighted MaxCut problem on 2-regular graphs, to explore its suitability %
\commentref{to} fully-connected graphs in which  %
long-range all to all interactions significantly amplify the  computational complexity with respect to local interactions. In this setting, our goal is twofold. On the one hand, we will show that the counterdiabatic variants of QAOA are also feasible for fully-connected models. On the other hand, we will show through an in-depth statistical study that the performance of QAOA and its variants depend on the spectral properties of the target Hamiltonian. We can in fact distinguish ``easy'' and ``hard'' Hamiltonians depending on the value of the gap between the target ground state and the first excited states, in a similar fashion to adiabatic quantum computing (AQC). In AQC, the minimum \textit{instantaneous} gap between the time-dependent ground state and first excited state classifies the hardness of the optimization problem as it determines the minimum time scale necessary for successful adiabatic state preparation~\cite{albash:2018}. In QAOA, instead, where continuous evolutions are replaced by digital steps and it is not possible to identify a continuous time-dependent generator of the quantum circuit, this role is played by the gap of the \textit{target} Hamiltonian. \commentref{This is a consequence of the fact that the variational procedure is more complicated for problem Hamiltonians with small gaps, as we are going to discuss later on.} %

The paper is organized as follows.  We introduce our model Hamiltonian in Sec.~\ref{sec:model} and review QAOA and its counterdiabatic corrections in Sec.~\ref{sec:qaoa}. In Sec.~\ref{sec:results}, we present the outcomes of our numerical analysis. Finally, we summarize our findings in Sec.~\ref{sec:conclusions}.

\section{MaxCut Model} \label{sec:model}

Combinatorial optimization problems are relevant  to economics and finance~\cite{White2021}, they include the Traveling Salesman Problem (TSP)~\cite{Hoffman2001}, the Minimum Spanning Tree Problem (MST)~\cite{Graham1985}, the Knapsack Problem~\cite{Salkin1975}, the MaxCut problem~\cite{papadimitriou1998combinatorial}, and many others~\cite{Trevisan2012,BARAHONA1983107,POLJAK1995249}.  In particular we here focus on quadratic unconstrained binary optimization (QUBO) that can be formally defined as follows: given a discrete set $K$  made of $N$ binary variables and a function $f(\vec{x})$ that maps each item $\vec{x}$ from $K$ to a continuous set, the objective is to find the optimal combination of elements, denoted by $\vec{x}^*$, that minimizes the function $f(\vec{x})$.

Given a graph $G(E,V)$, defined as a collection of vertices ($V$ set) interconnected by edges ($E$ set) the concept of a Maximum Cut arises, denoting a partition that cuts off the greatest number of edges between two separate sets of vertices. This partition, termed a \textit{cut}, is characterized by a size representing the count of cut edges, which must surpass the size of any alternative partition to qualify as the Maximum Cut. The pursuit of this Maximum Cut constitutes a computational challenge categorized under NP-hard problems~\cite{cook2000p} and is recognized as the MaxCut problem.
Classically we define the cost function $f(\vec{x})$ whose maximum is the solution of the MaxCut problem on a graph $G(E,V)$ as:
\begin{equation}
	f(\vec{x}) = \frac{1}{2}\sum_{i,j\in G(E,V)}\left(1-x_i x_j\right).
	\label{cost-classical-MaxCut}
\end{equation}
We suppose that the graph $G(E,V)$ has $N$ vertices. On each vertex $i$ the variable $x_i$ can assume either the value $x_i=+1$ or $x_i=-1$. The set of all possible states, in the form $(x_1,x_2,....,x_N)$ has dimension $2^N$. 
The MaxCut consists in  partitioning the graph into two subgraphs $A$ and $B$ such that all  the elements in $A$ assume the value $x_i=+1$, and all the elements in $B$ take the value $x_i=-1$ cutting the maximum number of edges.

Finding the maximum of  a cost function $f$ over a set of binary variables $\vec x$ is the same as calculating the ground state energy of a classical spin Hamiltonian. Classical spins can be set at the vertices of a graph whose edges depend on the Hamiltonian connectivity.
Therefore solving a QUBO is the same as finding the ground state of an Ising problem~\cite{lucas:2014}.  Various methodologies derived from statistical and quantum physics can be employed to address this challenge, including techniques such as Quantum and Simulated Annealing (QA and SA respectively)~\cite{kadowaki:1998,albash:2018,simulated-annealing-book,phegde:ga,gpassarelli:qa1,gpassarelli:qa2,gpassarelli:qa3,gpassarelli:qa4,gpassarelli:qa5,gpassarelli:qa6,gpassarelli:qa7,10.1088/1367-2630/ace547}, along with the QAOA. %

\subsection{Fully-connected spin model}
We associate a Pauli operator $\sigma_i^Z$ to each site $i$ of the graph. The (dimensionless) Hamiltonian describing the  MaxCut for a random fully-connected spin model %
is
\begin{equation}
    H_T=\sum_{i<j}^NJ_{ij}\sigma_i^Z\sigma_j^Z,
    \label{target-hamiltonian}
\end{equation}
where $\sigma_i^Z,\sigma_j^Z$ are the Pauli matrices acting on respectively the $i^\text{th}$ and $j^\text{th}$ spins and $J_{ij} = \mathcal{U}([-1, 1])$ are random couplings distributed according to the uniform distribution $\mathcal{U}$ in the given range. The cost function of QAOA, as discussed in Sec.~\ref{sec:qaoa}, is given by the expectation value of $H_T$ on a parametric quantum circuit.
As a proof of principle we focus on random instances with a small number of qubits, $N=5$ \commentref{(in Appendix~\ref{app:numerics}, we show that our results also hold for larger systems)}, and analyze $n=600$ samples (which gives an accurate sampling of the coupling distribution without requiring unfeasible computational resources) of the disordered interactions, each of them with different couplings $J_{ij}$. %
In the case $J_{ij}=1$ finding the ground state of the Hamiltonian of Eq.~\eqref{target-hamiltonian} is the same as finding the maximum of the cost function $f$ defined in Eq.~\eqref{cost-classical-MaxCut}. Due to the $Z_2$ symmetry  of the Hamiltonian of Eq.~\eqref{target-hamiltonian} all the eigenvalues are doubly degenerate.

\section{QAOA and variants}
\label{sec:qaoa}

QAOA is a hybrid quantum-classical algorithm proposed by Farhi et al~\cite{farhi2014quantum}. It  consists of a quantum part and a classical part. The objective of the algorithm is to find the ground state of a target Hamiltonian $H_T$. The algorithm works as follows. We consider two Hamiltonians: $H_T$, representing our target Hamiltonian whose ground state solves our optimization problem, and $H_X$, \commentref{a one-body Hamiltonian easy to diagonalize}, encoding quantum fluctuations:
\begin{equation}
    H_X = \sum_{i=1}^N \sigma_i^X,
    \label{mixer-hamiltonian}
\end{equation}
where $\sigma_i^X$ is a Pauli matrix associated with site $i$. 
The ground state of Eq.~\eqref{mixer-hamiltonian} is

\begin{equation}
    \ket{0} = \frac{1}{\sqrt{2^N}} \bigotimes_{i=1}^N \left( \ket{\uparrow}_i - \ket{\downarrow}_i \right),
    \label{start_state}
\end{equation}
where $\ket{\uparrow}_i$ and $\ket{\downarrow}_i$ are eigenstates of the $\sigma^Z_i$ operator associated with the $i^\text{th}$ site.

The algorithm consists in transforming $\ket{0}$ using a set of unitaries depending on $H_X$ and $H_T$ and on variational coefficients that must be optimized classically in order to minimize the expectation value of the target Hamiltonian on the trial state.

The trial wavefunction is chosen as
\begin{equation}
\ket{\psi^{(p)}(\vec{\beta},\vec{\gamma})} = \left(\prod_{k=1}^{p} U(\beta_k,H_X) U(\gamma_k,H_T)\right) \ket{0}
    \label{qaoa-state}
\end{equation}
where 
\begin{equation}
    U(\theta,H) = e^{-i\theta H}.
    \label{unitary}
\end{equation}
The wavefunction $|\psi^{(p)}\rangle$ depends on the $2p$  parameters  $\vec{\beta}=(\beta_1,\dots,\beta_p)$ and $\vec{\gamma}=(\gamma_1,\dots,\gamma_p)$, where  $p$ represents the number of QAOA steps.  The cost function to minimize is
\begin{equation}
    E^{(p)}(\vec{\beta},\vec{\gamma}) = \bra{\psi^{(p)}(\vec{\beta},\vec{\gamma})} H_T \ket{\psi^{(p)}(\vec{\beta},\vec{\gamma})}.
    \label{qaoa-function}
\end{equation}
QAOA consists in finding \textit{classically} (i.\,e., on a classical computer) $\vec{\beta}^*$ and $\vec{\gamma}^*$ that minimize the function in Eq.~\eqref{qaoa-function}. $\ket{\psi^{(p)}(\vec{\beta}^*,\vec{\gamma}^*)}$ represents an approximation of the true ground state of $H_T$ [Eq.~\eqref{target-hamiltonian}] after $p$ steps. %
Increasing the number $2p$  of variational parameters improves the ``expressivity'' of the trial wave function allowing for better and better approximations of the state we are looking for. %
In some cases~\cite{farhi:2000}, QAOA allows us to get the exact solution in a finite number of steps $p$. However, in most cases, there is only an asymptotic convergence to the exact ground state of $H_T$ by increasing the number of steps. %

Next, we describe digitized-counterdiabatic variants of QAOA, which allow improving the accuracy of QAOA at any circuit depth $p$~\cite{vizzuso2023}. This algorithm is inspired by the counterdiabatic potentials~\cite{delCampo2013Shortcuts}, included in the QAOA algorithm via next-order terms of the Baker-Hausdorff-Campbell (BHC) expansion~\cite{casas:zassenhaus}. One gets the so-called QAOA-CD~\cite{Chandarana:2022,Wurtz2022counterdiabaticity} by stopping at the first-order correction.
At each step $p$ of QAOA-CD, the QAOA state is not only transformed by the unitaries $U(\beta, H_X)$ and $U(\gamma, H_T)$ as in Eq.~\eqref{qaoa-state}, but also by an additional operator $U_\text{CD}$, as follows:
\begin{equation}   \ket{\psi_\text{CD}^{(p)}(\vec{\beta},\vec{\gamma},\vec{\alpha})} = \left(\prod_{k=1}^{p} U(\beta_k,H_X) U(\gamma_k,H_T)U_\text{CD}(\alpha_k)\right) \ket{0},
    \label{qaoa-cd-state}
\end{equation}
where we see another $p$-dimensional vector of parameters $\vec{\alpha}=(\alpha_1,\dots,\alpha_p)$ and
\begin{equation}
    U_\text{CD}(\alpha) = e^{-\alpha \left[H_X,H_T\right]}.
    \label{qaoa-cd-unitary}
\end{equation}
Looking at Eq.~\eqref{qaoa-function}, we can easily generalize a new cost function $E_\text{CD}^{(p)}(\vec{\beta},\vec{\gamma},\vec{\alpha})$ that counts $3p$ parameters per step. 

In QAOA-2CD, this process can be further improved adding the next term in the BHC expansion~\cite{vizzuso2023}. 
Therefore the trial wave function becomes 
\begin{align}  &\ket{\psi_\text{2CD}^{(p)}(\vec{\beta},\vec{\gamma},\vec{\alpha}, \vec{\delta}, \vec{\zeta})}= \notag \\ & \left(\prod_{k=1}^{p} U(\beta_k,H_X) U(\gamma_k,H_T)U_\text{CD}(\alpha_k) U_\text{2CD}(\delta_k, \zeta_k)\right)\ket{0},
   \label{psi-2cd-state}
\end{align}
where %
$\vec{\delta}=(\delta_1,\dots,\delta_p)$ and $\vec{\zeta}=(\zeta_1,\dots,\zeta_p)$ are $p$-dimensional vectors, and 
\begin{equation}
    U_\text{2CD}(\delta, \zeta) = e^{i\delta [H_X, [H_X, H_T]] - i\zeta [H_T, [H_X,H_T]]},
    \label{qaoa-2cd-unitary}
\end{equation}
and the new cost function is
$E_\text{2CD}(\vec{\beta},\vec{\gamma},\vec{\alpha}, \vec{\delta}, \vec{\zeta})$, depending on $5p$ parameters. The variants QAOA-CD and QAOA-2CD differ from QAOA only in the construction of the quantum part of the algorithm, the classical minimization changes only for the number of parameters of the cost function at any given step.

In order to analyze the accuracy of these algorithms,  we introduce the \textit{residual energy} and the \textit{fidelity} at each step $p$. The residual energy is defined as:
\begin{equation}
\varepsilon_\mathrm{res}^{(p)}(\vec{\beta}, \vec{\gamma})=\frac{E_p(\vec{\gamma},\vec{\beta})-E_\text{min}}{E_\text{max}-E_\text{min}},
\label{eres}
\end{equation}
where $E_{\text{min(max)}}$ is the minimum (maximum) eigenvalue of the Hamiltonian of Eq.~\eqref{target-hamiltonian}. This is trivially generalized to  QAOA-CD [$\varepsilon_{CD}^{(p)}(\vec{\beta}, \vec{\gamma}, \vec{\alpha})$] and QAOA-2CD [$\varepsilon_{2CD}^{(p)}(\vec{\beta}, \vec{\gamma}, \vec{\alpha}, \vec{\delta}, \vec{\zeta})$]. 

We can further introduce the fidelity, which  for QAOA is 
\begin{equation}
F^{(p)}(\vec{\beta}, \vec{\gamma}) = \left|\braket{\psi_T|\psi^{(p)}(\vec{\gamma},\vec{\beta})}\right|^2,
\label{fidelity}
\end{equation}
where $\ket{\psi_T}$ is the ground state of $H_T$ [see Eq.~\eqref{target-hamiltonian}]. We can easily generalize this equation to QAOA-CD [$F_{CD}^{(p)}(\vec{\beta}, \vec{\gamma}, \vec{\alpha})$] and QAOA-2CD [$F_{2CD}^{(p)}(\vec{\beta}, \vec{\gamma}, \vec{\alpha}, \vec{\delta}, \vec{\zeta})$].

\section{Results} \label{sec:results}

In this section, we discuss  results achieved by studying the disordered fully-connected spin model of Eq.~\eqref{target-hamiltonian} within the QAOA, QAOA-CD, and QAOA-2CD frameworks. \commentref{The classical optimization is performed using the L-BFGS-B method \cite{Byrd1995}. For each algorithm (QAOA, QAOA-CD, QAOA-2CD) and at each step, we start the optimization with $20$
randomly selected sets of initial angles, and select the best result from these optimizations. While this approach does not guarantee a global optimum, the optimization can be refined by increasing the number of initial angle choices when needed.}
Our results are averaged over a set of $n=600$ random instances. Disorder-averaged quantities are calculated as
\begin{align}
    \label{average}
     &\braket{K}=\frac{1}{n} \sum_{i=1}^n K_i\\ 
    &\sigma_K = \sqrt{\frac{1}{n} \sum_{i=1}^n  \left(K_i-\braket{K}\right)^2},
     \label{standard-deviation}
\end{align}
where $K_i$ is a generic observable as evaluated on a single instance characterized by a set of couplings $J^{(i)}_{jk}$ and $\braket{K}$ and $\sigma_K$ are respectively
the mean value and standard deviation of $K$.

The ground state of Eq.~\eqref{target-hamiltonian} has a double degeneracy, therefore the fidelity of Eq.~\eqref{fidelity} is computed as
\begin{equation}
    F^{(p)} = |\braket{\psi_T^1|\psi^{(p)}(\vec{\gamma^*},\vec{\beta^*})}|^2 + |\braket{\psi_T^2|\psi^{(p)}(\vec{\gamma^*},\vec{\beta^*})}|^2, 
    \label{fidelity-fully-connected}
\end{equation}
where $\psi_T^1$ and $\psi_T^2$ are the two degenerate eigenstates corresponding to the ground-state energy $E_{\text{min}}$  of
the Hamiltonian of Eq.~\eqref{target-hamiltonian}. 

The fidelity of Eq.~\eqref{fidelity-fully-connected} as well as the residual energy of Eq.~\eqref{eres} are used in the following to assess the performances of the algorithm and eventually connect them with the  spectral properties of $H_T$.
In quantum annealing~\cite{albash:2018,gpassarelli:qa1,santoro-martonak}, the minimal gap during the annealing dynamics determines the fidelity of the algorithm due to the occurrence of Landau-Zener transitions. In QAOA such strict relation between the gaps and the fidelity is not known nor well established, however a dense spectrum in the vicinity of the ground state,
makes the system more prone to be ``captured'' by local minima during the hybrid optimization procedure~\cite{LeoZhou2020}. To highlight the role of the gaps in the QAOA, in the following we will relate the QAOA performances for various instances keeping track of \commentref{the distance between ground and the first excited state. In this paper we define this quantity as $\Delta_{eg}$.}

\begin{figure*}
    \hspace{-0.2\textwidth}(a)\hspace{0.31\textwidth}(b)\hspace{0.31\textwidth}(c)\hfill\\
    \centering
    \includegraphics[width=0.295\textwidth]{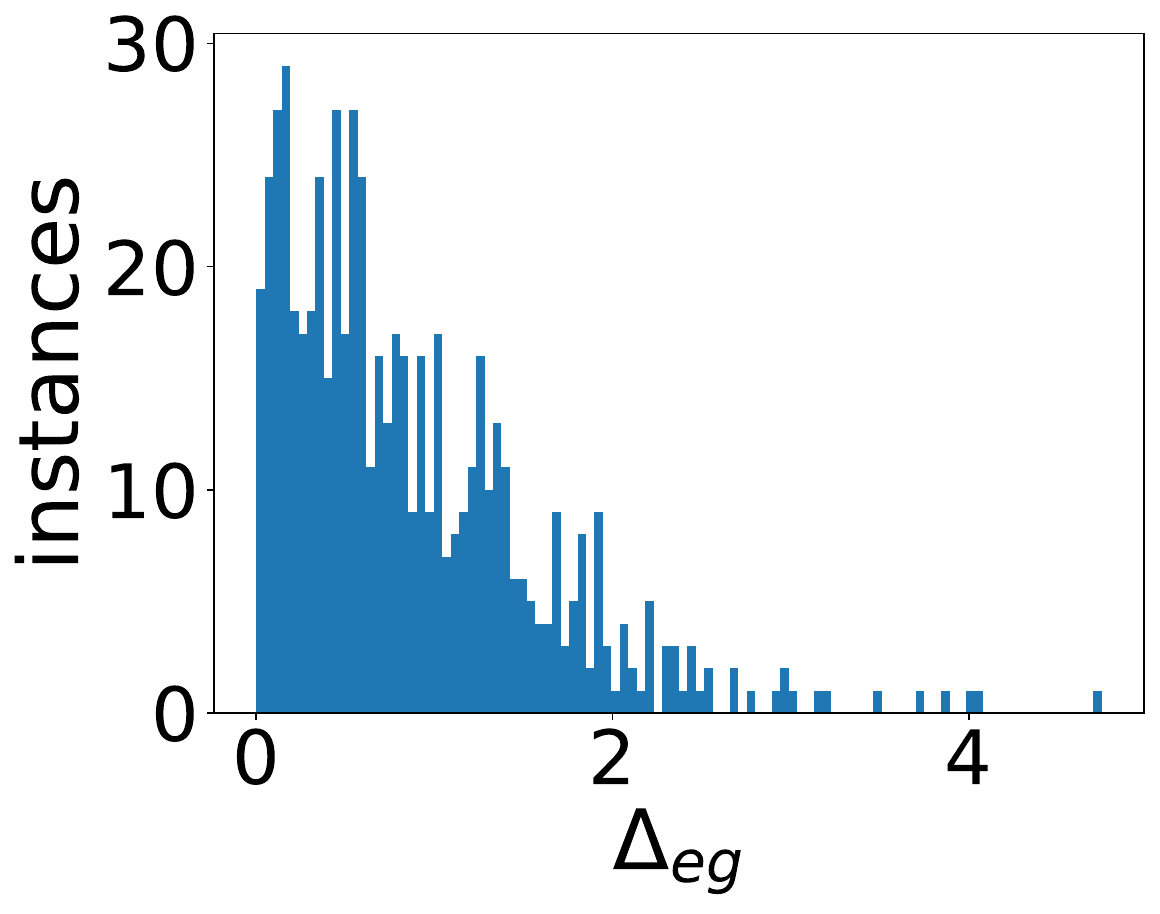}
    \includegraphics[width=0.32\textwidth]{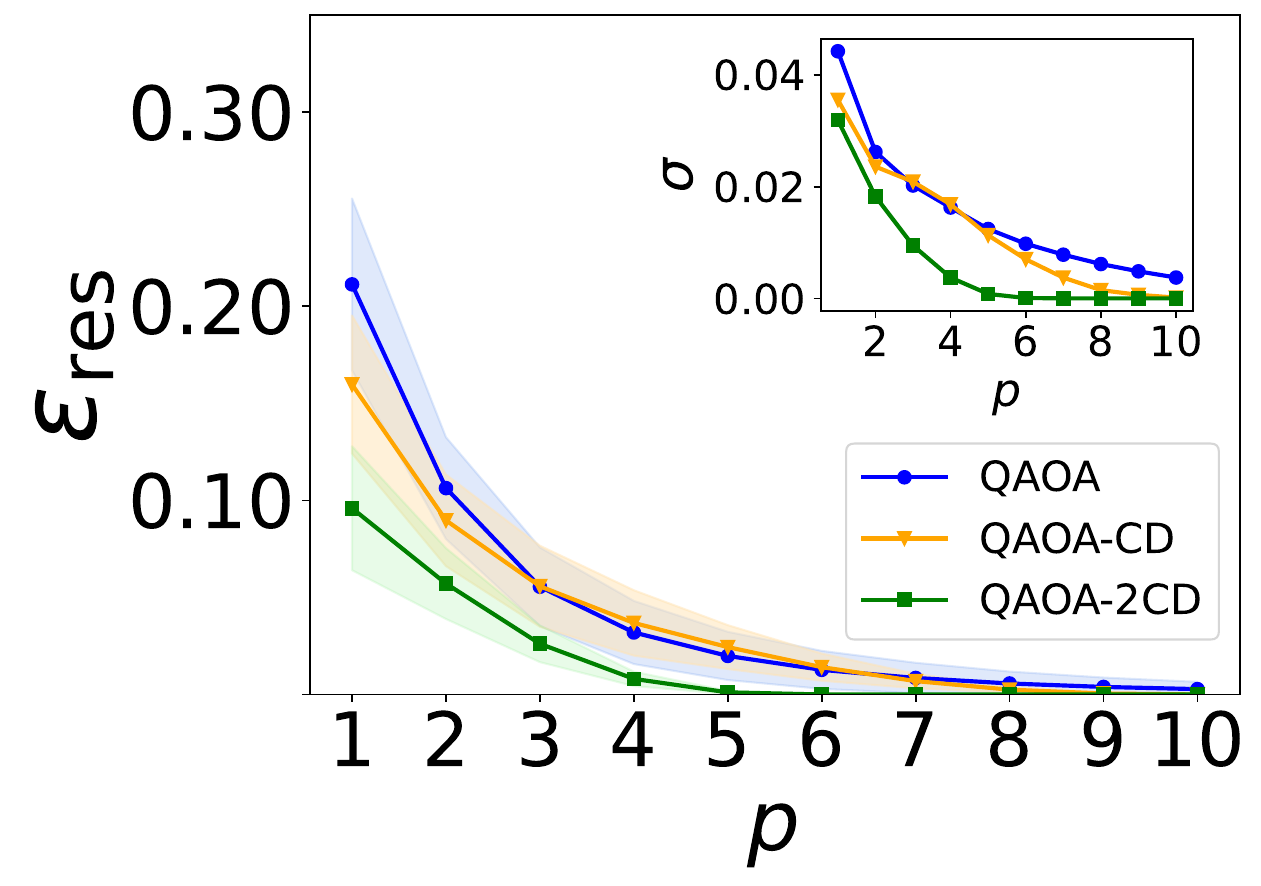}
    \includegraphics[width=0.32\textwidth]{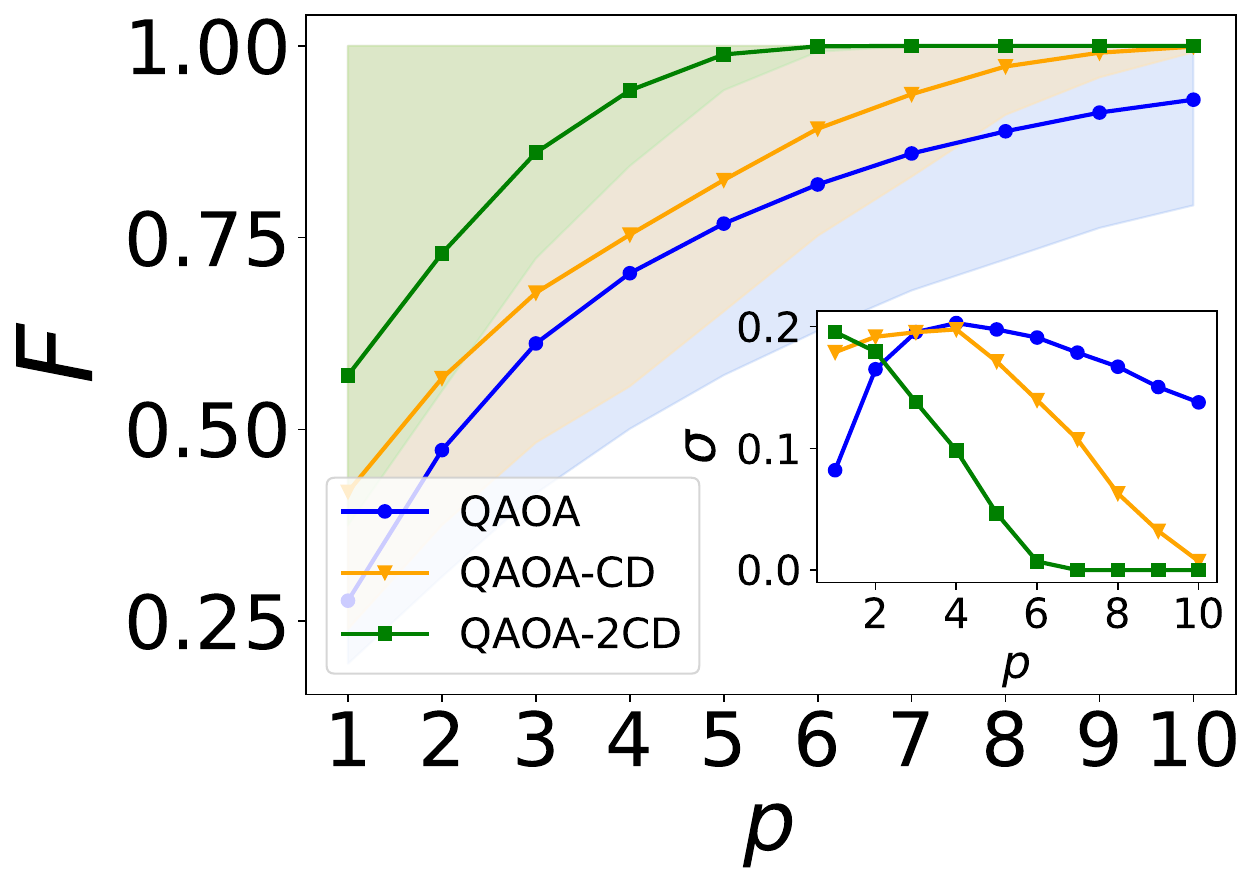}
    \caption{
    (a) Distribution of the gap between the ground state and the first excited state ($\Delta_{eg}$) among the $n=600$ instances.
(b) Residual energy and (c) fidelity in QAOA (blue curves), QAOA-CD (orange curves) and QAOA-2CD (green curves), as a function of the algorithms' step. The insets show the standard deviation for all the algorithms at each step $p$. At fixed step $p$, QAOA-2CD provides smaller residual energy and larger fidelity than the other methods.}
    \label{minima-gap-method-comparison}
\end{figure*}

First of all, in Fig.~\ref{minima-gap-method-comparison}(a),  we show $\Delta_{eg}$ for all the  $n=600$ instances, which will be used as a reference in the following. We see that the majority of random instances have small gaps, but the tail of the distribution extends up to $\Delta_{eg} \sim 5$.

Then in Figs.~\ref{minima-gap-method-comparison}(b,c) we show the residual energy and the fidelity as a function of $p$ (the number of steps) for the three considered algorithms: QAOA, QAOA-CD and QAOA-2CD. 
The full lines represent the averages over the random instances, while the shadowed areas represent the respective standard deviations. As expected, we see that in all algorithms the trial state gets closer and closer to the target ground state as the circuit depth is increased, as shown by both the residual energy decreasing towards zero and the fidelity going up to one. Overall, QAOA-2CD gives the best results not only in terms of average value but also in terms of statistical standard deviation, both for the residual energy $\varepsilon_\mathrm{res}$ and for the fidelity $F$.

In the insets of Figs.~\ref{minima-gap-method-comparison}(b,c) we also plot the standard deviations.
By increasing the circuit depth, the standard deviations always decrease, except for a nonmonotonic behavior of the standard deviation of the fidelity for QAOA and QAOA-CD at low $p$. After $p^*=6$ both the residual energy and the fidelity of the QAOA-2CD reach their limiting values, with negligible standard deviation, which means that we have reached the exact solution for almost all the instances.

\begin{figure}
    (a)\hspace{0.4\columnwidth}(b)\hspace*{0.4\columnwidth}\hfill\\
    \centering
    \includegraphics[width=0.49\columnwidth]{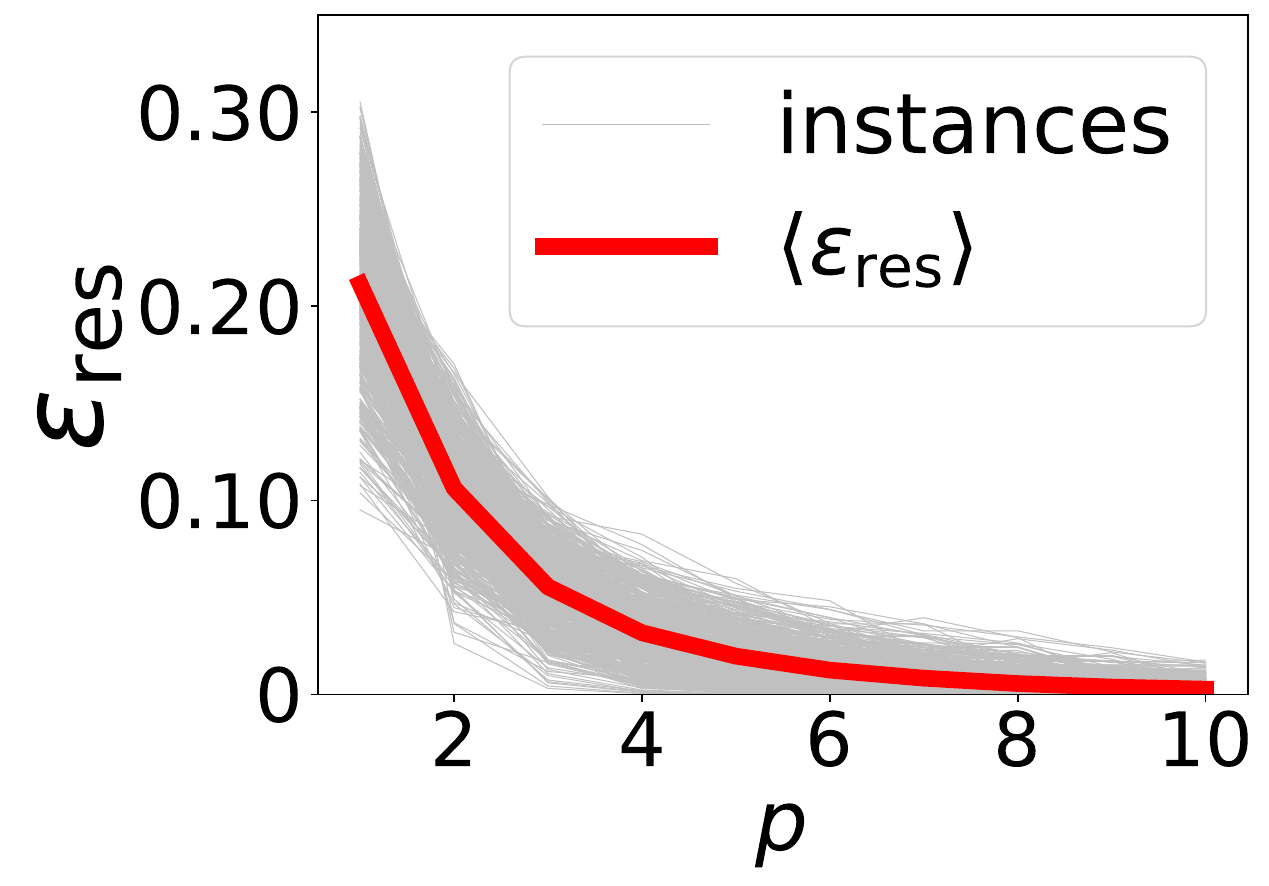}
    \includegraphics[width=0.49\columnwidth]{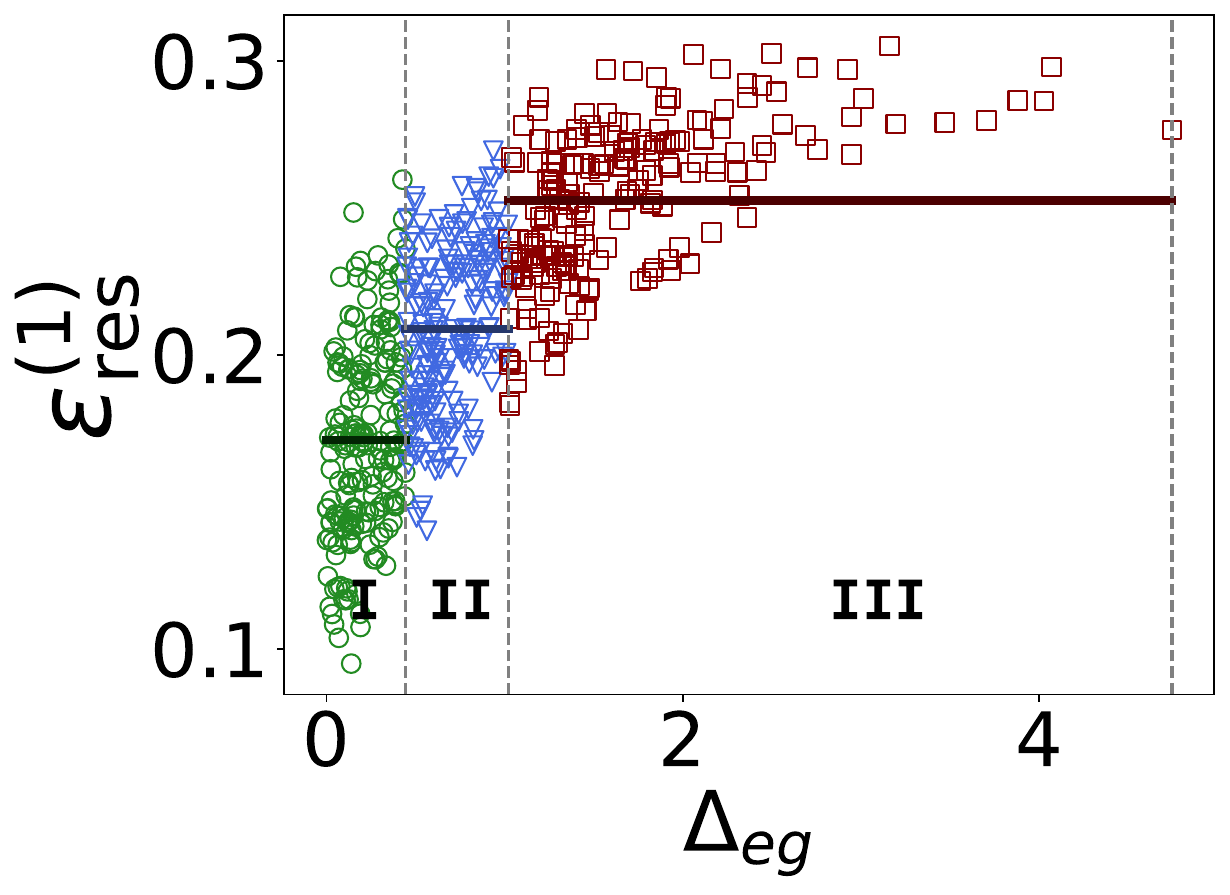}
    (c)\hspace{0.4\columnwidth}(d)\hspace*{0.3\columnwidth}\hfill\\
    \centering
    \includegraphics[width=0.49\columnwidth]{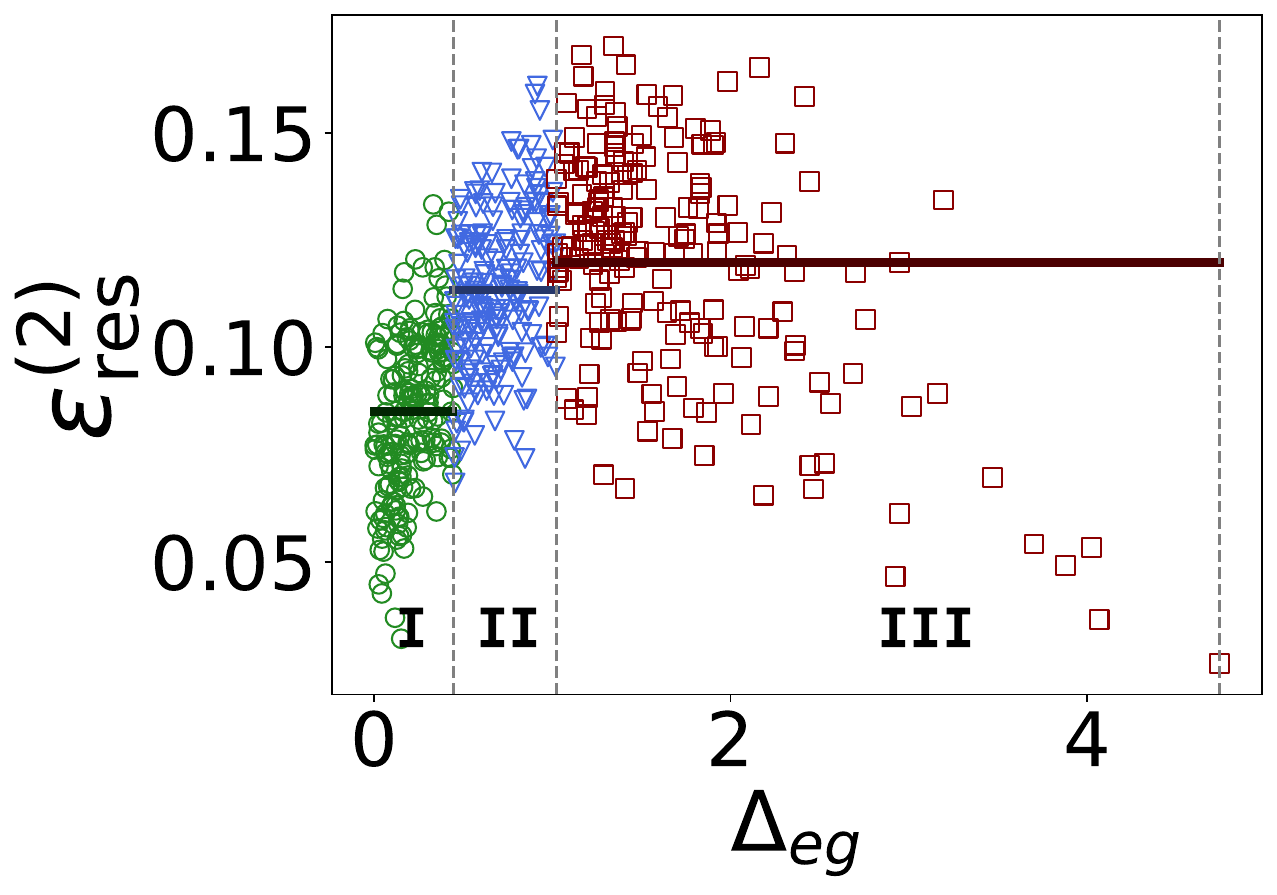}
    \includegraphics[width=0.49\columnwidth]{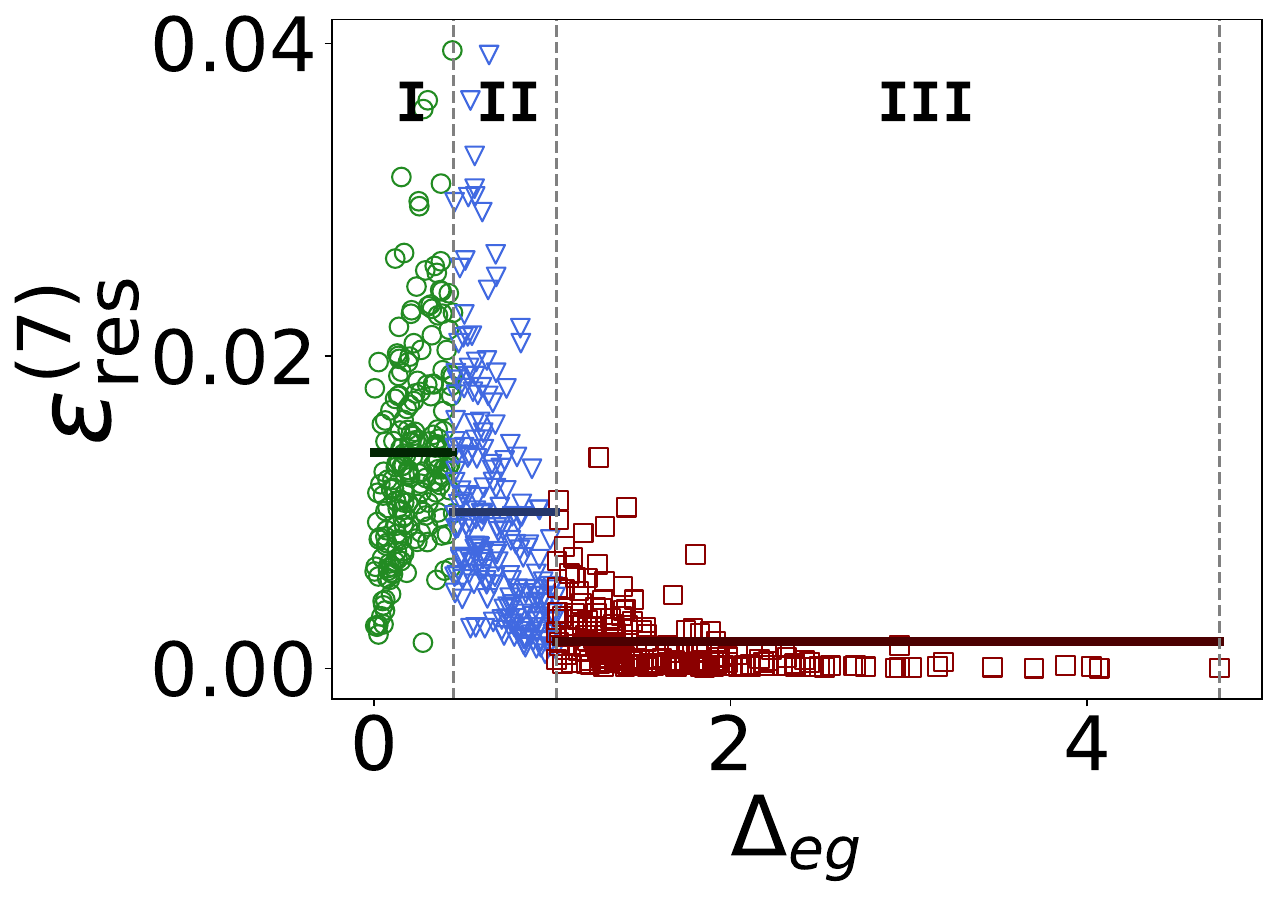}
\caption{(a) Residual energy versus QAOA step $p$. Grey lines represent $\varepsilon_{\mathrm{res,}i}^{(p)}$ of all the $n=600$ random instances and the red line their average value $\braket{\varepsilon_\mathrm{res}}$. (b-d) Residual energy $\varepsilon_{\mathrm{res,}i}^{(p)}$ as a function of the minimal gap $\Delta_{eg}$ for QAOA at steps (b) $p=1$, (c) $p=2$ and (d) $p=7$. Three distinct regions based on the energy gap separating the ground state from the first excited state can be singled out for each instance: \textbf{I}-zone {(green circles)}, corresponding to instances with $\Delta_{eg}\in\left[0,0.44\right]$, \textbf{II}-zone {(blue triangles)} for instances with $\Delta_{eg}\in\left[0.44,1.02\right]$, \textbf{III}-zone (red squares) for instances with $\Delta_{eg}\in\left[1.02,4.74\right]$. The horizontal line within each sector depicts the corresponding average value of that sector. We see that for $p=1$  these average values grow with increasing of $\Delta_{eg}$. This trend starts to change from step $p=2$ (panel c), whereas \textbf{III}-zone (panel d) shows a trend inversion (average value decreasing with the minimal gap). 
}
    \label{eres-QAOA}
\end{figure}

\begin{figure}[t]
    (a)\hspace{0.4\columnwidth}(b)\hspace*{0.3\columnwidth}\hfill\\
    \centering
    \includegraphics[width=0.49\columnwidth]{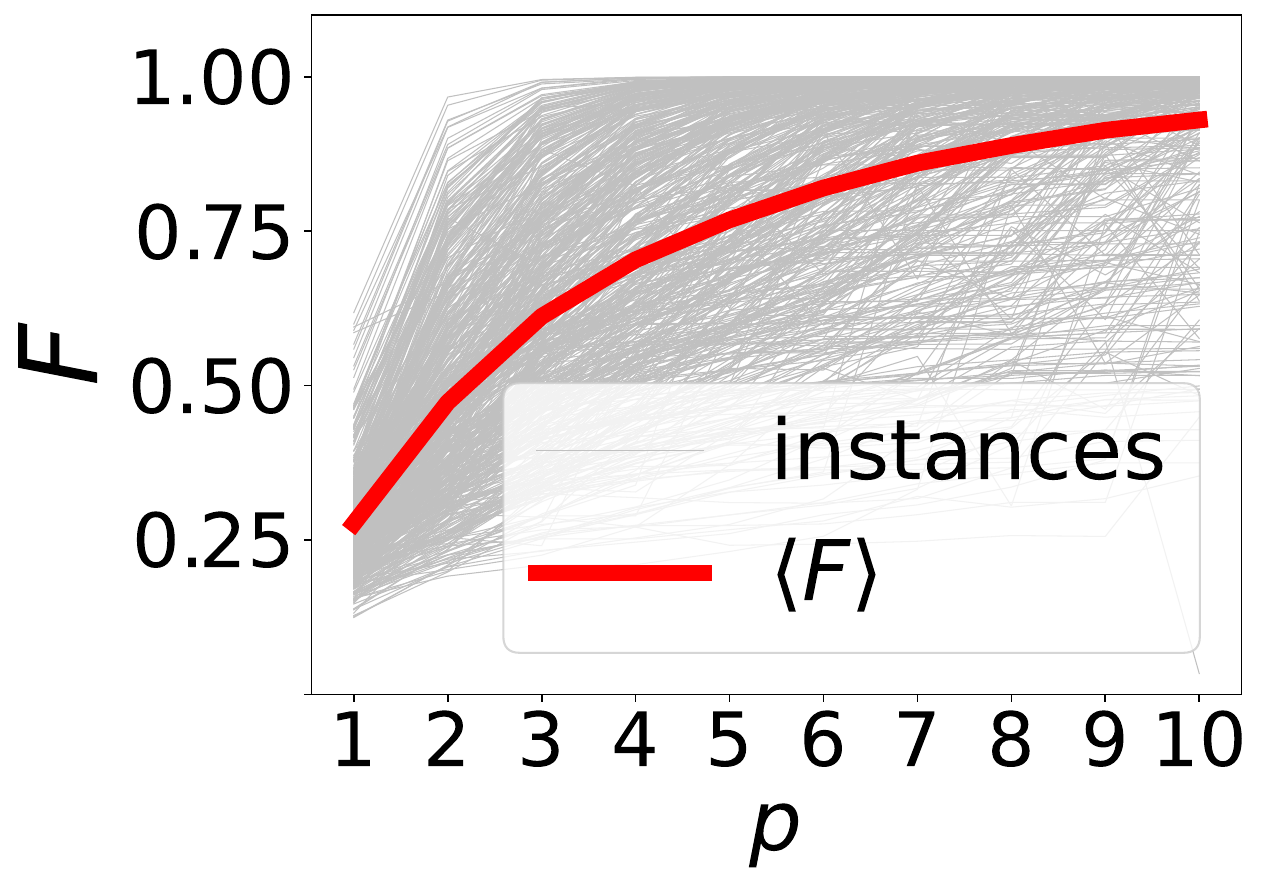}
    \includegraphics[width=0.49\columnwidth]{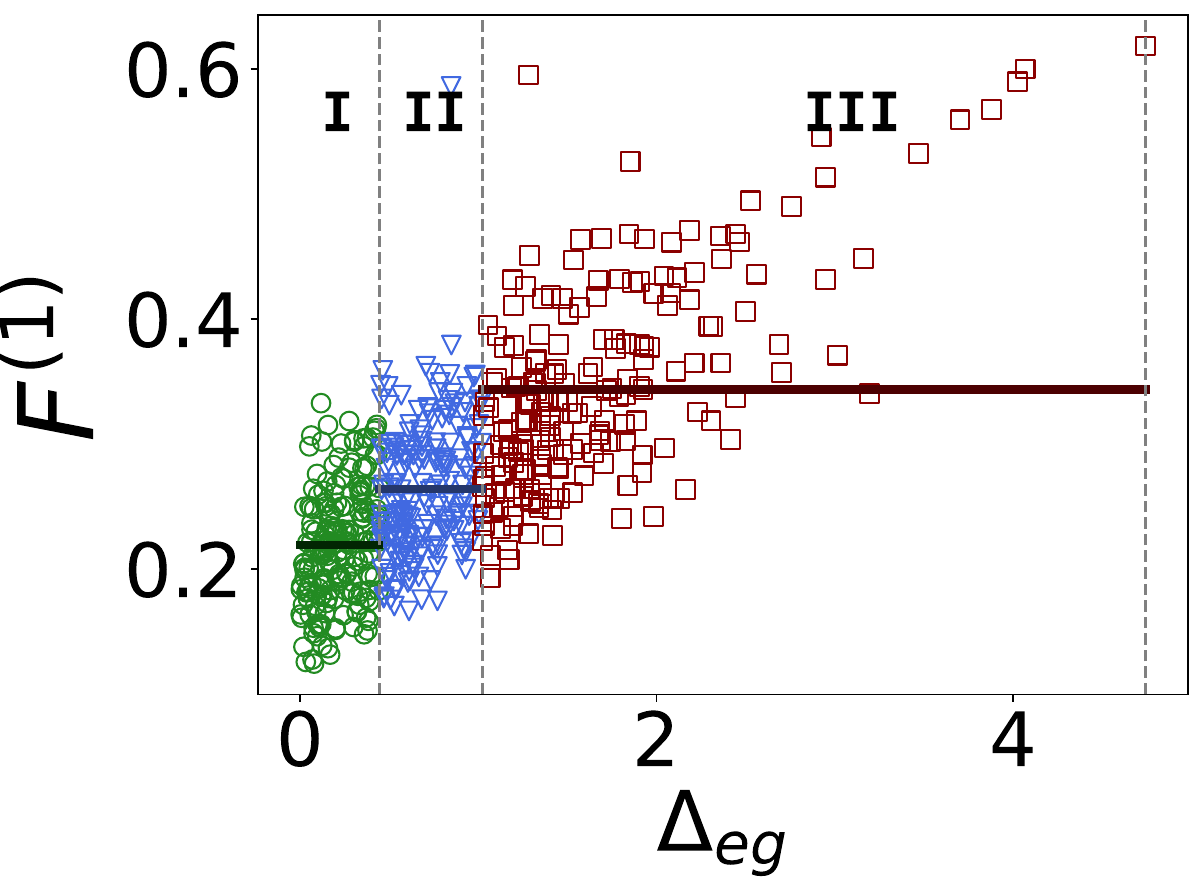}
    \\(c)\hspace{0.4\columnwidth}(d)\hspace*{0.3\columnwidth}\hfill\\
    \centering
    \includegraphics[width=0.49\columnwidth]{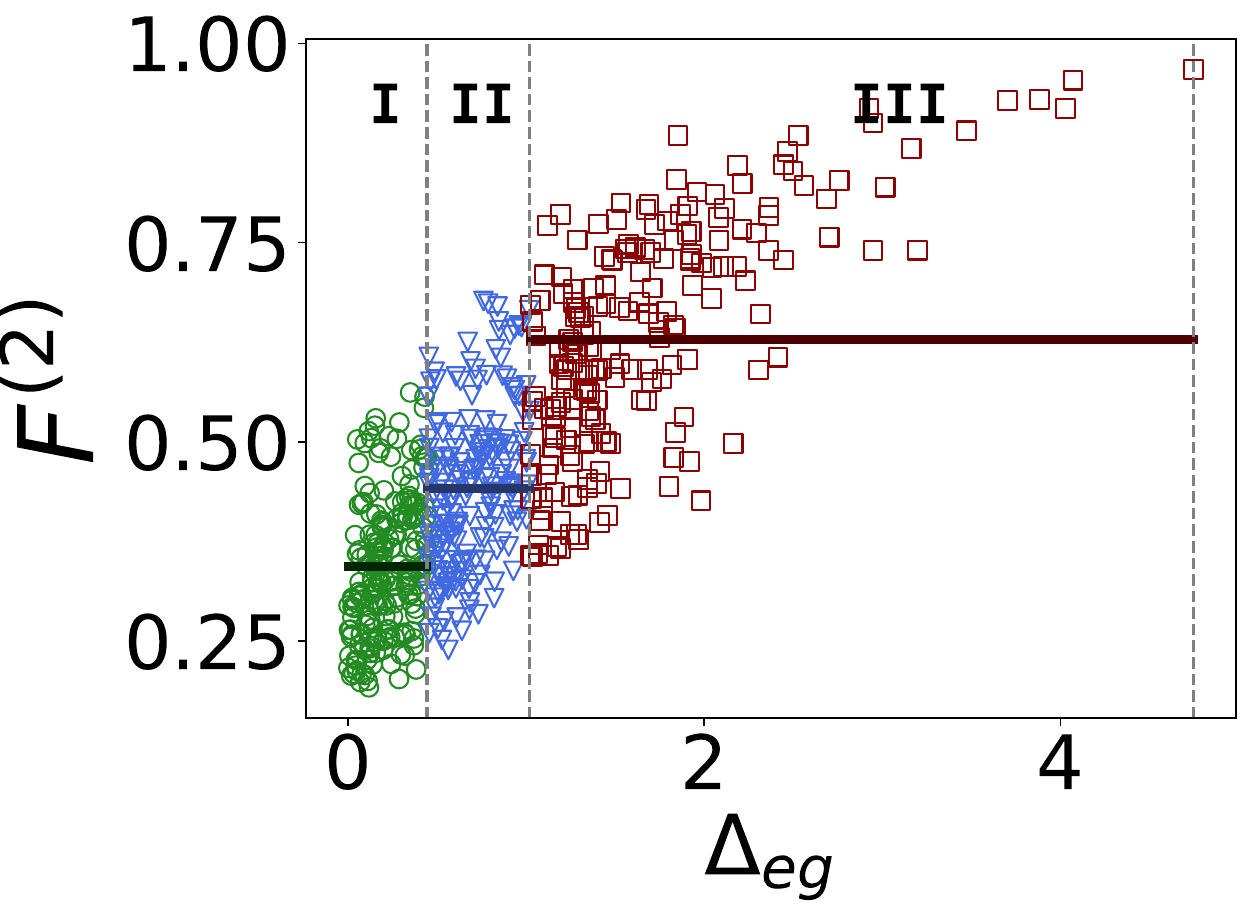}
    \includegraphics[width=0.49\columnwidth]{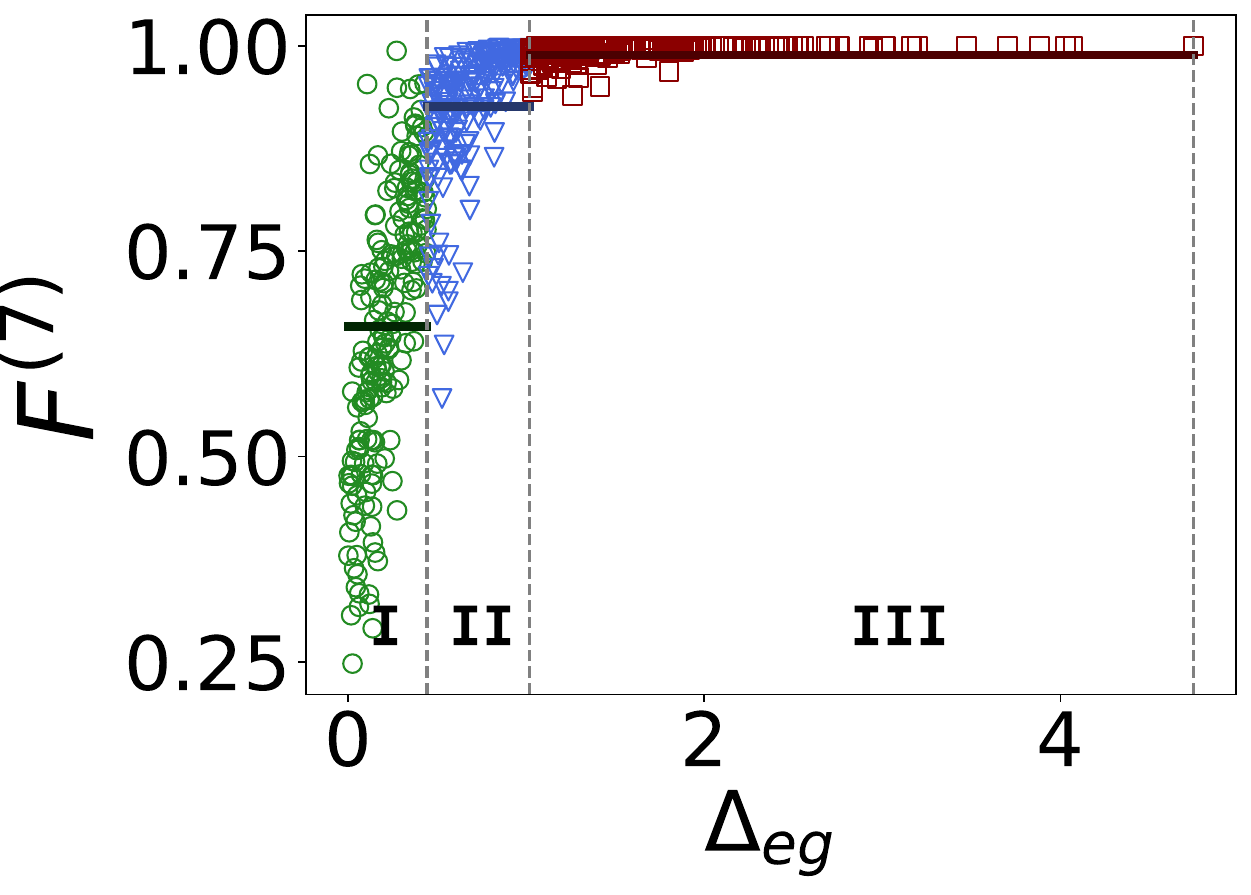}
    \caption{(a) Fidelity versus QAOA step $p$. Grey lines represent values $F_i^{(p)}$ of all $n=600$ random instances and the red line their average value $\braket{F}$. (b-d) Fidelity $F_i^{(p)}$ as a function of the minimal gap $\Delta_{eg}$ for QAOA at step (b) $p=1$, (c) $p=2$ and (d) $p=7$. In panels (b-d) the three distinct regions described in caption of Fig.~\ref{eres-QAOA} are evidenced. The horizontal line within each sector depicts the corresponding average value of that sector. For each step $p$ the fidelity closest to one is obtained for instances with $\Delta_{eg}$ belonging to \textbf{III}-zone. 
    }
    \label{fidelity-QAOA}
\end{figure}

\subsection{QAOA}
In Fig.~\ref{eres-QAOA}(a) we show the residual energy as a function of the number $p$ of iterations of QAOA. Each grey line corresponds to the residual energy $\varepsilon_{\mathrm{res,}i}^{(p)}$ for a random instance, while the red  bold curve indicates the average residual energy $\braket{\varepsilon_{\mathrm{res}}}$ over all the $n$ instances.  The average residual energy  at the step $p=10$ is of the order of $\braket{\varepsilon_\mathrm{res}^{(10)}}\simeq (2 \pm 3)\cdot 10^{-3}$,     
consistent with a zero residual energy at large $p$ which means that, on average, QAOA gives a good approximation of the exact solution at $p=10$.
However the finite standard deviation indicates that some of the instances have yet to converge to their respective ground states, justifying the need for counterdiabatic corrections at finite $p$. 

Is it possible to identify a common pattern among the instances that converge more slowly? In order to answer this question, we investigate the connection between the gap $\Delta_{eg}$ and complexity of the algorithm, following what is done in  Ref.~\cite{bishop2023set}  for quantum annealing. In Figs.~\ref{eres-QAOA}(b--d), we show scatter plots of the residual energy of each instance as a function of its gap $\Delta_{eg}$, at some fixed steps $p$. In particular, \commentref{we split the sorted gaps into three sets of equal cardinality (200 instances per set)}:
in \textbf{I} there are  problems with $\Delta_{eg}\in\left[0,0.44\right]$, in \textbf{II} problems with
$\Delta_{eg}\in\left[0.44,1.02\right]$ and in \textbf{III}  problems with $\Delta_{eg}\in\left[1.02,4.74\right]$. %
In panel (b) we show the scatter plots for  $p=1$, in (c) those for $p=2$ and in (d)  those for $p=7$.
In panel (b), with $p=1$, we see that the larger is the gap the bigger is the residual energy; this behavior is confirmed by the average value of residual energy in each region. \commentref{The reason is that, at the first steps of QAOA, the parameters are not sufficient to accurately describe the problem’s ground state. As a result, higher energy components of the QAOA variational guess over the computational basis states are detrimental for the algorithm.}
Starting from step $p=2$ we note that this trend reverses and becomes more and more marked as the number of steps $p$ increases, as we see comparing panels (c) and (d).
In panel (c) the average residual energies of the regions \textbf{II} and \textbf{III} have similar values, then in the panel (d), for $p=7$, the average
in the region \textbf{III} is  lower than that in region  \textbf{I}. The numerical values of the average residual energies in each region can be found in Appendix~\ref{app:numerics}.

The connection between $\Delta_{eg}$ and the performance of the algorithm becomes more evident if we look at the fidelity, shown in Fig.~\ref{fidelity-QAOA}.
In Fig.~\ref{fidelity-QAOA}(a) we show the fidelity of QAOA as a function of the step $p$, and in Figs.~\ref{fidelity-QAOA}(b--d) we show scatter plots of the fidelity as a function of the gap $\Delta_{eg}$ at fixed step.  In panel (a) we see that the average fidelity at $p=10$ is close to one, but there is a large standard deviation $\sigma_F$. This behavior differs from $\varepsilon_\mathrm{res}$,
for which QAOA at step $p=10$ provides a value much closer to its limit value and a much smaller standard deviation.

As we mentioned above, the residual energy measures the distance of the QAOA cost function from the true ground state energy, while the fidelity returns the overlap between the
trial state at the end of the optimization procedure and the exact ground state.
We can expect that the performance of the algorithm in terms of residual energy will appear to be generally better than that in terms of the fidelity
both in terms of averages and standard deviations at any fixed step.
Indeed, the algorithm is conceived to give the best answer for the residual energy, whereas the fidelity is computed after the optimal values $\vec{\beta}^*$ and $\vec{\gamma}^*$ at the given step $p$ are obtained.
Our results confirm such a picture, as shown in Appendix~\ref{app:numerics}.
A similar picture is also valid for the other analyzed QAOA variants. On the other hand, comparing the results of Figs.~\ref{eres-QAOA} and~\ref{fidelity-QAOA}  we can notice that in the case of the fidelity, unlike the residual energy, starting from $p=1$ the instances corresponding to larger $\Delta_{eg}$ yield a better fidelity, showing an increasing trend with respect to the value of $\Delta_{eg}$.

\begin{figure}
    (a)\hspace{0.45\columnwidth}(b)\hspace*{0.4\columnwidth}\hfill\\
    \centering
    \includegraphics[width=0.49\columnwidth]{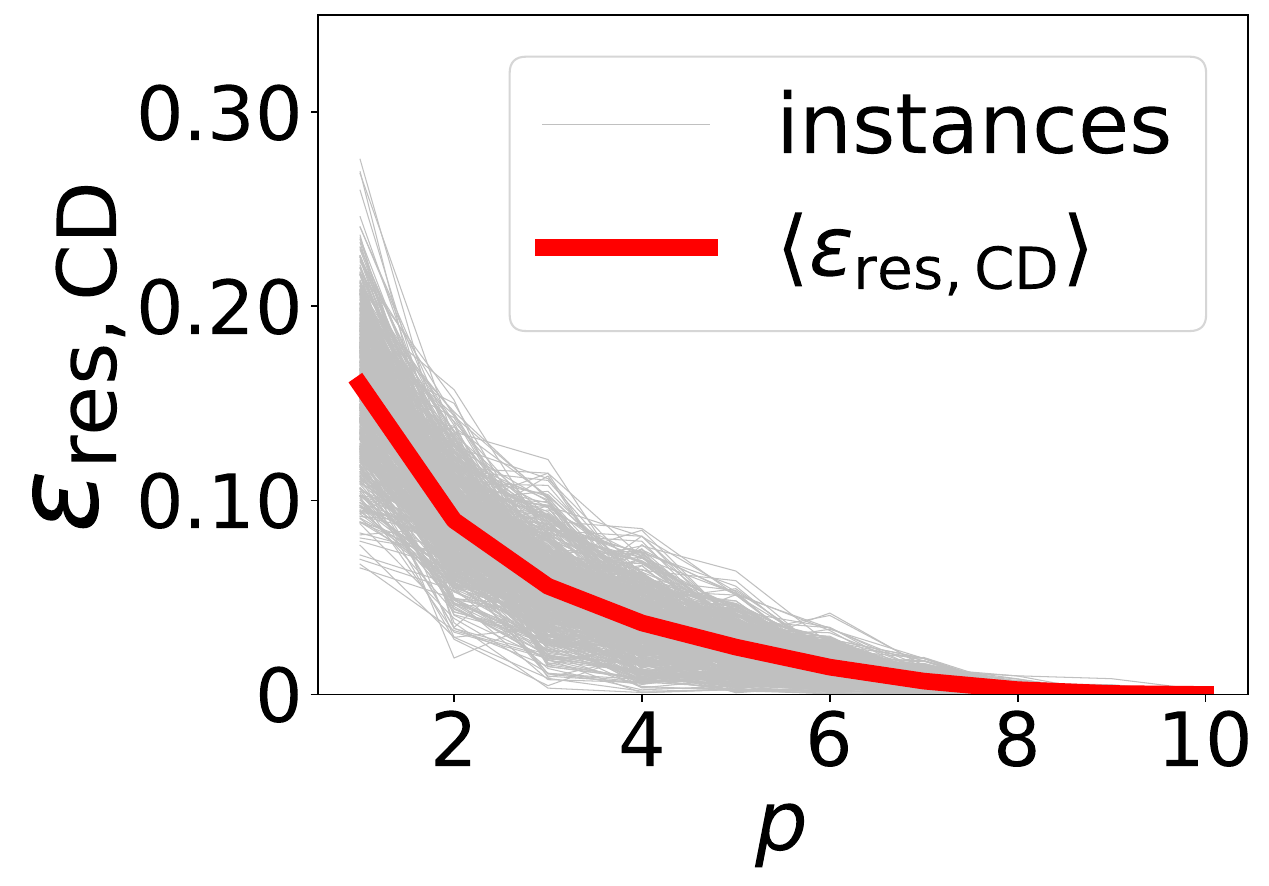}
    \includegraphics[width=0.49\columnwidth]{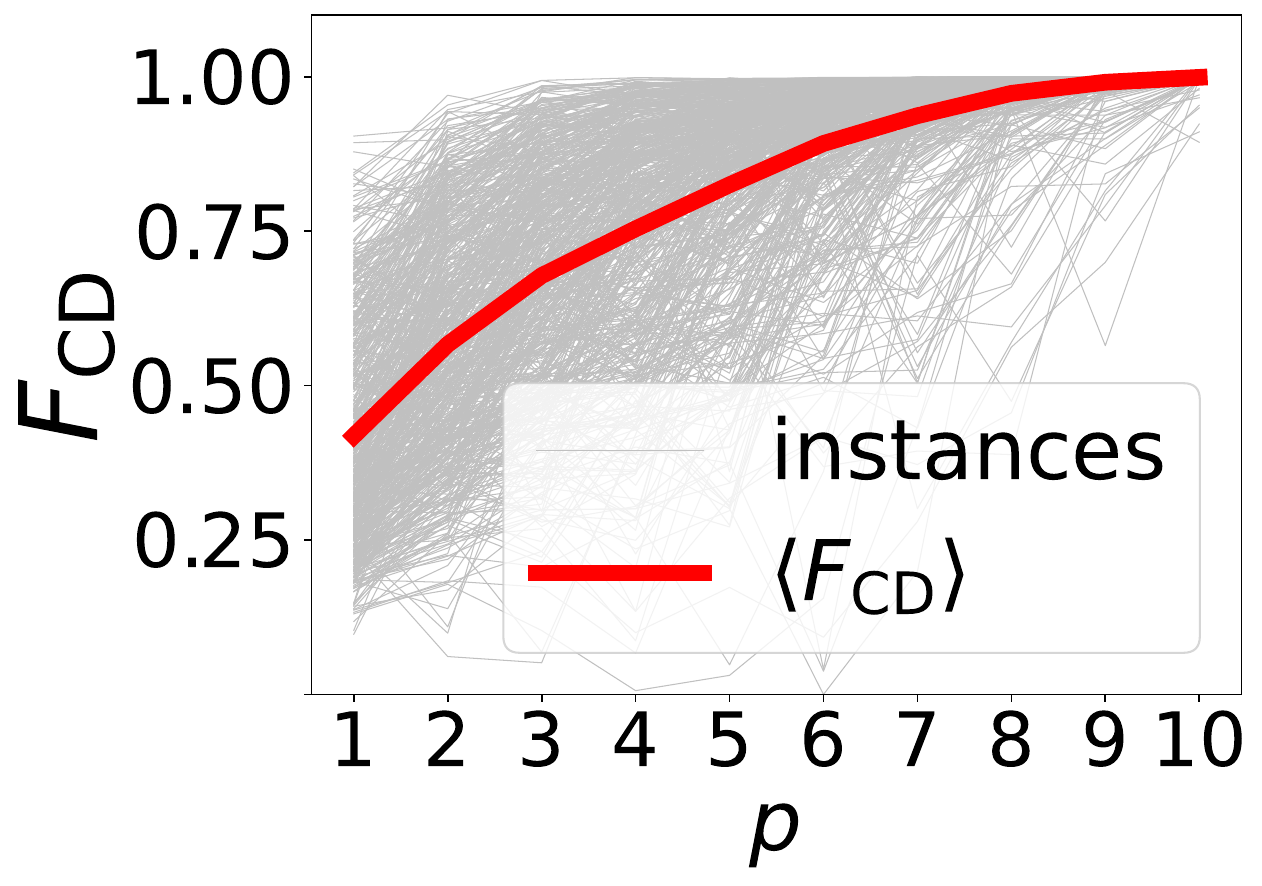}
    (c)\hspace{0.45\columnwidth}(d)\hspace*{0.4\columnwidth}\hfill\\
    \centering
    \includegraphics[width=0.49\columnwidth]{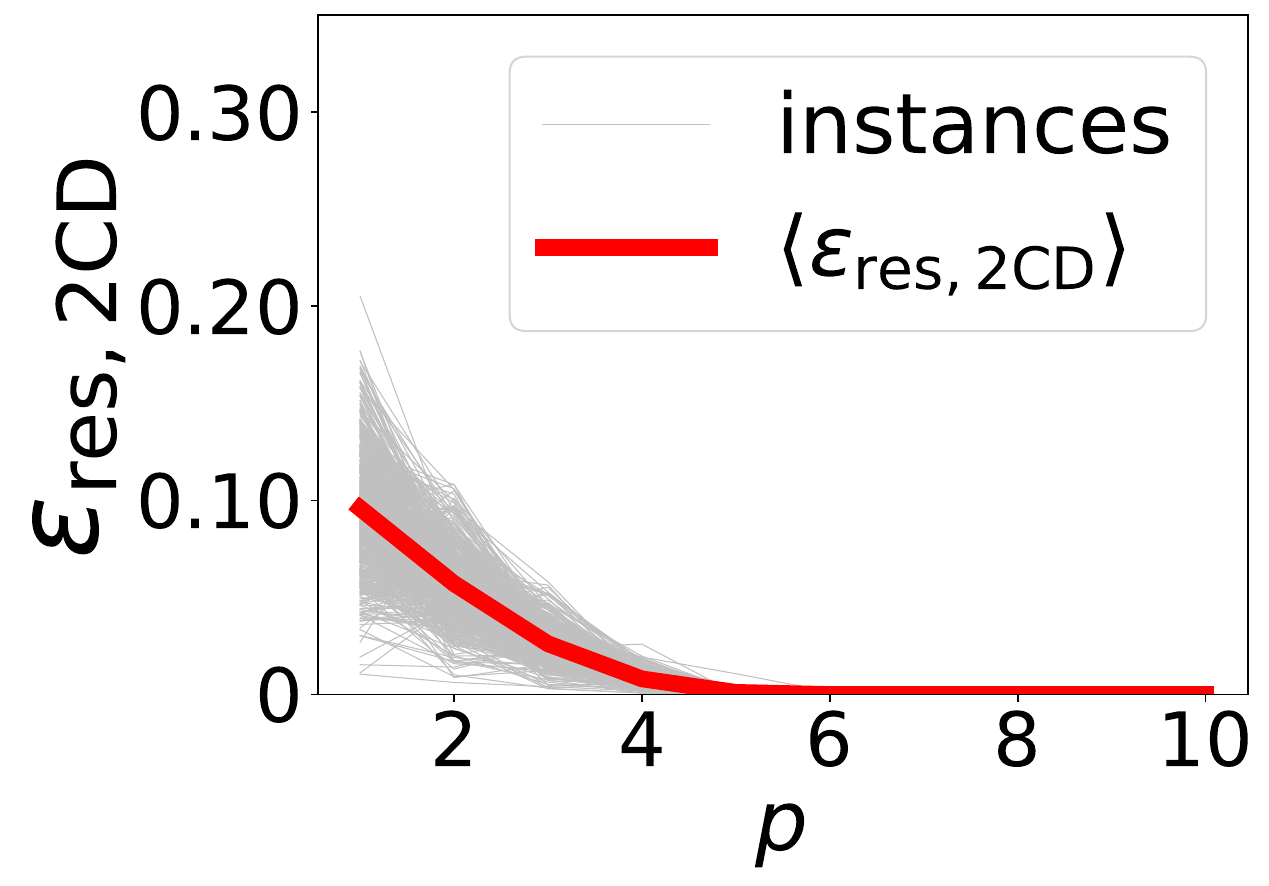}
    \includegraphics[width=0.49\columnwidth]{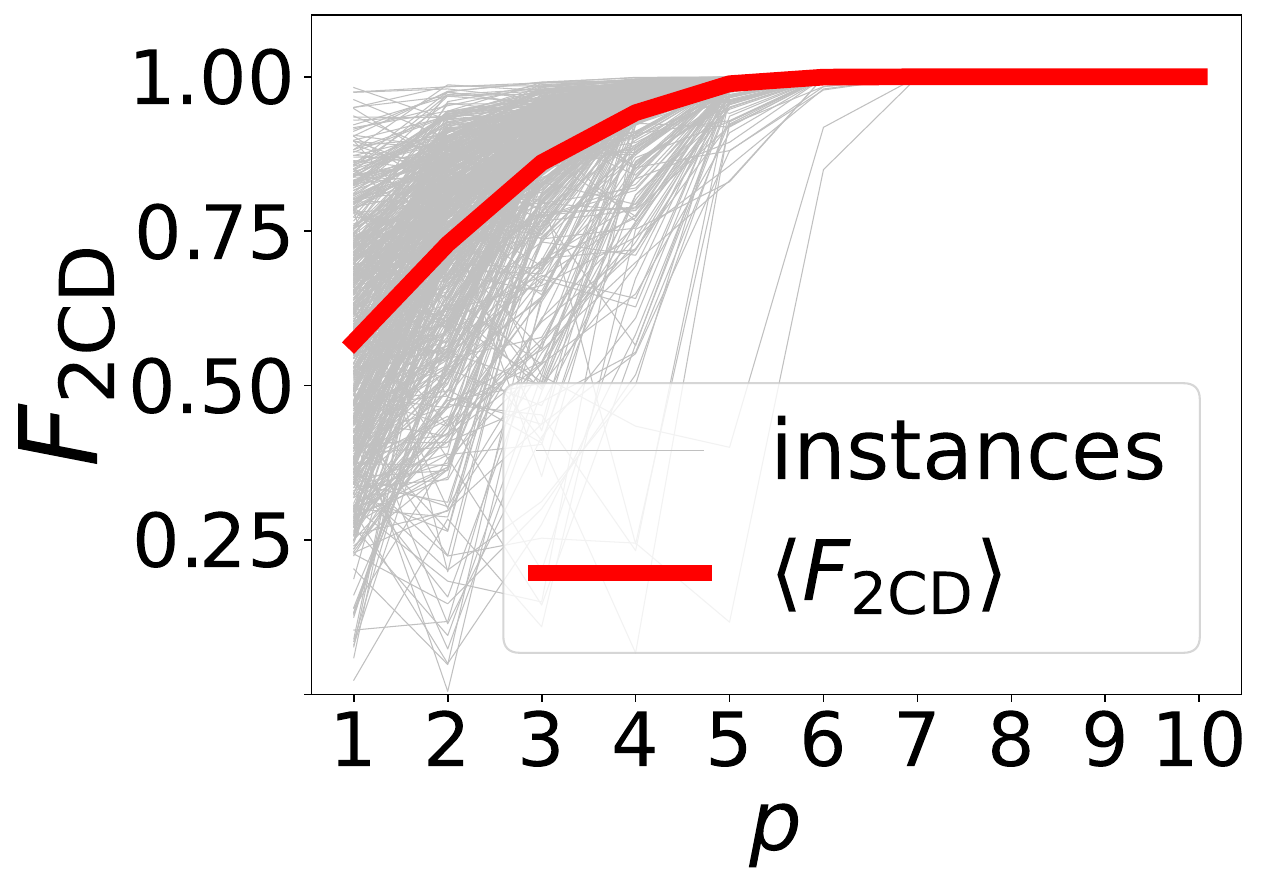}
    \caption{%
    (a) Residual energy and (b) fidelity of QAOA-CD versus step $p$. (c) Residual energy and (d) fidelity of QAOA-2CD versus $p$. The grey lines represent values of all the instances and the red line represents their respective averages.
    }
    \label{instances-QAOA-variants}
\end{figure}

In the context of Quantum Annealing (QA), the relationship between the minimum gap along the dynamics and the complexity of the algorithm is well established. This is due to the adiabatic theorem, as outlined for instance, in Refs.~\cite{kato1950,albash:2018}, which indicates that, during dynamics, the probability of having Landau-Zener transitions increases when the gap becomes smaller.
However, this result is unexpected for QAOA, where we are not making a real time dynamics, but we only care about the ``final time'' results.
Moreover, QAOA and QA are strictly related only in the $p\to \infty$ limit, which is not the case here.  Therefore, in principle, we cannot fully explain the relationship between $\Delta_{eg}$ and the complexity of the algorithm in terms of the connection between QA and QAOA.

However, the correlation between $\Delta_{eg}$ and the values of residual energy and fidelity in QAOA could be related to the fact that in the presence of a dense spectrum of the Hamiltonian $H_T$ there is a large likelihood of identifying local minima in close proximity to the ground state energy. \commentref{To better understand why small gaps influence the performance of QAOA, we analyzed the path followed by our optimizer routine for three instances characterized by different values of the spectral gap, one per each zone, at the first step of QAOA, where the parameter space is two-dimensional and can be visualized. Our analysis shows that in all three cases the minimizer can quickly find the global minimum regardless of the gap, but the variational ansatz itself is a worse approximation of the true ground state when the minimum gap is smaller.} \commentref{In order to better understand how the gap is important in the minimizer, the end of section \ref{sub:QAOA-CD and QAOA-2CD} focuses on different roles played by the residual energy and the fidelity in finding the solution for QAOA and QAOA-2CD.}

\subsection{QAOA-CD and QAOA-2CD}\label{sub:QAOA-CD and QAOA-2CD}

In this section, we will investigate whether QAOA-CD and QAOA-2CD confirm the dependence on the $H_T$ minimum gap found in QAOA. Furthermore, these algorithms will be compared to find out which one of them yields the optimal results in terms of average value and standard deviation for the residual energy and the fidelity.  
In order to make a fair comparison, we apply  QAOA-CD and QAOA-2CD to the same instances of the fully-connected spin model analyzed for QAOA.

\begin{figure*}
  \includegraphics[width=\textwidth]{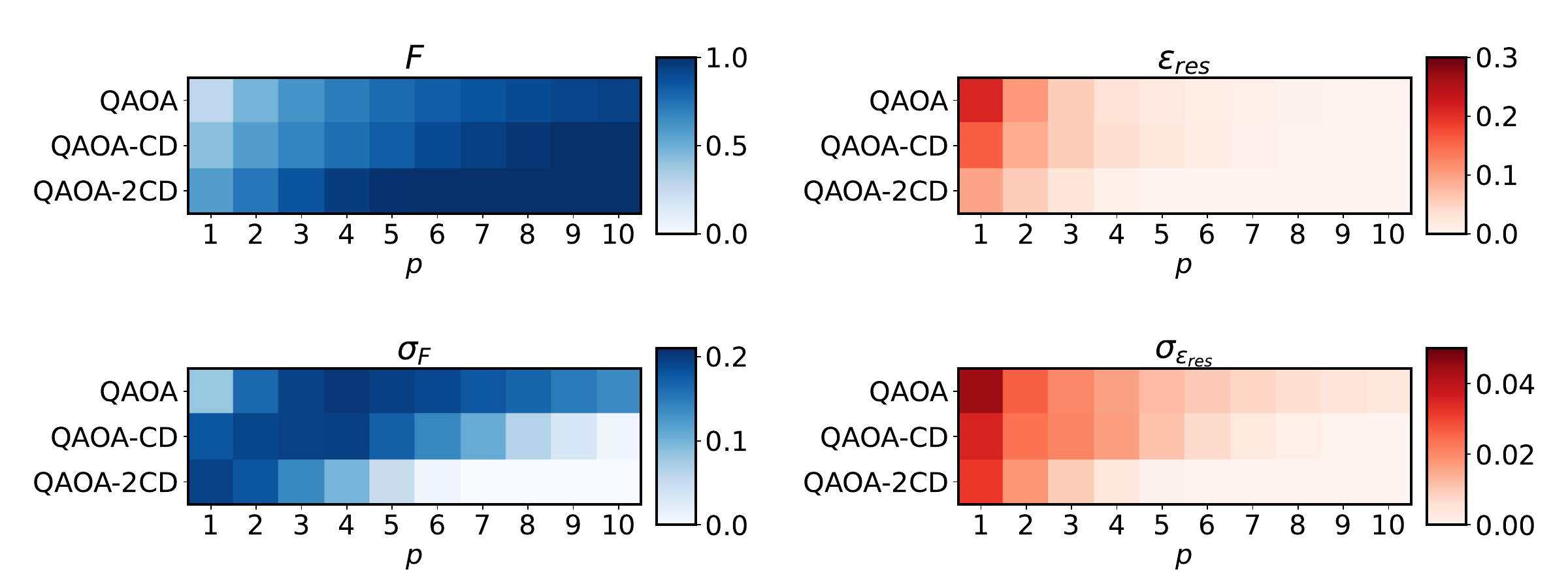} 
  \caption{Fidelity $F$ (top left), its standard deviation $\sigma_F$ (bottom left), residual energy $\varepsilon_\text{res}$ (top right) and its standard deviation $\sigma_{\varepsilon_\text{res}}$ (bottom right) as a function
  of the step $p$, as calculated with all three methods (QAOA, QAOA-CD, QAOA-2CD) after averaging over the $n= 600$ random instances. We clearly see the residual energy decreasing towards zero and the fidelity increasing
  towards one as the step increases. The best performances are obtained with QAOA-2CD.}
  \label{fig:F_and_eres}
\end{figure*}

In Fig.~\ref{instances-QAOA-variants}, we present the residual energies and fidelities as a function of $p$ for both QAOA-CD and QAOA-2CD. The grey lines represent the single instances, while the red ones represent the average values. Fig.~\ref{instances-QAOA-variants}(a,b) shows the performance of QAOA-CD. We observe that there is a step $p=8$ in the residual energy plot after which the standard deviation is negligible.
This feature has not been observed in standard QAOA, see Fig.~\ref{eres-QAOA}. The QAOA-CD fidelity at $p=10$ has a smaller standard deviation than the QAOA one and converges to a value very close to one.

In Fig.~\ref{instances-QAOA-variants}(c,d) we present the residual energy and the fidelity for QAOA-2CD. Similarly to QAOA-CD, there is also a step $p$ for the residual energy in QAOA-2CD beyond which the standard deviation becomes close to zero. In this case, specifically, this happens at $p = 5$, earlier than for QAOA-CD. 

Concerning the fidelity, in contrast to QAOA and QAOA-CD, where the standard deviation always remains nonzero up to $p=10$, the fidelity in QAOA-2CD has a negligible error beyond the $p=6$ step.
In particular, before convergence, at the $p = 6$ step it is about ten times smaller than that found at the same step in QAOA and QAOA-CD. 

\commentref{Instances with $\Delta_{eg} \sim 0$ exhibit quasi-degeneracy between the ground and first excited states, leading to the QAOA state converging to a solution that reflects an overlap between these states [see Fig.~\ref{fidelity-QAOA}(a)]. This behavior contrasts with the QAOA-2CD state, where, by step $p=7$, the fidelity reaches $1$ [see Fig.~\ref{instances-QAOA-variants}(d)]. From this observation, it appears that the role of the counterdiabatic correction in QAOA is analogous to that of the counterdiabatic potential in quantum annealing: both prevent transitions from the ground state to excited states, regardless of the system’s gap $\Delta_{eg}$. The key distinction is that in QA, gaps are dynamical, while in QAOA, $\Delta_{eg}$ pertains to the gap of the target Hamiltonian $H_T$.}

The residual energy decreases by increasing the order of the BHC expansion, following the opposite behavior as the fidelity. However, before convergence, its standard deviation is smaller than the one of the fidelity, as shown explicitly in Appendix~\ref{app:numerics}, where we report the numerical values of the average fidelities and residual energies in the three algorithms. Here in the main text we instead  summarize these results visually in Fig.~\ref{fig:F_and_eres}, where we plot the fidelity, the residual energy and their respective standard deviations in the form of heatmaps, for all analyzed algorithms and steps $p$. We generally see that QAOA-2CD is the algorithm yielding the best performances in terms of both metrics. For instance, the darker area in the top left panel, representing fidelities close to one, and the lighter area in the top right panel, showing small values of the residual energy, both occur at large steps $p$ in correspondence of the QAOA-2CD line, where, correspondingly, their respective standard deviations vanish.

\begin{figure}[b]
    (a)\hspace{0.4\columnwidth}(b)\hspace*{0.3\columnwidth}\hfill\\
    \centering
    \includegraphics[width=0.45\columnwidth]{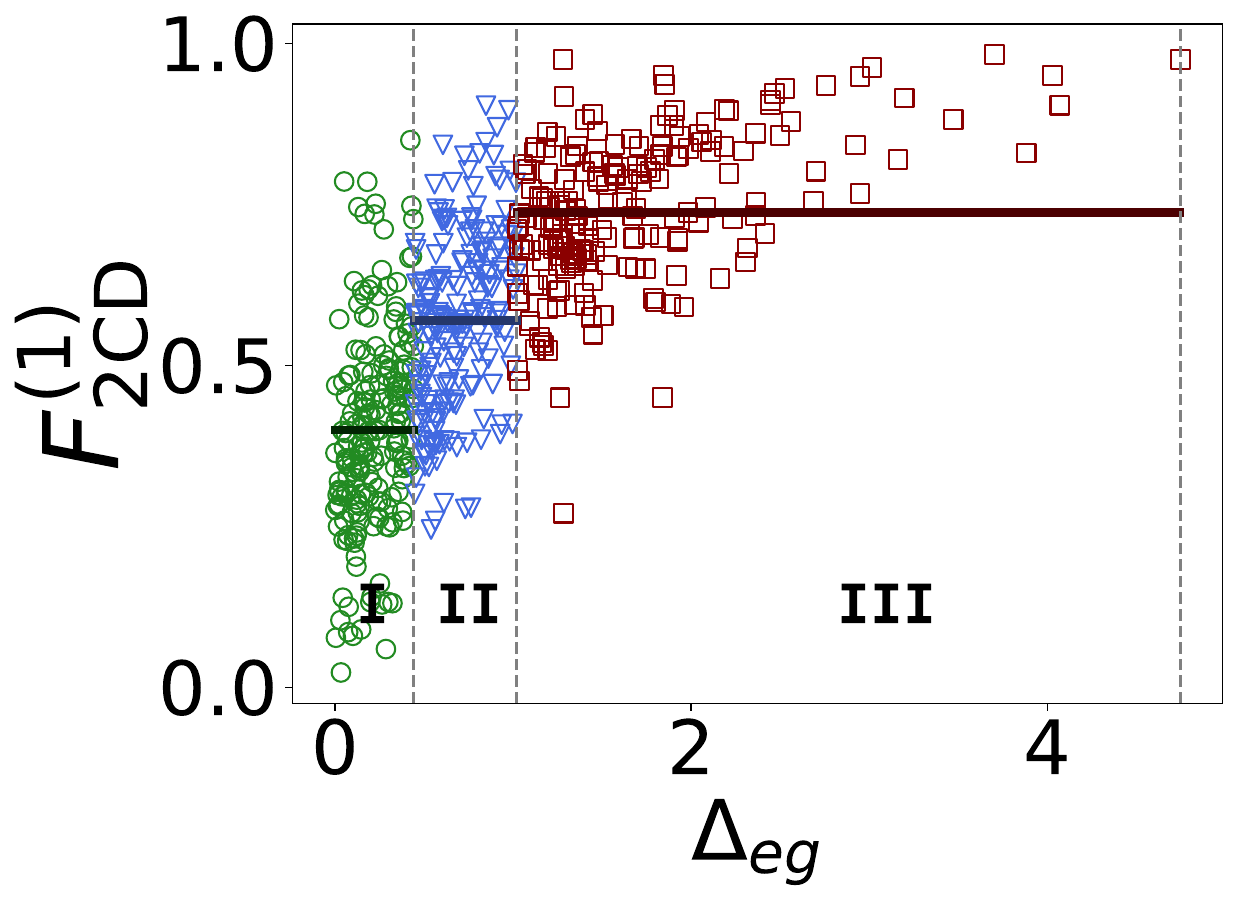}
    \includegraphics[width=0.45\columnwidth]{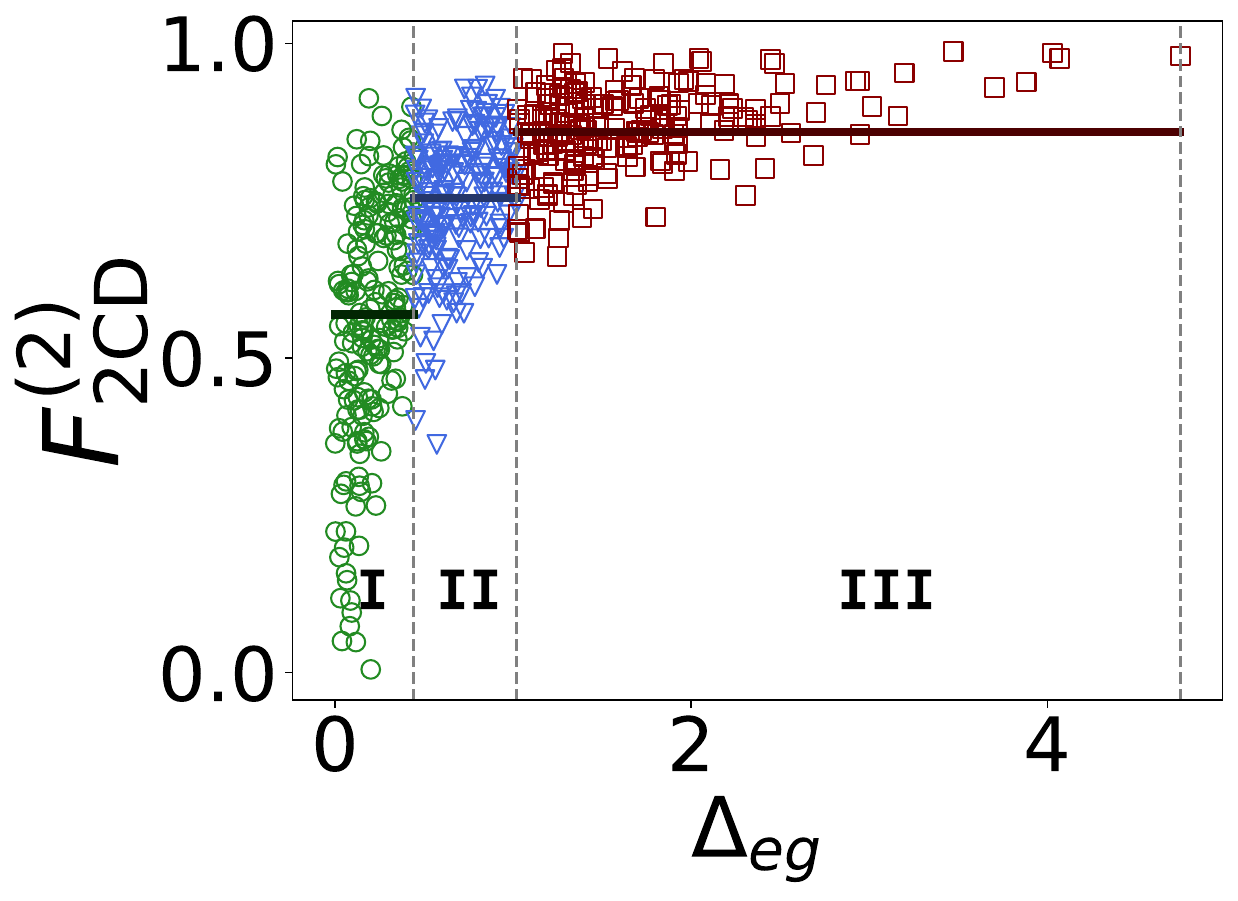}
    \centering
    \\(c)\hspace{0.4\columnwidth}(d)\hspace*{0.3\columnwidth}\hfill\\
    \includegraphics[width=0.45\columnwidth]{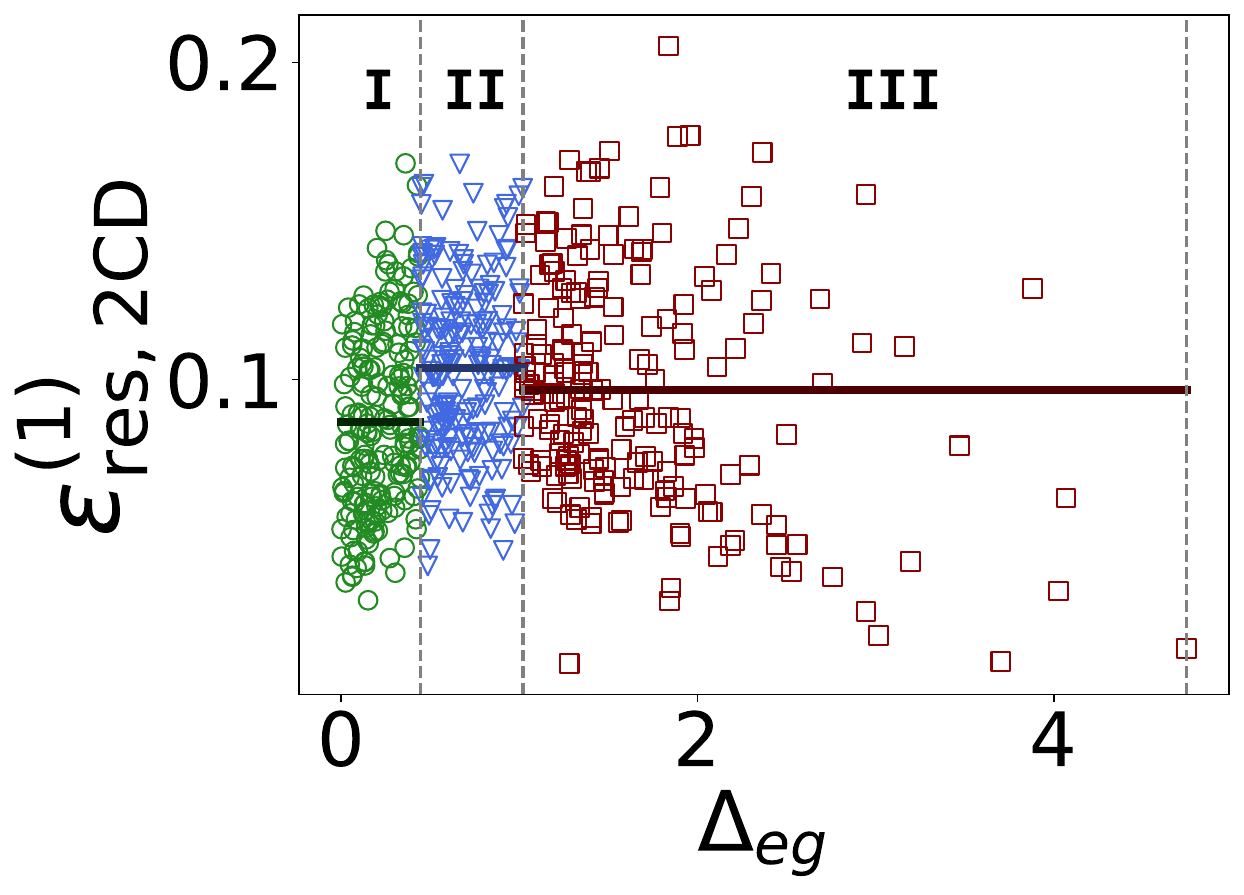}
    \includegraphics[width=0.45\columnwidth]{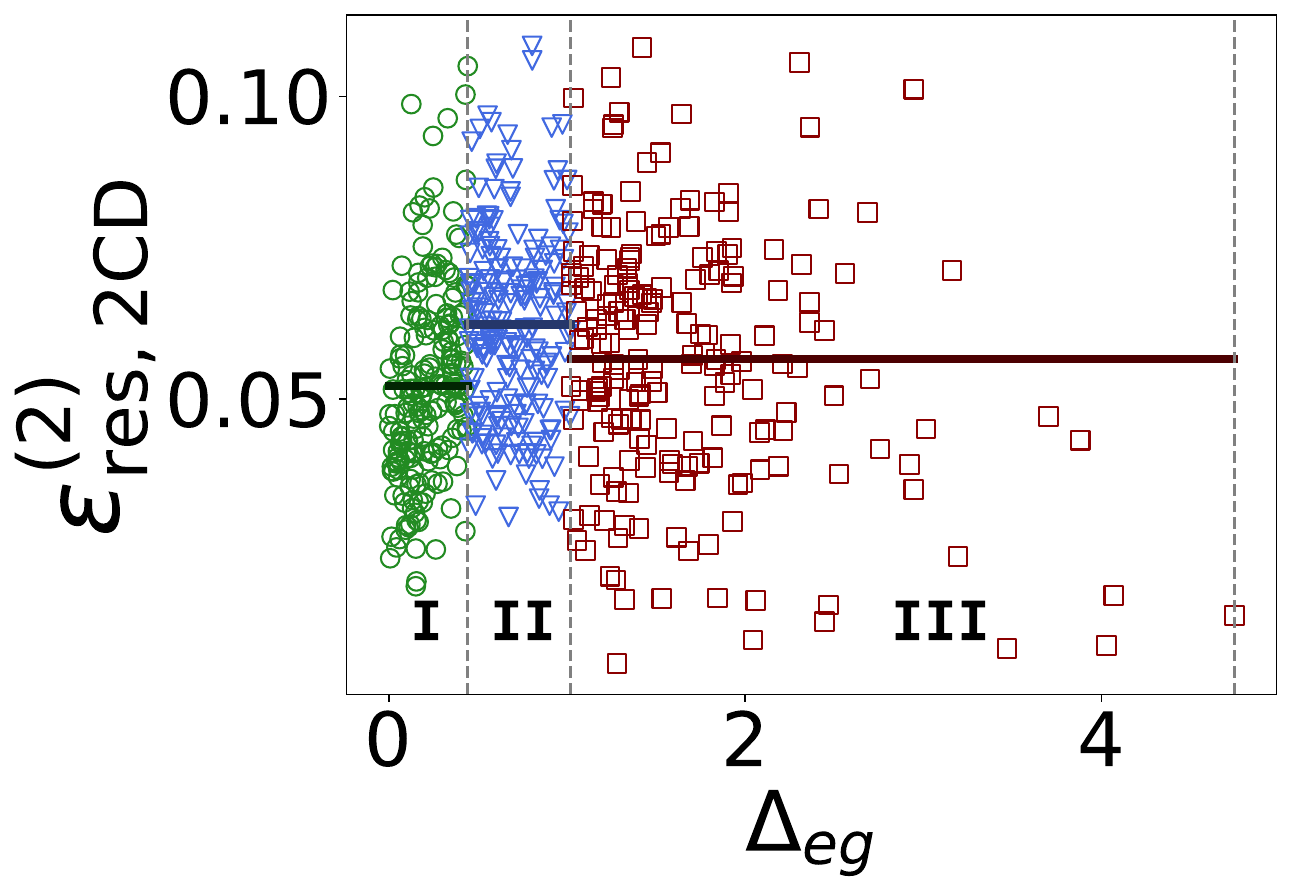}
    \caption{ Fidelity $F_i^{(p)}$ of QAOA-2CD for all $n = $ 600 random instances versus $\Delta_{eg}$ for $p=1$ (a) and $p=2$ (b). Each point in the scatter plot represents the fidelity $F_\mathrm{2CD,i}^{(p)}$ for a single instance. Residual energy $\varepsilon_\text{2CD,i}^{(p)}$ for all $n=$ 600 different instances vs $\Delta_{eg}$ of QAOA-2CD
    for $p=1$ (c) and $p=2$ (d).
    Graphs delineate three distinct regions based on the energy gap $\Delta_{eg}$, as illustrated in caption of Fig.~\ref{eres-QAOA}.     
    The horizontal line within each sector depicts the corresponding average value of that sector. Fidelity in QAOA-2CD increases with $\Delta_{eg}$ as for QAOA. It is not possible to appreciate a specific behavior for residual energy.
}
    \label{eres-QAOA-2CD-vs-gaps}
\end{figure}

Finally, we study the performance of QAOA-2CD in each of the three separate zones identified by the ranges of the final gaps $\Delta_{eg}$, in terms of residual energy and fidelity. We report our results in Fig.~\ref{eres-QAOA-2CD-vs-gaps}, where we study these quantities for two fixed steps of the algorithm: $p = 1$, panels (a,c), and $p = 2$, panels (b,d). 

Concerning the fidelity in Figs.~\ref{eres-QAOA-2CD-vs-gaps}(a,b), we confirm the same trend seen in QAOA: the larger the gap, the closer the fidelity is to one. The solutions found in region \textbf{III} are closer to the real target state than the solutions found in region \textbf{I} for both steps $p$ displayed. In fact, for steps $p=1,2$, the average values show a growth trend that brings the average fidelity very close to the solution in region \textbf{III}. 

Regarding the residual energy, see Figs.~\ref{eres-QAOA-2CD-vs-gaps}(c,d), we observe a non monotonic dependence of the average residual energy as a function of $\Delta_{eg}$, shown by the fact that, in region \textbf{II}, the average residual energy is higher than in region \textbf{I}. This pattern is found both for $p = 1$ and $p = 2$. This is in contrast with the QAOA results, shown in Fig.~\ref{eres-QAOA}, which only display the same pattern starting from later steps (for instance, $p = 2$).

These observations are also valid for all other steps before convergence, as detailed in Appendix~\ref{app:numerics}, where we report the numerical values obtained by our analysis for other steps $p$, for the fidelity and the residual energy, in the three zones. At fixed $p$, the average value of the fidelity of QAOA-2CD in each region is generally closer to one than the average fidelity of QAOA. Besides, the standard deviation of the fidelity for all algorithms is smaller for problems characterized by a larger final gap, and also convergence requires fewer steps. %

\begin{figure}[b]
    (a)\hspace{0.4\columnwidth}(b)\hspace*{0.3\columnwidth}\hfill\\
    \centering
    \includegraphics[width=0.45\columnwidth]{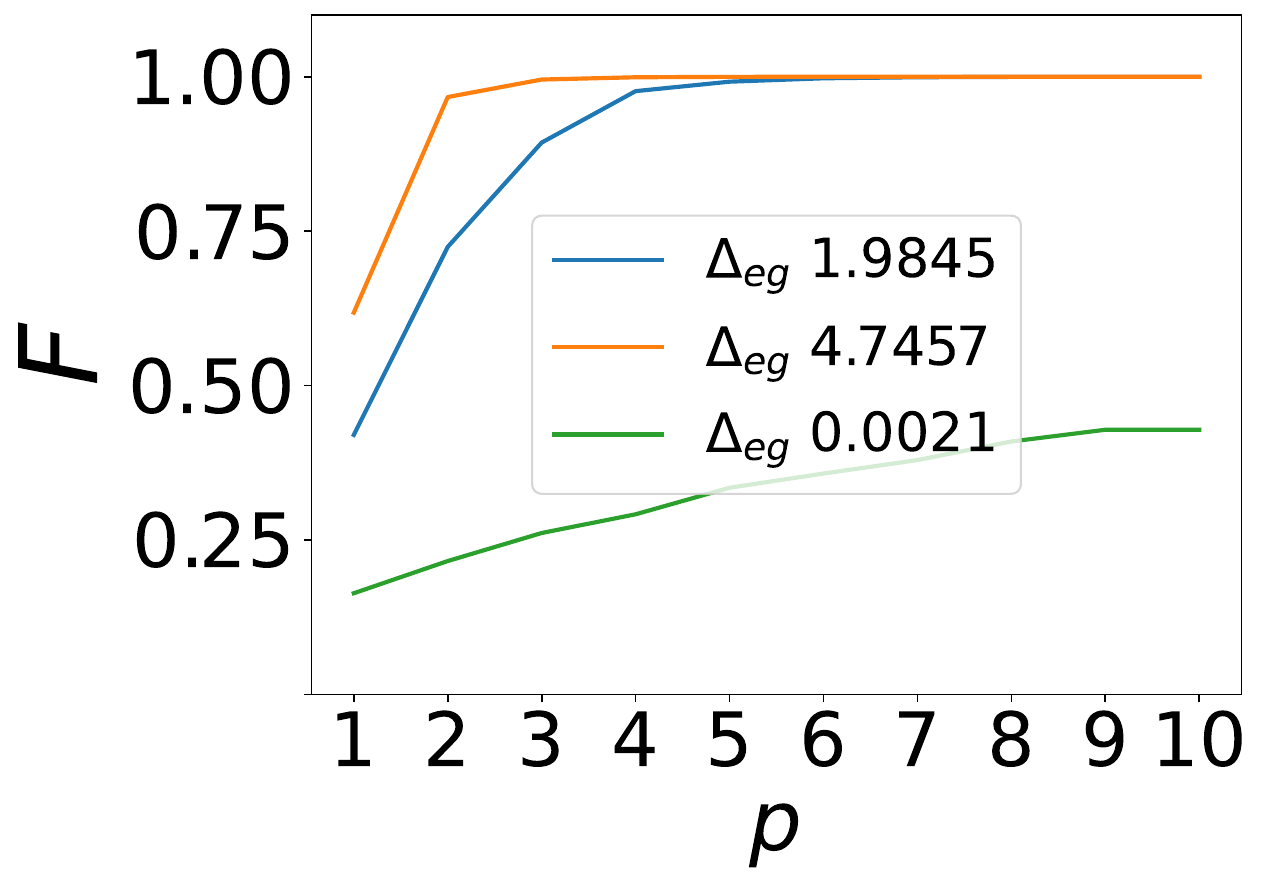}
    \includegraphics[width=0.45\columnwidth]{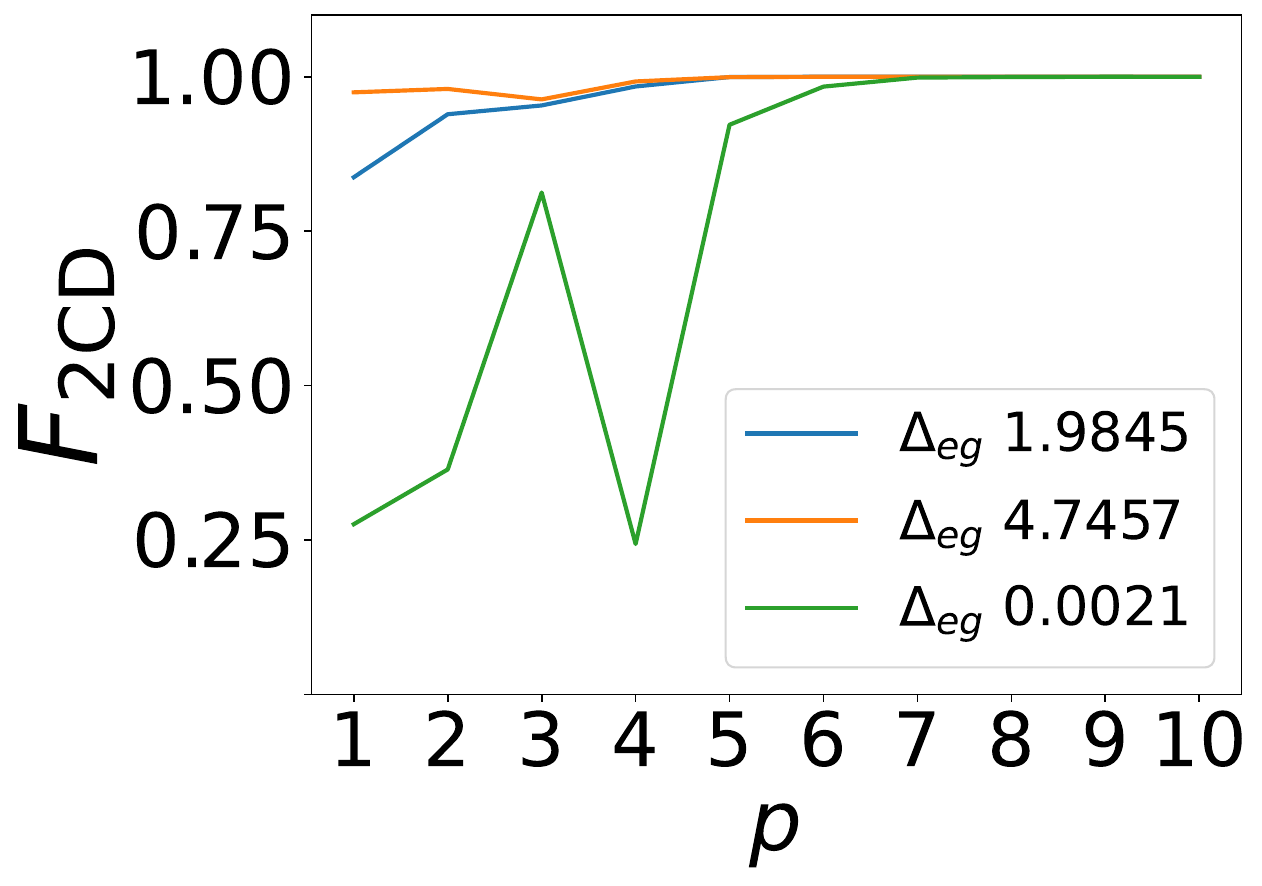}
    \centering
    \\(c)\hspace{0.4\columnwidth}(d)\hspace*{0.3\columnwidth}\hfill\\
    \includegraphics[width=0.45\columnwidth]{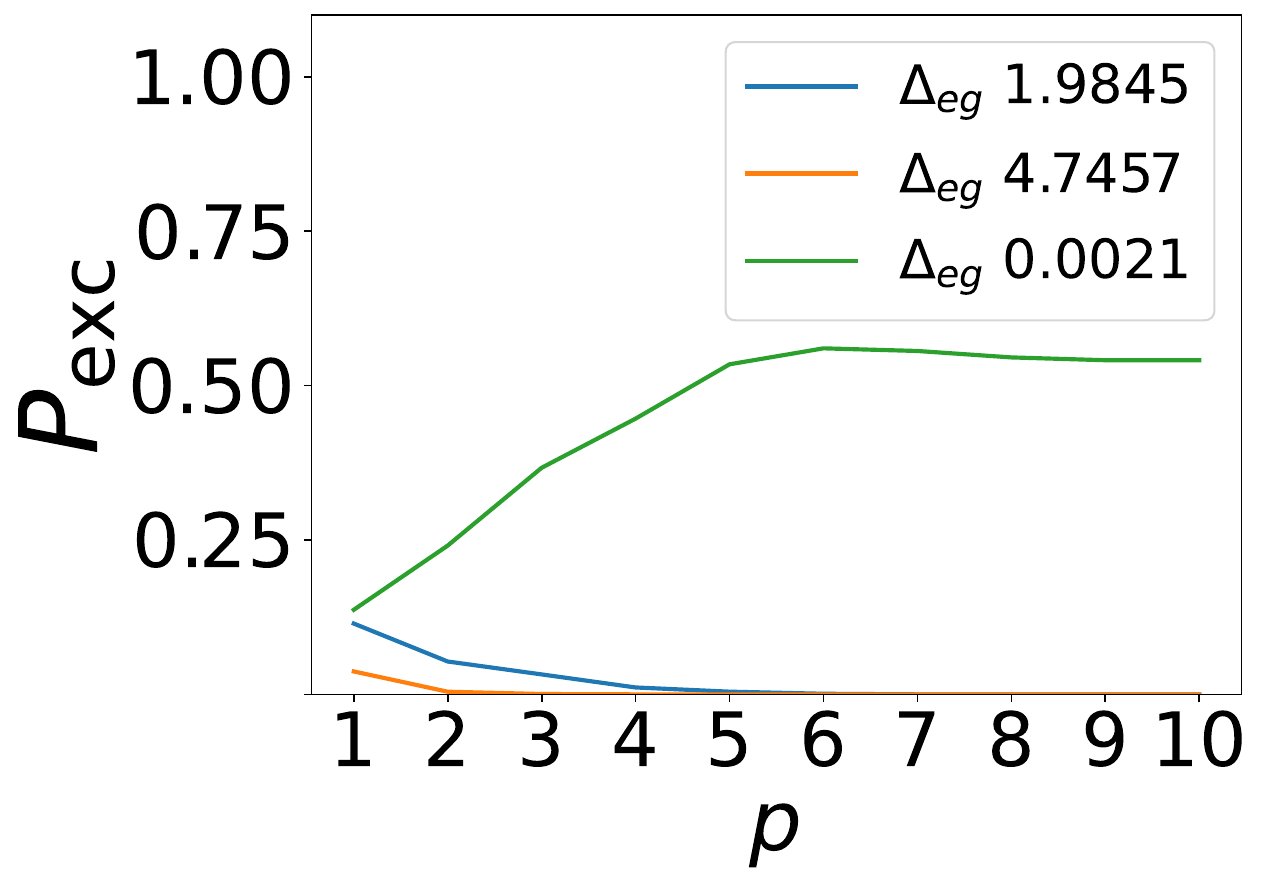}
    \includegraphics[width=0.45\columnwidth]{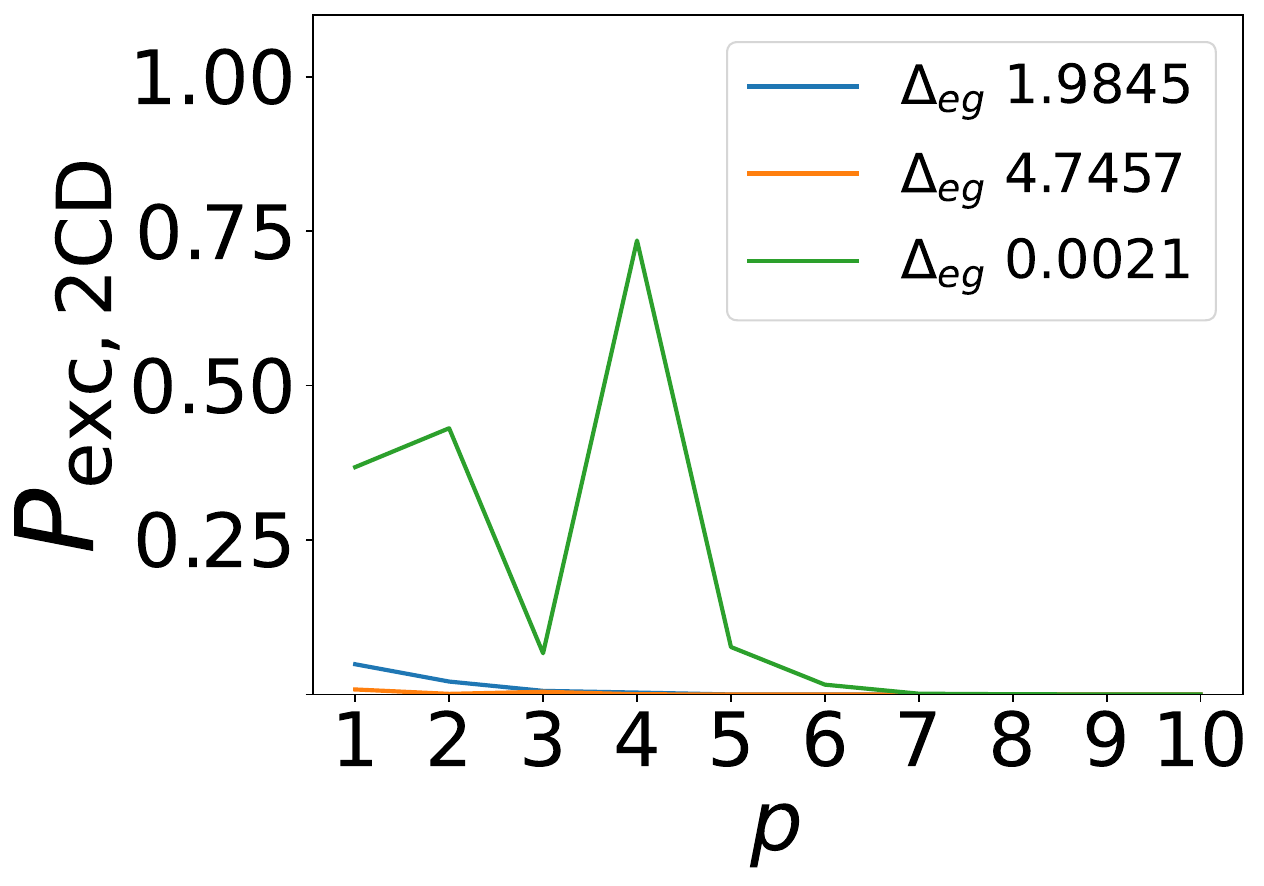}
    \caption{(a) Fidelity between the QAOA state and the true ground state. (b) Excited-state probability of the optimal QAOA state. (c) Fidelity between the QAOA-2CD state and the true ground state. (d) Excited-state probability of the optimal QAOA-2CD state. Results are shown for three instances with different gap values $\Delta_{eg}$.}
    \label{fidelity-different-states}
\end{figure}

\commentref{In Fig.~\ref{fidelity-different-states} we examine three instances from distinct zones. In particular we focus on the instance with $\Delta_{eg}=0.0021$ from \textbf{I}-zone (green curve), which has the smallest $\Delta_{eg}$ among all the instances analyzed; instance with $\Delta_{eg}=1.9845$ from \textbf{II}-zone (blue curve); and instance with $\Delta_{eg}=4.7457$ from \textbf{III}-zone (orange curve), the instance with the largest value of $\Delta_{eg}$. In particular Fig.~\ref{fidelity-different-states}(a,\,b) shows the fidelity of these three instances with the ground state of $H_T$, $\ket{\psi_T}$ for QAOA and QAOA-2CD respectively. In Fig.~\ref{fidelity-different-states}(c,\,d) we represent the overlap between QAOA and QAOA-2CD respectively and the first excited state. }

\commentref{By comparing panels (a) and (c), we observe that, at convergence, the instance with the smallest gap (orange curve) in QAOA exhibits an overlap between the ground and first excited states. This is because, with a very small $\Delta_{eg}$, the ground and first excited states are quasi-degenerate, so QAOA effectively finds the solution in the subspace spanned by both states. However, in panels (b) and (d), we see that counterdiabaticity can resolve this quasi-degeneracy: despite requiring more steps $p$ than other instances, the instance with the smallest gap in QAOA-2CD converges to the true ground state, despite the degeneracy.}

\begin{table}[t]
\setlength{\tabcolsep}{6pt}
    \centering
    \begin{tabular}{cccc}
    \toprule
    $p$ & $\braket{F}$   & $\braket{F_\mathrm{CD}}$ & $\braket{F_\mathrm{2CD}}$\\
    \midrule
    $1$     & $0.27\pm 0.08$   &  $0.41\pm 0.17$  & $ 0.57\pm0.19$ \\
    $2$     &   $0.47\pm0.16$  &  $0.56 \pm 0.19$ &  $0.73\pm0.18$ \\
    $3$     &   $0.61\pm0.19$  & $0.68\pm0.19$  &  $0.86\pm0.14$ \\
    $4$     &  $0.70\pm0.19$   & $0.76\pm0.19$  & $0.94\pm0.10$  \\
    $5$     &  $0.77\pm0.19$   & $0.83\pm0.17$  & $0.99\pm0.04$  \\
    $6$     &    $0.82\pm0.19$ & $0.89\pm0.14$ & $1.00\pm0.01$ \\
    $7$     &  $0.86\pm0.17$  & $0.94\pm0.11$  & $\sim 1$  \\
    $8$     &  $0.89\pm0.17$   & $0.97\pm0.06$ &  $\sim 1$ \\
    $9$     &  $0.91\pm0.15$  & $0.99\pm0.03$  & $\sim 1$ \\
    $10$     &   $0.93\pm0.14$  & $\sim 1$ & $\sim 1$ \\
    \bottomrule
    \end{tabular}
    \caption{Average fidelities $\braket{F}$, $\braket{F_\mathrm{CD}}$ and $\braket{F_\mathrm{2CD}}$ for  QAOA, QAOA-CD and QAOA-2CD, respectively, together with their respective
    standard deviations for all the analyzed $p$ steps. For each fixed $p$, the QAOA-2CD algorithm outperforms both the QAOA-CD and QAOA algorithms. Values denoted
    as $\sim1$ have standard deviations $\leq 10^{-4}$.}
    \label{results-fidelity-table}
\end{table}

\begin{table}[b!]
    \setlength{\tabcolsep}{6pt}
    \centering
    \begin{tabular}{cccc}
    \toprule
     $p$& $\braket{\varepsilon_\mathrm{res}}$   & $\braket{\varepsilon_\mathrm{res,CD}}$ & $\braket{\varepsilon_\mathrm{res,2CD}}$\\
    \midrule
    $1$     & $0.21\pm 0.04$   &  $0.15\pm 0.03$  & $ 0.09\pm0.03$ \\
    $2$     &   $0.11\pm0.03$  &  $0.08 \pm 0.02$ &  $0.05\pm0.02$ \\
    $3$     &  $0.05\pm0.02$   &  $0.05\pm0.02$ &  $0.026\pm0.009$ \\
    $4$     &  $0.03\pm0.02$   &  $0.03\pm0.02$ & $0.008\pm0.003$  \\
    $5$     &  $0.02\pm0.01$   & $0.02\pm0.01$  & $\sim 0$  \\
    $6$     &    $0.01\pm0.01$ & $0.014\pm0.006$ & $\sim 0$ \\
    $7$     &   $0.008\pm0.008$  & $0.006\pm0.004$  &  $\sim 0$ \\
    $8$     &  $0.006\pm0.006$   & $0.002\pm0.001$  &  $\sim 0$ \\
    $9$     &   $0.004\pm0.004$  &  $\sim 0$ &  $\sim 0$ \\
    $10$     &   $ 0.02\pm0.04$  & $\sim 0$ & $\sim 0$ \\
    \bottomrule
    \end{tabular}
    \caption{Average residual energies  $\braket{\varepsilon_\mathrm{res}}$, $\braket{\varepsilon_\mathrm{res,CD}}$ and $\braket{\varepsilon_\mathrm{res,2CD}}$
    for QAOA, QAOA-CD and QAOA-2CD, respectively, together with their respective standard deviations for all the analyzed $p$ steps. For each fixed $p$, the QAOA-2CD
    algorithm outperforms both the QAOA-CD and QAOA algorithms. Value denoted as $\sim 0$ have standard deviations $\leq 10^{-4}$.}
    \label{results-eres-table}
\end{table}

\section{Conclusions}\label{sec:conclusions}

To summarize, in this manuscript we have applied digitized counterdiabatic QAOA algorithms to study fully-connected spin models. In particolar we have compared the performances of QAOA, QAOA-CD, QAOA-2CD on 
a fully-connected disordered MaxCut problem with $N=5$ vertices. We analyzed $n=600$ random instances 
quantifying the quality of the algorithm with the residual energy (i.\,e., distance between the energy of the calculated solution and the exact ground-state energy) and the fidelity (i.\,e., the overlap between the exact ground state and the calculated state). %
Consistently with our prior findings~\cite{vizzuso2023}, we observe that the QAOA-2CD overperforms the other variants, at the price of an increased number of classical parameters to optimize at fixed \textit{p}. \commentref{On the other hand, QAOA-CD and QAOA-2CD perform similarly to QAOA for a fixed number of variational parameters (not shown)~\cite{vizzuso2023}.}

We also analyzed the performances of the algorithms in connection with the gap between the ground state manifold and the first excited state, showing that in general, large gap instances are easier to be solved with this approach. While the role of spectral gaps are well known in QA, in QAOA their role is more subtle and this result is somehow nontrivial. 
\commentref{The underlying idea, however, is that a dense spectrum in the vicinity of the ground state makes either the system more prone to be “captured” by local minima during the hybrid optimization procedure or, more likely, the overlap of the variational ground state with the excited states of $H_T$ not negligible.}
This approach could be used as a starting point to train a neural network with datasets sorted by spectral gaps to approximate efficiently the QAOA angles as done in Ref.~\cite{bishop2023set} for QA.

\begin{acknowledgments}
 G.\,P. and P.\,L.\ acknowledge financial support from PNRR MUR Project PE0000023-NQSTI. G.\,P.\ acknowledges computational resources
from the CINECA award under the ISCRA initiative. M.\,V.\ and G.\,P.\ ackknowledge computational resources from MUR, PON “Ricerca e Innovazione 2014-2020”, under Grant No. PIR01\_00011
- (I.Bi.S.Co.). This work was supported by PNRR
MUR project PE0000023 - NQSTI, by the European
Union’s Horizon 2020 research and innovation programme under Grant Agreement No 101017733, by
the MUR project CN\_00000013-ICSC (P.\,L.), and by
the QuantERA II Programme STAQS project that
has received funding from the European Union’s Horizon 2020 research and innovation program.
\end{acknowledgments}

\appendix
\section{Numerical results}\label{app:numerics}

The numerical results used for Fig.~\ref{fig:F_and_eres} are reported in Tables~\ref{results-fidelity-table} and~\ref{results-eres-table}. Instead, in Tables~\ref{fidelities} and~\ref{residual-energies} we report the zone-resolved fidelities and residual energies, respectively. 

\begin{table*}[b!]
\setlength{\tabcolsep}{3pt}
    \centering
  \begin{tabular}{c@{\hskip 6mm}ccc@{\hskip 6mm}ccc@{\hskip 6mm}ccc}
    \toprule
     &
      \multicolumn{3}{c}{$\mathbf{I}$-zone} &
      \multicolumn{3}{c}{$\mathbf{II}$-zone} &
      \multicolumn{3}{c}{$\mathbf{III}$-zone} \\[1ex]
    $p$ &QAOA& QAOA-CD& QAOA-2CD &QAOA& QAOA-CD& QAOA-2CD&QAOA& QAOA-CD& QAOA-2CD\\
    \midrule
    $1$ &  $0.22\pm 0.05$   &  $0.29\pm0.10$  & $0.40\pm0.15$ 
    &$0.26\pm 0.05$   &  $0.40\pm 0.15$  & $0.57\pm 0.14$  
    &$0.35\pm 0.08$   &  $0.56\pm 0.16$  & $0.74\pm 0.11$ \\
     
     $2$&   $0.34\pm0.09$  & $0.40\pm0.12$ &  $0.56\pm0.19$  
     &$0.44\pm 0.10$   &  $0.55\pm 0.13$  & $0.75\pm 0.10$ 
     &$0.63\pm 0.14$   &  $0.75\pm 0.12$  & $0.86\pm 0.07$\\
     
     $6$& $0.60\pm0.14$ & $0.77\pm0.19$ & $0.99\pm0.01$ 
     &$0.87\pm 0.10$   &  $0.93\pm 0.04$  & $\sim 1$   
     &  $0.98\pm0.02$  & $0.97\pm0.02$&$\sim 1$ \\
    
    $10$&   $0.80\pm0.18$  & $0.99\pm0.01$ & $\sim 1$  
    &$0.99\pm 0.02$   &  $\sim 1$  & $\sim 1$ 
    &$0.999\pm0.002$   &  $\sim 1$ & $\sim 1$\\
    \bottomrule
  \end{tabular}
  \caption{
  Fidelities for $p=1,2,6,10$, in the three zones identified in the main text by the range of the minimum gap $\Delta_{eg}$. We compare QAOA, QAOA-CD and QAOA-2CD. Values 
  denoted as $\sim 1$ have standard deviations $\sigma\leq10^{-4}$. 
  }
  \label{fidelities}
\end{table*}

\begin{table*}[t]
\setlength{\tabcolsep}{2pt}
  \begin{tabular}{c@{\hskip 4mm}ccc@{\hskip 4mm}ccc@{\hskip 4mm}ccc}
    \toprule
     &
      \multicolumn{3}{c}{$\mathbf{I}$-zone} &
      \multicolumn{3}{c}{$\mathbf{II}$-zone} &
      \multicolumn{3}{c}{$\mathbf{III}$-zone} \\[1ex]
    $p$&QAOA& QAOA-CD& QAOA-2CD &QAOA& QAOA-CD& QAOA-2CD&QAOA& QAOA-CD& QAOA-2CD\\
    \midrule
    $1$ &  $0.17\pm 0.03$   &  $0.14\pm0.03$  & $0.09\pm0.03$ &$0.21\pm0.03$   &  $0.16\pm0.03$  & $0.10\pm0.03$  
    &$0.25\pm0.03$   &  $0.18\pm0.04$  & $0.10\pm0.04$ \\
     
     $2$ &  $0.08\pm 0.02$   &  $0.08\pm0.02$  & $0.05\pm0.02$ &$0.11\pm0.02$   &  $0.10\pm0.02$  & $0.06\pm0.02$  
     &$0.12\pm0.03$   &  $0.09\pm0.03$  & $0.06\pm0.02$ \\
     
     $6$&  $0.017\pm 0.008$   &  $0.014\pm0.006$  & $\sim 0$ &$0.017\pm0.009$   &  $0.016\pm0.006$  & $\sim 0$  
     &$0.003\pm0.004$   &  $0.011\pm0.006$  & $\sim 0$ \\
    
    $10$&  $ 0.006\pm0.003$   &  $\sim 0$  & $\sim 0$ 
    &$0.002\pm0.003$   &  $\sim 0$  & $\sim 0$  
    &$\sim 0$   &  $\sim 0$  & $\sim 0$ \\
    \bottomrule
  \end{tabular}
  \caption{Residual energies for $p=1,2,6,10$, in the three zones identified in the main text by the range of the minimum gap $\Delta_{eg}$. We compare QAOA, QAOA-CD and QAOA-2CD.
  Values denoted as $\sim 0$ have standard deviations $\sigma\leq10^{-4}$. %
  }
  \label{residual-energies}
\end{table*}

\commentref{In order to provide a more quantitative analysis, Tab.~\ref{step-comparing} compares the number of steps required to achieve an average fidelity of approximately $1$, with a tolerance of $10^{-2}$ for each algorithm and zone. We see that in $\mathbf{III}$-zone (where $\Delta_{eg}$ is larger), for all three algorithms, we need fewer steps to reach this tolerance. The lowest value in the table is for QAOA-2CD.}

\begin{table}[b!]
    \setlength{\tabcolsep}{6pt}
    \centering
    \begin{tabular}{cccc}
    \toprule
     & QAOA   & QAOA-CD & QAOA-2CD\\
    \midrule
     $\mathbf{I}$-zone  & $p>10$   &  $p=10$  & $p=6$ \\
    $\mathbf{II}$-zone     &   $p=10$  &  $p=6$ & $p=4$  \\
    $\mathbf{III}$-zone  &  $p=6$   &  $p=6$ &  $p=2$ \\
    \bottomrule
    \end{tabular}
    \caption{The number of steps required to achieve an average fidelity of approximately $1$ with a tolerance of $10^{-2}$. The values are then compared with respect to the different algorithms and different zones, depending on the value of $\Delta_{eg}$. The lowest value is observed for QAOA-2CD in the $\mathbf{III}$-zone.}
    \label{step-comparing}
\end{table}

\begin{figure}[b]
    (a)\hspace{0.4\columnwidth}(b)\hspace*{0.3\columnwidth}\hfill\\
    \centering
    \includegraphics[width=0.45\columnwidth]{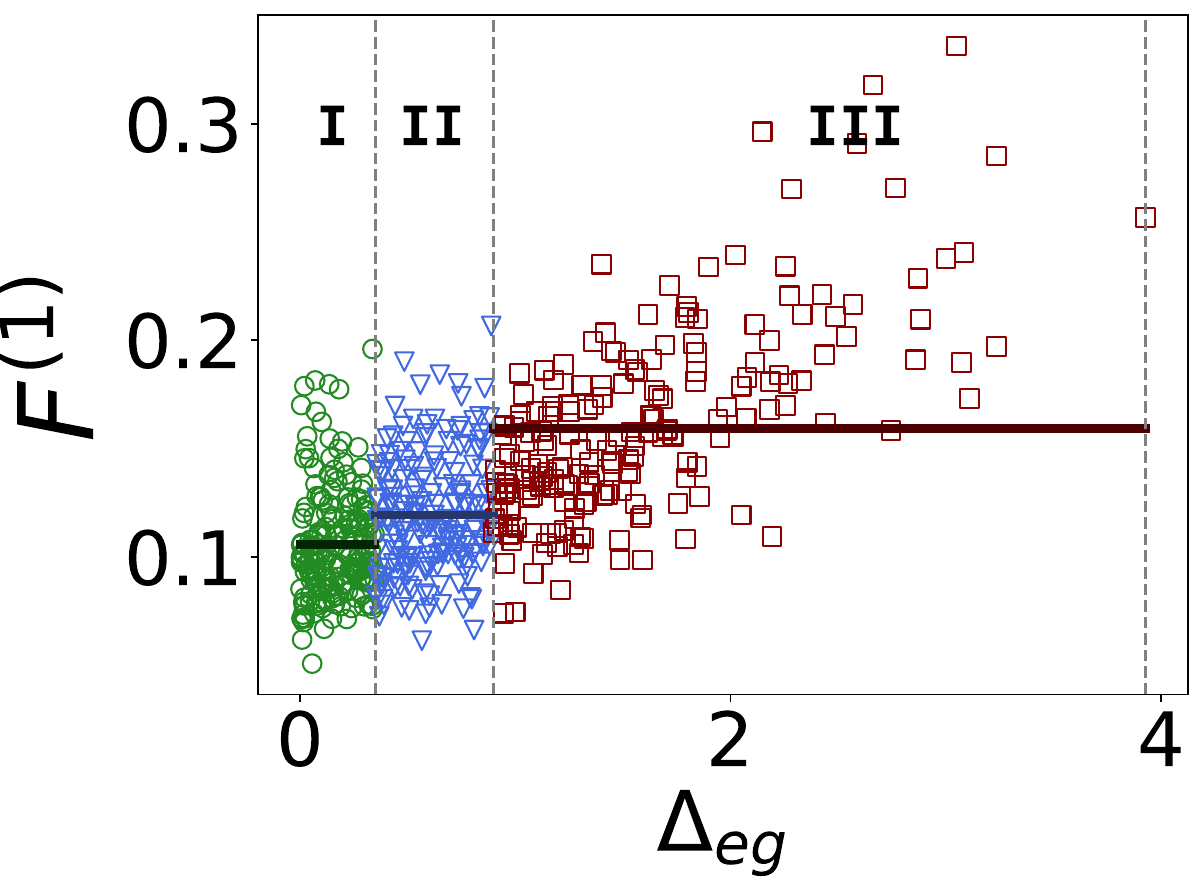}
    \includegraphics[width=0.45\columnwidth]{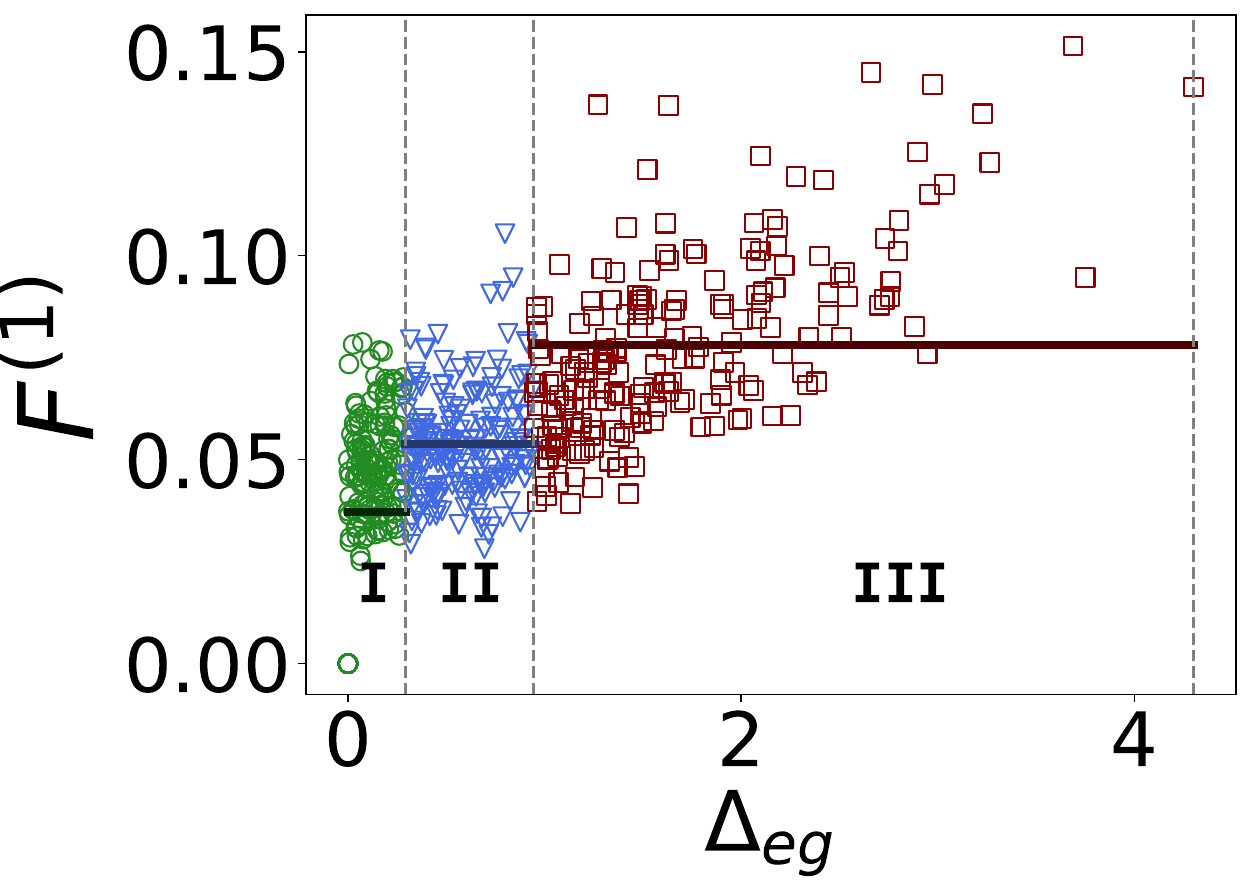}
    \centering
    \\(c)\hspace{0.4\columnwidth}(d)\hspace*{0.3\columnwidth}\hfill\\
    \includegraphics[width=0.45\columnwidth]{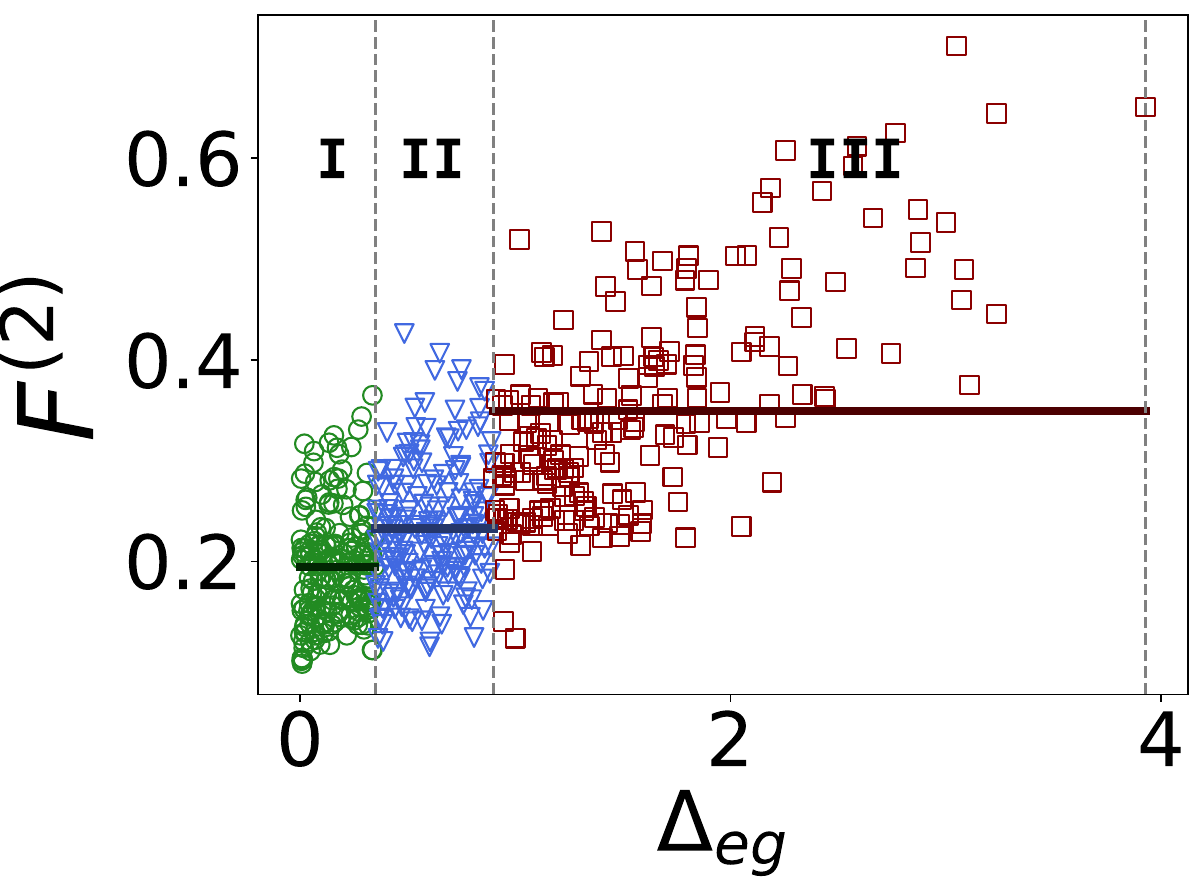}
    \includegraphics[width=0.45\columnwidth]{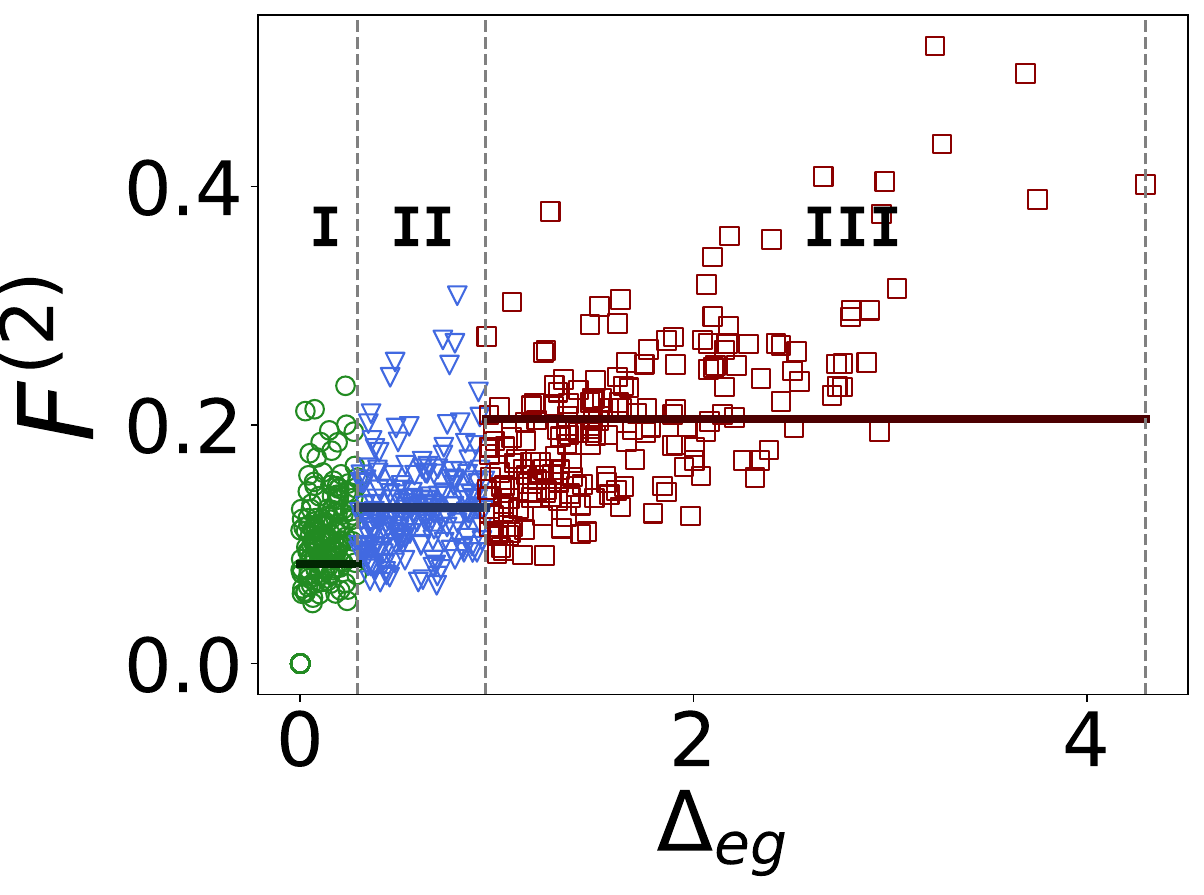}
    \\(e)\hspace{0.4\columnwidth}(f)\hspace*{0.3\columnwidth}\hfill\\
    \includegraphics[width=0.45\columnwidth]{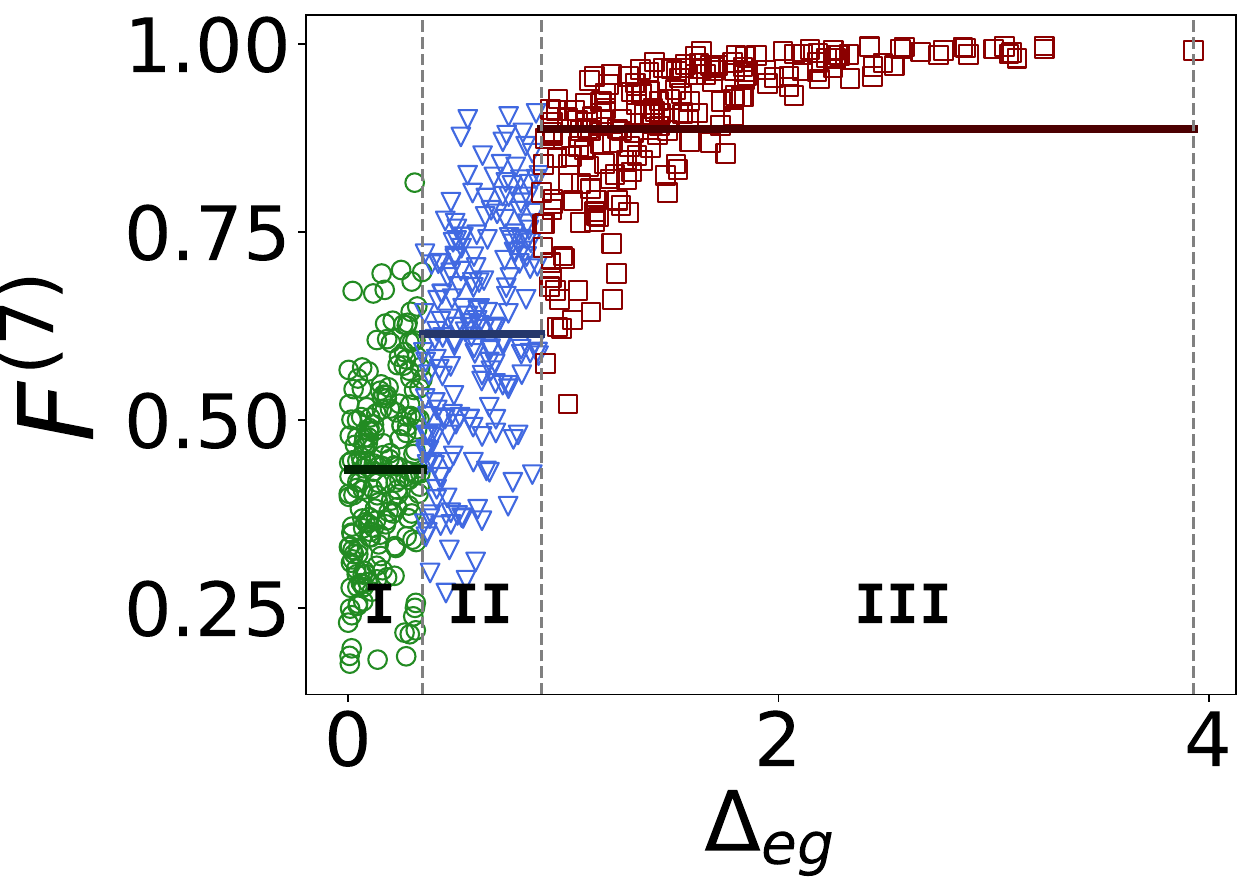}
    \includegraphics[width=0.45\columnwidth]{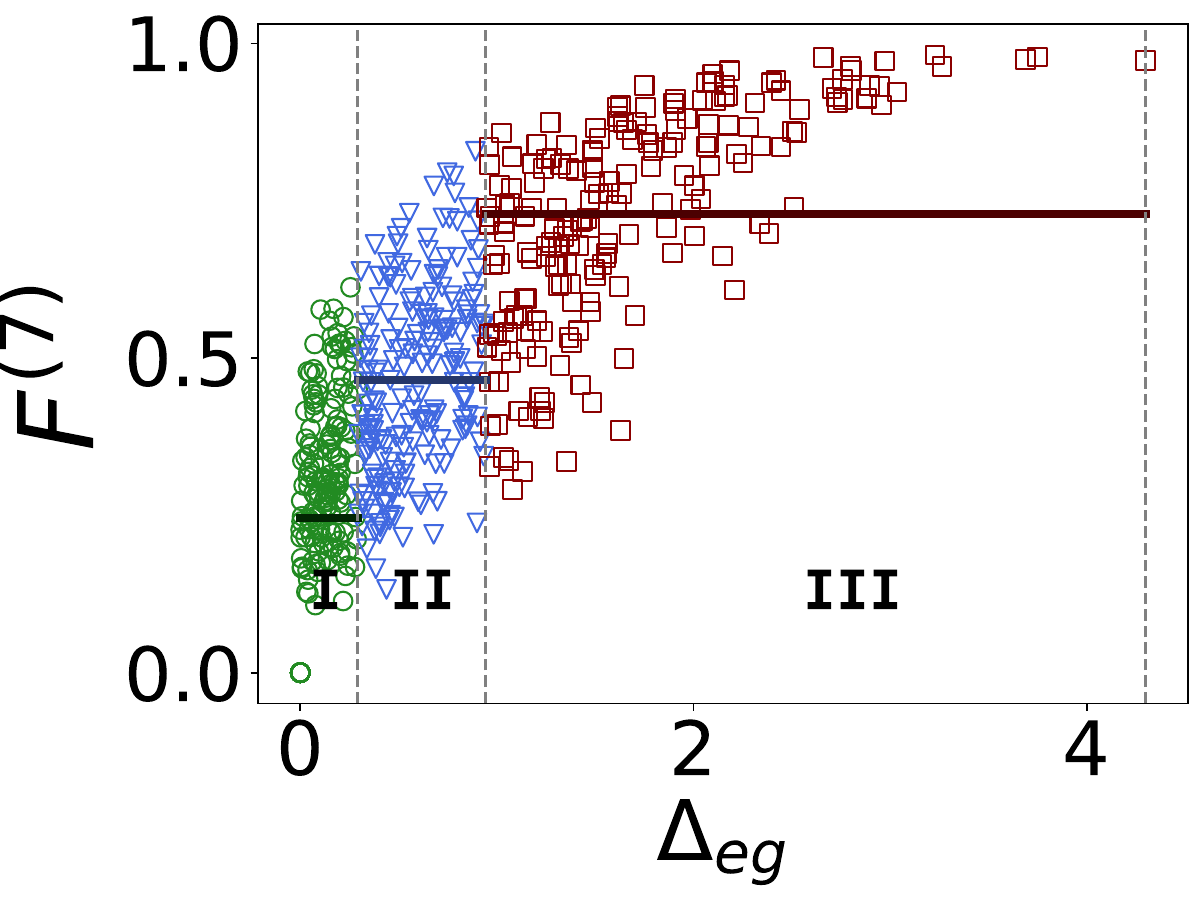}
    \caption{Fidelity $F_i^{(p)}$ as a function of the minimal gap $\Delta_{eg}$ for QAOA. Panels (a,\,c,\,e) are for $N=7$ spins, while panels (b,\,d,\,f) are for $N = 9$. The different rows correspond to $p=1$, $p=2$, $p=7$, respectively. Three distinct regions based on the energy gap separating the ground state from the first excited state can be singled as described in the main text. The horizontal line within each sector depicts the corresponding average value of that sector.}
    \label{eres-QAOA-2CD-vs-gaps-spin-7}
\end{figure}

\commentref{Fig.s \ref{eres-QAOA-2CD-vs-gaps-spin-7} illustrate the fidelity $F_i^{(p)}$ for $n=600$ random instances of the QAOA for steps $p=1, 2$ and $7$, respectively, for a fully-connected system with $N=7$ and $N=9$ spins.
In both systems, we split the sorted gaps into three sets of equal cardinalities (200 instances per set). For $N = 7$, the three regions are $\Delta_{eg}\in\left[0,0.35\right]$ (\textbf{I}), $\Delta_{eg}\in\left[0.35,0.90\right]$ (\textbf{II}), and $\Delta_{eg}\in\left[0.90,3.93\right]$ (\textbf{III}). For $N=9$ spins, the three regions are $\Delta_{eg}\in\left[0,0.29\right]$ (\textbf{I}), $\Delta_{eg}\in\left[0.29,0.94\right]$ (\textbf{II}), and $\Delta_{eg}\in\left[0.94,4.30\right]$ (\textbf{III}). For both systems, we observe a trend consistent with that found for $N=5$: as the energy gap increases, the algorithm’s performance correspondingly improves. It is evident that as the number of spins in the system increases, the algorithm encounters some challenges in converging to the true ground state. This trend is particularly pronounced for step sizes $p=1$ and $p=2$, where the system tends to exhibit smaller average values compared to the case with $N=5$ spins. Furthermore, the variances observed in these cases are significantly larger than those observed for the $N=5$ system. These findings suggest that the increased system size introduces additional challenges in accurately determining the ground state, both in terms of mean values and the variability of the results. This behavior highlights the growing complexity of the problem as the number of spins increases.}

\section{Minimization}\label{app:minimization}

\commentref{In this section, we examine the classical minimization aspect of our approach in detail. The cost function, as defined in Eq.~\eqref{qaoa-function}, presents a challenging landscape characterized by high complexity and the presence of numerous local minima. This complexity poses a significant challenge for the minimization process, as the minimizer may easily become trapped within one of these local minima, hindering convergence to a global minimum.}

\commentref{To mitigate this risk and enhance the overall effectiveness of the algorithm, we employ a strategy involving multiple independent minimization attempts. Specifically, at each step, we initiate $S=20$ separate minimizations of the cost function from Eq.~\eqref{qaoa-function}, each starting from a random set of initial angles. By adopting this approach, we aim to increase the chances of achieving a result closer to the global minimum, as the multiple initializations help explore different regions of the solution space more effectively.}

\commentref{Moreover, during our analysis, we identified instances with non-monotonic behavior as a function of $p$. Upon increasing the value of $S$, we observed a shift toward a monotonic behavior in the resulting curves. This observation suggests that a larger number of initializations with distinct starting angles can improve the stability of the minimization process and contribute to more consistent convergence patterns.
}

\begin{figure*}[b]
    \hspace{-0.2\textwidth}(a)\hspace{0.31\textwidth}(b)\hspace{0.31\textwidth}(c)\hfill\\
    \centering
    \includegraphics[width=0.3\textwidth]{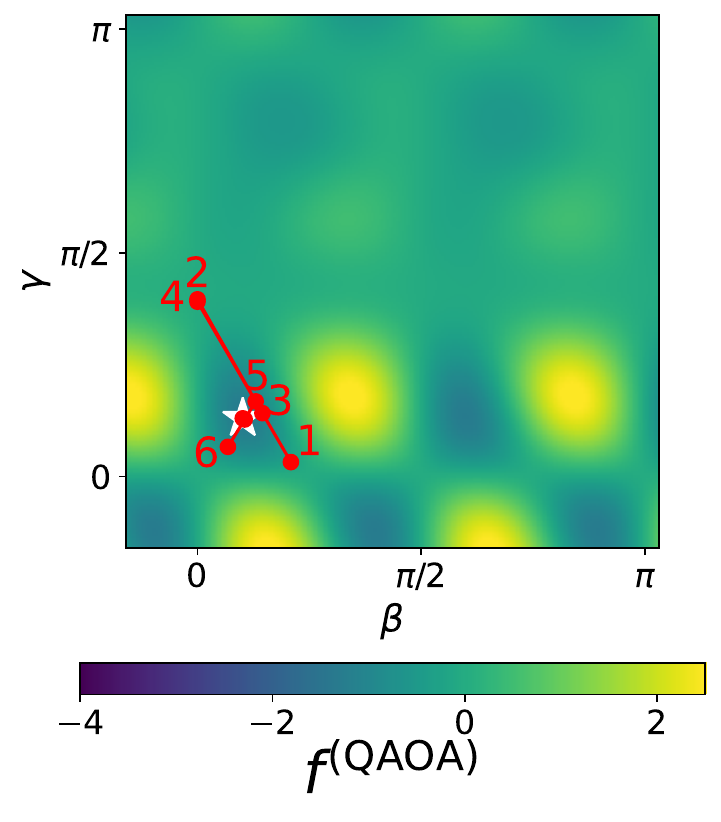}
    \includegraphics[width=0.3\textwidth]{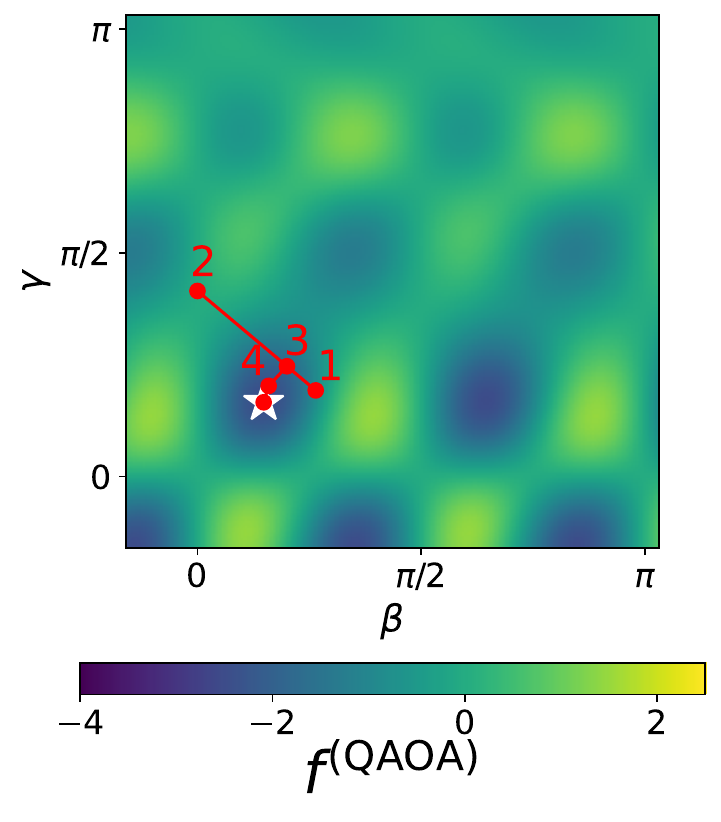}
    \includegraphics[width=0.3\textwidth]{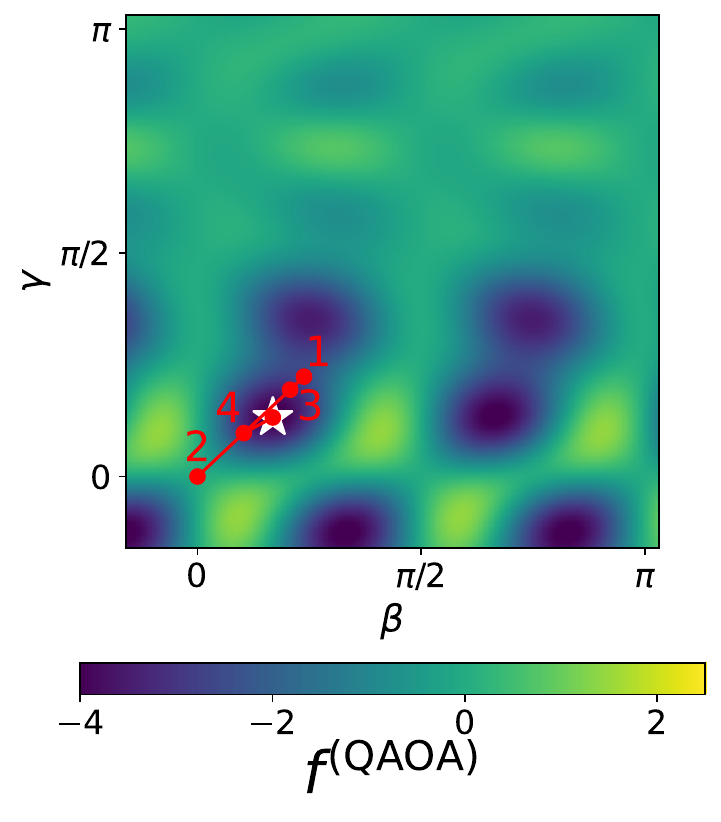}
    \caption{QAOA cost function (see Eq.~\eqref{qaoa-function}) at step $p=1$ for spin $N=5$ for three different instances. Panel a) shows cost function of the instance with lowest $\Delta_{eg}$ ($\Delta_{eg}=0.002$) among all random instances analyzed, panel b) an instance with $\Delta_{eg} = 1.98$ and in panel c) cost function of the instance with the largest $\Delta_{eg}$ among the instances analyzed $\Delta_{eg} = 4.745$. The red segments illustrate the minimization path taken from a selected set of random initial angles. These particular angles were chosen based on a posterior analysis, as they led to the best minimum found in each case, marked by a white star in the figures.}
    \label{iter-minimization}
\end{figure*}

\commentref{Fig.~\ref{iter-minimization} shows the cost function of QAOA at step $p=1$ in Eq.~\eqref{qaoa-function} for a spin system of size $N=5$, for three specific instances. Panel (a) illustrates the instance with the smallest energy gap, $\Delta_{eg}$, among those studied, with $\Delta_{eg} = 0.002$. Panel (b) shows the cost function for an instance characterized by $\Delta_{eg} = 1.98$. Lastly, panel (c) displays the cost function for the instance with the largest observed gap, where $\Delta_{eg} = 4.745$.}

\commentref{For each of these cost functions, the figure includes the minimization path in red, illustrating the trajectory followed during minimization for a specific set of initial angles. Notably, we selected the initial angles that, as determined through posterior analysis, led to the best minimum (white star) achieved for each case.}

\clearpage


\begin{thebibliography}{69}%
\makeatletter
\providecommand \@ifxundefined [1]{%
 \@ifx{#1\undefined}
}%
\providecommand \@ifnum [1]{%
 \ifnum #1\expandafter \@firstoftwo
 \else \expandafter \@secondoftwo
 \fi
}%
\providecommand \@ifx [1]{%
 \ifx #1\expandafter \@firstoftwo
 \else \expandafter \@secondoftwo
 \fi
}%
\providecommand \natexlab [1]{#1}%
\providecommand \enquote  [1]{``#1''}%
\providecommand \bibnamefont  [1]{#1}%
\providecommand \bibfnamefont [1]{#1}%
\providecommand \citenamefont [1]{#1}%
\providecommand \href@noop [0]{\@secondoftwo}%
\providecommand \href [0]{\begingroup \@sanitize@url \@href}%
\providecommand \@href[1]{\@@startlink{#1}\@@href}%
\providecommand \@@href[1]{\endgroup#1\@@endlink}%
\providecommand \@sanitize@url [0]{\catcode `\\12\catcode `\$12\catcode
  `\&12\catcode `\#12\catcode `\^12\catcode `\_12\catcode `\%12\relax}%
\providecommand \@@startlink[1]{}%
\providecommand \@@endlink[0]{}%
\providecommand \url  [0]{\begingroup\@sanitize@url \@url }%
\providecommand \@url [1]{\endgroup\@href {#1}{\urlprefix }}%
\providecommand \urlprefix  [0]{URL }%
\providecommand \Eprint [0]{\href }%
\providecommand \doibase [0]{https://doi.org/}%
\providecommand \selectlanguage [0]{\@gobble}%
\providecommand \bibinfo  [0]{\@secondoftwo}%
\providecommand \bibfield  [0]{\@secondoftwo}%
\providecommand \translation [1]{[#1]}%
\providecommand \BibitemOpen [0]{}%
\providecommand \bibitemStop [0]{}%
\providecommand \bibitemNoStop [0]{.\EOS\space}%
\providecommand \EOS [0]{\spacefactor3000\relax}%
\providecommand \BibitemShut  [1]{\csname bibitem#1\endcsname}%
\let\auto@bib@innerbib\@empty
\bibitem [{\citenamefont {Cerezo}\ \emph {et~al.}(2021)\citenamefont {Cerezo},
  \citenamefont {Arrasmith}, \citenamefont {Babbush}, \citenamefont {Benjamin},
  \citenamefont {Endo}, \citenamefont {Fujii}, \citenamefont {McClean},
  \citenamefont {Mitarai}, \citenamefont {Yuan}, \citenamefont {Cincio} \emph
  {et~al.}}]{cerezo2021variational}%
  \BibitemOpen
  \bibfield  {author} {\bibinfo {author} {\bibfnamefont {M.}~\bibnamefont
  {Cerezo}}, \bibinfo {author} {\bibfnamefont {A.}~\bibnamefont {Arrasmith}},
  \bibinfo {author} {\bibfnamefont {R.}~\bibnamefont {Babbush}}, \bibinfo
  {author} {\bibfnamefont {S.~C.}\ \bibnamefont {Benjamin}}, \bibinfo {author}
  {\bibfnamefont {S.}~\bibnamefont {Endo}}, \bibinfo {author} {\bibfnamefont
  {K.}~\bibnamefont {Fujii}}, \bibinfo {author} {\bibfnamefont {J.~R.}\
  \bibnamefont {McClean}}, \bibinfo {author} {\bibfnamefont {K.}~\bibnamefont
  {Mitarai}}, \bibinfo {author} {\bibfnamefont {X.}~\bibnamefont {Yuan}},
  \bibinfo {author} {\bibfnamefont {L.}~\bibnamefont {Cincio}}, \emph
  {et~al.},\ }\href@noop {} {\bibfield  {journal} {\bibinfo  {journal} {Nature
  Reviews Physics}\ }\textbf {\bibinfo {volume} {3}},\ \bibinfo {pages} {625}
  (\bibinfo {year} {2021})}\BibitemShut {NoStop}%
\bibitem [{\citenamefont {Colless}\ \emph {et~al.}(2018)\citenamefont
  {Colless}, \citenamefont {Ramasesh}, \citenamefont {Dahlen}, \citenamefont
  {Blok}, \citenamefont {Kimchi-Schwartz}, \citenamefont {McClean},
  \citenamefont {Carter}, \citenamefont {de~Jong},\ and\ \citenamefont
  {Siddiqi}}]{Colless2018}%
  \BibitemOpen
  \bibfield  {author} {\bibinfo {author} {\bibfnamefont {J.~I.}\ \bibnamefont
  {Colless}}, \bibinfo {author} {\bibfnamefont {V.~V.}\ \bibnamefont
  {Ramasesh}}, \bibinfo {author} {\bibfnamefont {D.}~\bibnamefont {Dahlen}},
  \bibinfo {author} {\bibfnamefont {M.~S.}\ \bibnamefont {Blok}}, \bibinfo
  {author} {\bibfnamefont {M.~E.}\ \bibnamefont {Kimchi-Schwartz}}, \bibinfo
  {author} {\bibfnamefont {J.~R.}\ \bibnamefont {McClean}}, \bibinfo {author}
  {\bibfnamefont {J.}~\bibnamefont {Carter}}, \bibinfo {author} {\bibfnamefont
  {W.~A.}\ \bibnamefont {de~Jong}},\ and\ \bibinfo {author} {\bibfnamefont
  {I.}~\bibnamefont {Siddiqi}},\ }\href
  {https://doi.org/10.1103/PhysRevX.8.011021} {\bibfield  {journal} {\bibinfo
  {journal} {Phys. Rev. X}\ }\textbf {\bibinfo {volume} {8}},\ \bibinfo {pages}
  {011021} (\bibinfo {year} {2018})}\BibitemShut {NoStop}%
\bibitem [{\citenamefont {Grimsley}\ \emph {et~al.}(2019)\citenamefont
  {Grimsley}, \citenamefont {Economou}, \citenamefont {Barnes},\ and\
  \citenamefont {Mayhall}}]{grimsley2019adaptive}%
  \BibitemOpen
  \bibfield  {author} {\bibinfo {author} {\bibfnamefont {H.~R.}\ \bibnamefont
  {Grimsley}}, \bibinfo {author} {\bibfnamefont {S.~E.}\ \bibnamefont
  {Economou}}, \bibinfo {author} {\bibfnamefont {E.}~\bibnamefont {Barnes}},\
  and\ \bibinfo {author} {\bibfnamefont {N.~J.}\ \bibnamefont {Mayhall}},\
  }\href@noop {} {\bibfield  {journal} {\bibinfo  {journal} {Nature
  communications}\ }\textbf {\bibinfo {volume} {10}},\ \bibinfo {pages} {3007}
  (\bibinfo {year} {2019})}\BibitemShut {NoStop}%
\bibitem [{\citenamefont {Kandala}\ \emph {et~al.}(2017)\citenamefont
  {Kandala}, \citenamefont {Mezzacapo}, \citenamefont {Temme}, \citenamefont
  {Takita}, \citenamefont {Brink}, \citenamefont {Chow},\ and\ \citenamefont
  {Gambetta}}]{kandala2017hardware}%
  \BibitemOpen
  \bibfield  {author} {\bibinfo {author} {\bibfnamefont {A.}~\bibnamefont
  {Kandala}}, \bibinfo {author} {\bibfnamefont {A.}~\bibnamefont {Mezzacapo}},
  \bibinfo {author} {\bibfnamefont {K.}~\bibnamefont {Temme}}, \bibinfo
  {author} {\bibfnamefont {M.}~\bibnamefont {Takita}}, \bibinfo {author}
  {\bibfnamefont {M.}~\bibnamefont {Brink}}, \bibinfo {author} {\bibfnamefont
  {J.~M.}\ \bibnamefont {Chow}},\ and\ \bibinfo {author} {\bibfnamefont
  {J.~M.}\ \bibnamefont {Gambetta}},\ }\href@noop {} {\bibfield  {journal}
  {\bibinfo  {journal} {nature}\ }\textbf {\bibinfo {volume} {549}},\ \bibinfo
  {pages} {242} (\bibinfo {year} {2017})}\BibitemShut {NoStop}%
\bibitem [{\citenamefont {Yao}\ \emph {et~al.}(2021)\citenamefont {Yao},
  \citenamefont {Lin},\ and\ \citenamefont {Bukov}}]{Yao2021}%
  \BibitemOpen
  \bibfield  {author} {\bibinfo {author} {\bibfnamefont {J.}~\bibnamefont
  {Yao}}, \bibinfo {author} {\bibfnamefont {L.}~\bibnamefont {Lin}},\ and\
  \bibinfo {author} {\bibfnamefont {M.}~\bibnamefont {Bukov}},\ }\href
  {https://doi.org/10.1103/PhysRevX.11.031070} {\bibfield  {journal} {\bibinfo
  {journal} {Phys. Rev. X}\ }\textbf {\bibinfo {volume} {11}},\ \bibinfo
  {pages} {031070} (\bibinfo {year} {2021})}\BibitemShut {NoStop}%
\bibitem [{\citenamefont {Zhang}\ \emph {et~al.}(2023)\citenamefont {Zhang},
  \citenamefont {Xu}, \citenamefont {Zhang}, \citenamefont {Yung},
  \citenamefont {Huang},\ and\ \citenamefont {Liu}}]{Zhang2023}%
  \BibitemOpen
  \bibfield  {author} {\bibinfo {author} {\bibfnamefont {H.}~\bibnamefont
  {Zhang}}, \bibinfo {author} {\bibfnamefont {X.}~\bibnamefont {Xu}}, \bibinfo
  {author} {\bibfnamefont {C.}~\bibnamefont {Zhang}}, \bibinfo {author}
  {\bibfnamefont {M.-H.}\ \bibnamefont {Yung}}, \bibinfo {author}
  {\bibfnamefont {T.}~\bibnamefont {Huang}},\ and\ \bibinfo {author}
  {\bibfnamefont {Y.}~\bibnamefont {Liu}},\ }\href
  {https://doi.org/10.1103/PhysRevA.108.042611} {\bibfield  {journal} {\bibinfo
   {journal} {Phys. Rev. A}\ }\textbf {\bibinfo {volume} {108}},\ \bibinfo
  {pages} {042611} (\bibinfo {year} {2023})}\BibitemShut {NoStop}%
\bibitem [{\citenamefont {Anschuetz}\ \emph {et~al.}(2019)\citenamefont
  {Anschuetz}, \citenamefont {Olson}, \citenamefont {Aspuru-Guzik},\ and\
  \citenamefont {Cao}}]{Anschuetz2019}%
  \BibitemOpen
  \bibfield  {author} {\bibinfo {author} {\bibfnamefont {E.}~\bibnamefont
  {Anschuetz}}, \bibinfo {author} {\bibfnamefont {J.}~\bibnamefont {Olson}},
  \bibinfo {author} {\bibfnamefont {A.}~\bibnamefont {Aspuru-Guzik}},\ and\
  \bibinfo {author} {\bibfnamefont {Y.}~\bibnamefont {Cao}},\ }in\ \href@noop
  {} {\emph {\bibinfo {booktitle} {Quantum Technology and Optimization
  Problems}}},\ \bibinfo {editor} {edited by\ \bibinfo {editor} {\bibfnamefont
  {S.}~\bibnamefont {Feld}}\ and\ \bibinfo {editor} {\bibfnamefont
  {C.}~\bibnamefont {Linnhoff-Popien}}}\ (\bibinfo  {publisher} {Springer
  International Publishing},\ \bibinfo {address} {Cham},\ \bibinfo {year}
  {2019})\ pp.\ \bibinfo {pages} {74--85}\BibitemShut {NoStop}%
\bibitem [{\citenamefont {Karamlou}\ \emph {et~al.}(2021)\citenamefont
  {Karamlou}, \citenamefont {Simon}, \citenamefont {Katabarwa}, \citenamefont
  {Scholten}, \citenamefont {Peropadre},\ and\ \citenamefont
  {Cao}}]{karamlou2021analyzing}%
  \BibitemOpen
  \bibfield  {author} {\bibinfo {author} {\bibfnamefont {A.~H.}\ \bibnamefont
  {Karamlou}}, \bibinfo {author} {\bibfnamefont {W.~A.}\ \bibnamefont {Simon}},
  \bibinfo {author} {\bibfnamefont {A.}~\bibnamefont {Katabarwa}}, \bibinfo
  {author} {\bibfnamefont {T.~L.}\ \bibnamefont {Scholten}}, \bibinfo {author}
  {\bibfnamefont {B.}~\bibnamefont {Peropadre}},\ and\ \bibinfo {author}
  {\bibfnamefont {Y.}~\bibnamefont {Cao}},\ }\href@noop {} {\bibfield
  {journal} {\bibinfo  {journal} {npj Quantum Information}\ }\textbf {\bibinfo
  {volume} {7}},\ \bibinfo {pages} {156} (\bibinfo {year} {2021})}\BibitemShut
  {NoStop}%
\bibitem [{\citenamefont {Farhi}\ \emph {et~al.}(2014)\citenamefont {Farhi},
  \citenamefont {Goldstone},\ and\ \citenamefont {Gutmann}}]{farhi2014quantum}%
  \BibitemOpen
  \bibfield  {author} {\bibinfo {author} {\bibfnamefont {E.}~\bibnamefont
  {Farhi}}, \bibinfo {author} {\bibfnamefont {J.}~\bibnamefont {Goldstone}},\
  and\ \bibinfo {author} {\bibfnamefont {S.}~\bibnamefont {Gutmann}},\
  }\href@noop {} {\bibinfo {title} {A quantum approximate optimization
  algorithm}} (\bibinfo {year} {2014}),\ \Eprint
  {https://arxiv.org/abs/1411.4028} {arXiv:1411.4028 [quant-ph]} \BibitemShut
  {NoStop}%
\bibitem [{\citenamefont {Venkatesh}\ \emph {et~al.}(2024)\citenamefont
  {Venkatesh}, \citenamefont {Macaluso}, \citenamefont {Nuske}, \citenamefont
  {Klusch},\ and\ \citenamefont {Dengel}}]{venkatesh2024qubitefficient}%
  \BibitemOpen
  \bibfield  {author} {\bibinfo {author} {\bibfnamefont {S.~M.}\ \bibnamefont
  {Venkatesh}}, \bibinfo {author} {\bibfnamefont {A.}~\bibnamefont {Macaluso}},
  \bibinfo {author} {\bibfnamefont {M.}~\bibnamefont {Nuske}}, \bibinfo
  {author} {\bibfnamefont {M.}~\bibnamefont {Klusch}},\ and\ \bibinfo {author}
  {\bibfnamefont {A.}~\bibnamefont {Dengel}},\ }\href@noop {} {\bibinfo {title}
  {Qubit-efficient variational quantum algorithms for image segmentation}}
  (\bibinfo {year} {2024}),\ \Eprint {https://arxiv.org/abs/2405.14405}
  {arXiv:2405.14405 [cs.CV]} \BibitemShut {NoStop}%
\bibitem [{\citenamefont {Sarmina}\ \emph {et~al.}(2024)\citenamefont
  {Sarmina}, \citenamefont {Sun},\ and\ \citenamefont
  {Dong}}]{sarmina2024parameter}%
  \BibitemOpen
  \bibfield  {author} {\bibinfo {author} {\bibfnamefont {B.~G.}\ \bibnamefont
  {Sarmina}}, \bibinfo {author} {\bibfnamefont {G.-H.}\ \bibnamefont {Sun}},\
  and\ \bibinfo {author} {\bibfnamefont {S.-H.}\ \bibnamefont {Dong}},\
  }\href@noop {} {\bibinfo {title} {Parameter optimization comparison in qaoa
  using stochastic hill climbing with random re-starts and local search with
  entangled and non-entangled mixing operators}} (\bibinfo {year} {2024}),\
  \Eprint {https://arxiv.org/abs/2405.08941} {arXiv:2405.08941 [quant-ph]}
  \BibitemShut {NoStop}%
\bibitem [{\citenamefont {Montanez-Barrera}\ and\ \citenamefont
  {Michielsen}(2024)}]{montanezbarrera2024universal}%
  \BibitemOpen
  \bibfield  {author} {\bibinfo {author} {\bibfnamefont {J.~A.}\ \bibnamefont
  {Montanez-Barrera}}\ and\ \bibinfo {author} {\bibfnamefont {K.}~\bibnamefont
  {Michielsen}},\ }\href@noop {} {\bibinfo {title} {Towards a universal qaoa
  protocol: Evidence of quantum advantage in solving combinatorial optimization
  problems}} (\bibinfo {year} {2024}),\ \Eprint
  {https://arxiv.org/abs/2405.09169} {arXiv:2405.09169 [quant-ph]} \BibitemShut
  {NoStop}%
\bibitem [{\citenamefont {Tsvelikhovskiy}\ \emph {et~al.}(2024)\citenamefont
  {Tsvelikhovskiy}, \citenamefont {Safro},\ and\ \citenamefont
  {Alexeev}}]{tsvelikhovskiy2024equivariant}%
  \BibitemOpen
  \bibfield  {author} {\bibinfo {author} {\bibfnamefont {B.}~\bibnamefont
  {Tsvelikhovskiy}}, \bibinfo {author} {\bibfnamefont {I.}~\bibnamefont
  {Safro}},\ and\ \bibinfo {author} {\bibfnamefont {Y.}~\bibnamefont
  {Alexeev}},\ }\href@noop {} {\bibinfo {title} {Equivariant qaoa and the duel
  of the mixers}} (\bibinfo {year} {2024}),\ \Eprint
  {https://arxiv.org/abs/2405.07211} {arXiv:2405.07211 [quant-ph]} \BibitemShut
  {NoStop}%
\bibitem [{\citenamefont {Boy}\ and\ \citenamefont {Wales}(2024)}]{Boy2024}%
  \BibitemOpen
  \bibfield  {author} {\bibinfo {author} {\bibfnamefont {C.}~\bibnamefont
  {Boy}}\ and\ \bibinfo {author} {\bibfnamefont {D.~J.}\ \bibnamefont
  {Wales}},\ }\href {https://doi.org/10.1103/PhysRevA.109.062602} {\bibfield
  {journal} {\bibinfo  {journal} {Phys. Rev. A}\ }\textbf {\bibinfo {volume}
  {109}},\ \bibinfo {pages} {062602} (\bibinfo {year} {2024})}\BibitemShut
  {NoStop}%
\bibitem [{\citenamefont {Lucas}(2014)}]{lucas:2014}%
  \BibitemOpen
  \bibfield  {author} {\bibinfo {author} {\bibfnamefont {A.}~\bibnamefont
  {Lucas}},\ }\bibfield  {journal} {\bibinfo  {journal} {Frontiers in Physics}\
  }\textbf {\bibinfo {volume} {2}},\ \href
  {https://doi.org/10.3389/fphy.2014.00005} {10.3389/fphy.2014.00005} (\bibinfo
  {year} {2014})\BibitemShut {NoStop}%
\bibitem [{\citenamefont {Hadfield}\ \emph {et~al.}(2019)\citenamefont
  {Hadfield}, \citenamefont {Wang}, \citenamefont {O’Gorman}, \citenamefont
  {Rieffel}, \citenamefont {Venturelli},\ and\ \citenamefont
  {Biswas}}]{Hadfield2019}%
  \BibitemOpen
  \bibfield  {author} {\bibinfo {author} {\bibfnamefont {S.}~\bibnamefont
  {Hadfield}}, \bibinfo {author} {\bibfnamefont {Z.}~\bibnamefont {Wang}},
  \bibinfo {author} {\bibfnamefont {B.}~\bibnamefont {O’Gorman}}, \bibinfo
  {author} {\bibfnamefont {E.~G.}\ \bibnamefont {Rieffel}}, \bibinfo {author}
  {\bibfnamefont {D.}~\bibnamefont {Venturelli}},\ and\ \bibinfo {author}
  {\bibfnamefont {R.}~\bibnamefont {Biswas}},\ }\bibfield  {journal} {\bibinfo
  {journal} {Algorithms}\ }\textbf {\bibinfo {volume} {12}},\ \href
  {https://doi.org/10.3390/a12020034} {10.3390/a12020034} (\bibinfo {year}
  {2019})\BibitemShut {NoStop}%
\bibitem [{\citenamefont {Bärtschi}\ and\ \citenamefont
  {Eidenbenz}(2020)}]{Bartschi2020}%
  \BibitemOpen
  \bibfield  {author} {\bibinfo {author} {\bibfnamefont {A.}~\bibnamefont
  {Bärtschi}}\ and\ \bibinfo {author} {\bibfnamefont {S.}~\bibnamefont
  {Eidenbenz}},\ }in\ \href {https://doi.org/10.1109/QCE49297.2020.00020}
  {\emph {\bibinfo {booktitle} {2020 IEEE International Conference on Quantum
  Computing and Engineering (QCE)}}}\ (\bibinfo {year} {2020})\ pp.\ \bibinfo
  {pages} {72--82}\BibitemShut {NoStop}%
\bibitem [{\citenamefont {Villalba-Diez}\ \emph {et~al.}(2022)\citenamefont
  {Villalba-Diez}, \citenamefont {González-Marcos},\ and\ \citenamefont
  {Ordieres-Meré}}]{Villalba-Diez2022}%
  \BibitemOpen
  \bibfield  {author} {\bibinfo {author} {\bibfnamefont {J.}~\bibnamefont
  {Villalba-Diez}}, \bibinfo {author} {\bibfnamefont {A.}~\bibnamefont
  {González-Marcos}},\ and\ \bibinfo {author} {\bibfnamefont {J.~B.}\
  \bibnamefont {Ordieres-Meré}},\ }\bibfield  {journal} {\bibinfo  {journal}
  {Sensors}\ }\textbf {\bibinfo {volume} {22}},\ \href
  {https://doi.org/10.3390/s22010244} {10.3390/s22010244} (\bibinfo {year}
  {2022})\BibitemShut {NoStop}%
\bibitem [{\citenamefont {Golden}\ \emph {et~al.}(2021)\citenamefont {Golden},
  \citenamefont {Bärtschi}, \citenamefont {O’Malley},\ and\ \citenamefont
  {Eidenbenz}}]{Golden2021}%
  \BibitemOpen
  \bibfield  {author} {\bibinfo {author} {\bibfnamefont {J.}~\bibnamefont
  {Golden}}, \bibinfo {author} {\bibfnamefont {A.}~\bibnamefont {Bärtschi}},
  \bibinfo {author} {\bibfnamefont {D.}~\bibnamefont {O’Malley}},\ and\
  \bibinfo {author} {\bibfnamefont {S.}~\bibnamefont {Eidenbenz}},\ }in\ \href
  {https://doi.org/10.1109/QCE52317.2021.00030} {\emph {\bibinfo {booktitle}
  {2021 IEEE International Conference on Quantum Computing and Engineering
  (QCE)}}}\ (\bibinfo {year} {2021})\ pp.\ \bibinfo {pages}
  {137--147}\BibitemShut {NoStop}%
\bibitem [{\citenamefont {Fuchs}\ \emph {et~al.}(2022)\citenamefont {Fuchs},
  \citenamefont {Lye}, \citenamefont {Møll~Nilsen}, \citenamefont {Stasik},\
  and\ \citenamefont {Sartor}}]{Fuchs2022}%
  \BibitemOpen
  \bibfield  {author} {\bibinfo {author} {\bibfnamefont {F.~G.}\ \bibnamefont
  {Fuchs}}, \bibinfo {author} {\bibfnamefont {K.~O.}\ \bibnamefont {Lye}},
  \bibinfo {author} {\bibfnamefont {H.}~\bibnamefont {Møll~Nilsen}}, \bibinfo
  {author} {\bibfnamefont {A.~J.}\ \bibnamefont {Stasik}},\ and\ \bibinfo
  {author} {\bibfnamefont {G.}~\bibnamefont {Sartor}},\ }\bibfield  {journal}
  {\bibinfo  {journal} {Algorithms}\ }\textbf {\bibinfo {volume} {15}},\ \href
  {https://doi.org/10.3390/a15060202} {10.3390/a15060202} (\bibinfo {year}
  {2022})\BibitemShut {NoStop}%
\bibitem [{\citenamefont {Egger}\ \emph {et~al.}(2021)\citenamefont {Egger},
  \citenamefont {Mare{\v{c}}ek},\ and\ \citenamefont {Woerner}}]{egger2021}%
  \BibitemOpen
  \bibfield  {author} {\bibinfo {author} {\bibfnamefont {D.~J.}\ \bibnamefont
  {Egger}}, \bibinfo {author} {\bibfnamefont {J.}~\bibnamefont
  {Mare{\v{c}}ek}},\ and\ \bibinfo {author} {\bibfnamefont {S.}~\bibnamefont
  {Woerner}},\ }\href {https://doi.org/10.22331/q-2021-06-17-479} {\bibfield
  {journal} {\bibinfo  {journal} {Quantum}\ }\textbf {\bibinfo {volume} {5}},\
  \bibinfo {pages} {479} (\bibinfo {year} {2021})}\BibitemShut {NoStop}%
\bibitem [{\citenamefont {Yoshioka}\ \emph {et~al.}(2023)\citenamefont
  {Yoshioka}, \citenamefont {Sasada}, \citenamefont {Nakano},\ and\
  \citenamefont {Fujii}}]{Yoshioka2023}%
  \BibitemOpen
  \bibfield  {author} {\bibinfo {author} {\bibfnamefont {T.}~\bibnamefont
  {Yoshioka}}, \bibinfo {author} {\bibfnamefont {K.}~\bibnamefont {Sasada}},
  \bibinfo {author} {\bibfnamefont {Y.}~\bibnamefont {Nakano}},\ and\ \bibinfo
  {author} {\bibfnamefont {K.}~\bibnamefont {Fujii}},\ }\href
  {https://doi.org/10.1103/PhysRevResearch.5.023071} {\bibfield  {journal}
  {\bibinfo  {journal} {Phys. Rev. Res.}\ }\textbf {\bibinfo {volume} {5}},\
  \bibinfo {pages} {023071} (\bibinfo {year} {2023})}\BibitemShut {NoStop}%
\bibitem [{\citenamefont {Wurtz}\ and\ \citenamefont {Love}(2021)}]{Wurtz2021}%
  \BibitemOpen
  \bibfield  {author} {\bibinfo {author} {\bibfnamefont {J.}~\bibnamefont
  {Wurtz}}\ and\ \bibinfo {author} {\bibfnamefont {P.~J.}\ \bibnamefont
  {Love}},\ }\href {https://doi.org/10.1109/TQE.2021.3122568} {\bibfield
  {journal} {\bibinfo  {journal} {IEEE Transactions on Quantum Engineering}\
  }\textbf {\bibinfo {volume} {2}},\ \bibinfo {pages} {1} (\bibinfo {year}
  {2021})}\BibitemShut {NoStop}%
\bibitem [{\citenamefont {{Gomez Cadavid}}\ \emph {et~al.}(2023)\citenamefont
  {{Gomez Cadavid}}, \citenamefont {{Montalban}}, \citenamefont {{Dalal}},
  \citenamefont {{Solano}},\ and\ \citenamefont {{Hegade}}}]{cadavid2023}%
  \BibitemOpen
  \bibfield  {author} {\bibinfo {author} {\bibfnamefont {A.}~\bibnamefont
  {{Gomez Cadavid}}}, \bibinfo {author} {\bibfnamefont {I.}~\bibnamefont
  {{Montalban}}}, \bibinfo {author} {\bibfnamefont {A.}~\bibnamefont
  {{Dalal}}}, \bibinfo {author} {\bibfnamefont {E.}~\bibnamefont {{Solano}}},\
  and\ \bibinfo {author} {\bibfnamefont {N.~N.}\ \bibnamefont {{Hegade}}},\
  }\href@noop {} {\bibfield  {journal} {\bibinfo  {journal} {arXiv e-prints}\ }
  (\bibinfo {year} {2023})},\ \Eprint {https://arxiv.org/abs/2308.15475}
  {2308.15475 [quant-ph]} \BibitemShut {NoStop}%
\bibitem [{\citenamefont {Ji}\ \emph {et~al.}(2023)\citenamefont {Ji},
  \citenamefont {Koenig},\ and\ \citenamefont {Polian}}]{ji2023improving}%
  \BibitemOpen
  \bibfield  {author} {\bibinfo {author} {\bibfnamefont {Y.}~\bibnamefont
  {Ji}}, \bibinfo {author} {\bibfnamefont {K.~F.}\ \bibnamefont {Koenig}},\
  and\ \bibinfo {author} {\bibfnamefont {I.}~\bibnamefont {Polian}},\
  }\href@noop {} {\bibinfo {title} {Improving the performance of digitized
  counterdiabatic quantum optimization via algorithm-oriented qubit mapping}}
  (\bibinfo {year} {2023}),\ \Eprint {https://arxiv.org/abs/2311.14624}
  {arXiv:2311.14624 [quant-ph]} \BibitemShut {NoStop}%
\bibitem [{\citenamefont {Yu}\ \emph {et~al.}(2022)\citenamefont {Yu},
  \citenamefont {Cao}, \citenamefont {Dewey}, \citenamefont {Wang},
  \citenamefont {Shannon},\ and\ \citenamefont {Joynt}}]{Yu2022}%
  \BibitemOpen
  \bibfield  {author} {\bibinfo {author} {\bibfnamefont {Y.}~\bibnamefont
  {Yu}}, \bibinfo {author} {\bibfnamefont {C.}~\bibnamefont {Cao}}, \bibinfo
  {author} {\bibfnamefont {C.}~\bibnamefont {Dewey}}, \bibinfo {author}
  {\bibfnamefont {X.-B.}\ \bibnamefont {Wang}}, \bibinfo {author}
  {\bibfnamefont {N.}~\bibnamefont {Shannon}},\ and\ \bibinfo {author}
  {\bibfnamefont {R.}~\bibnamefont {Joynt}},\ }\href
  {https://doi.org/10.1103/PhysRevResearch.4.023249} {\bibfield  {journal}
  {\bibinfo  {journal} {Phys. Rev. Res.}\ }\textbf {\bibinfo {volume} {4}},\
  \bibinfo {pages} {023249} (\bibinfo {year} {2022})}\BibitemShut {NoStop}%
\bibitem [{\citenamefont {Zhu}\ \emph {et~al.}(2022)\citenamefont {Zhu},
  \citenamefont {Tang}, \citenamefont {Barron}, \citenamefont
  {Calderon-Vargas}, \citenamefont {Mayhall}, \citenamefont {Barnes},\ and\
  \citenamefont {Economou}}]{zhu2022}%
  \BibitemOpen
  \bibfield  {author} {\bibinfo {author} {\bibfnamefont {L.}~\bibnamefont
  {Zhu}}, \bibinfo {author} {\bibfnamefont {H.~L.}\ \bibnamefont {Tang}},
  \bibinfo {author} {\bibfnamefont {G.~S.}\ \bibnamefont {Barron}}, \bibinfo
  {author} {\bibfnamefont {F.~A.}\ \bibnamefont {Calderon-Vargas}}, \bibinfo
  {author} {\bibfnamefont {N.~J.}\ \bibnamefont {Mayhall}}, \bibinfo {author}
  {\bibfnamefont {E.}~\bibnamefont {Barnes}},\ and\ \bibinfo {author}
  {\bibfnamefont {S.~E.}\ \bibnamefont {Economou}},\ }\href
  {https://doi.org/10.1103/PhysRevResearch.4.033029} {\bibfield  {journal}
  {\bibinfo  {journal} {Phys. Rev. Res.}\ }\textbf {\bibinfo {volume} {4}},\
  \bibinfo {pages} {033029} (\bibinfo {year} {2022})}\BibitemShut {NoStop}%
\bibitem [{\citenamefont {Bravyi}\ \emph {et~al.}(2020)\citenamefont {Bravyi},
  \citenamefont {Kliesch}, \citenamefont {Koenig},\ and\ \citenamefont
  {Tang}}]{Bravyi2020}%
  \BibitemOpen
  \bibfield  {author} {\bibinfo {author} {\bibfnamefont {S.}~\bibnamefont
  {Bravyi}}, \bibinfo {author} {\bibfnamefont {A.}~\bibnamefont {Kliesch}},
  \bibinfo {author} {\bibfnamefont {R.}~\bibnamefont {Koenig}},\ and\ \bibinfo
  {author} {\bibfnamefont {E.}~\bibnamefont {Tang}},\ }\href
  {https://doi.org/10.1103/PhysRevLett.125.260505} {\bibfield  {journal}
  {\bibinfo  {journal} {Phys. Rev. Lett.}\ }\textbf {\bibinfo {volume} {125}},\
  \bibinfo {pages} {260505} (\bibinfo {year} {2020})}\BibitemShut {NoStop}%
\bibitem [{\citenamefont {Zou}(2023)}]{zou2023multiscale}%
  \BibitemOpen
  \bibfield  {author} {\bibinfo {author} {\bibfnamefont {P.}~\bibnamefont
  {Zou}},\ }\href@noop {} {\bibinfo {title} {Multiscale quantum approximate
  optimization algorithm}} (\bibinfo {year} {2023}),\ \Eprint
  {https://arxiv.org/abs/2312.06181} {arXiv:2312.06181 [quant-ph]} \BibitemShut
  {NoStop}%
\bibitem [{\citenamefont {Magann}\ \emph {et~al.}(2022)\citenamefont {Magann},
  \citenamefont {Rudinger}, \citenamefont {Grace},\ and\ \citenamefont
  {Sarovar}}]{Magann2022}%
  \BibitemOpen
  \bibfield  {author} {\bibinfo {author} {\bibfnamefont {A.~B.}\ \bibnamefont
  {Magann}}, \bibinfo {author} {\bibfnamefont {K.~M.}\ \bibnamefont
  {Rudinger}}, \bibinfo {author} {\bibfnamefont {M.~D.}\ \bibnamefont
  {Grace}},\ and\ \bibinfo {author} {\bibfnamefont {M.}~\bibnamefont
  {Sarovar}},\ }\href {https://doi.org/10.1103/PhysRevLett.129.250502}
  {\bibfield  {journal} {\bibinfo  {journal} {Phys. Rev. Lett.}\ }\textbf
  {\bibinfo {volume} {129}},\ \bibinfo {pages} {250502} (\bibinfo {year}
  {2022})}\BibitemShut {NoStop}%
\bibitem [{\citenamefont {Blekos}\ \emph {et~al.}(2024)\citenamefont {Blekos},
  \citenamefont {Brand}, \citenamefont {Ceschini}, \citenamefont {Chou},
  \citenamefont {Li}, \citenamefont {Pandya},\ and\ \citenamefont
  {Summer}}]{BLEKOS20241}%
  \BibitemOpen
  \bibfield  {author} {\bibinfo {author} {\bibfnamefont {K.}~\bibnamefont
  {Blekos}}, \bibinfo {author} {\bibfnamefont {D.}~\bibnamefont {Brand}},
  \bibinfo {author} {\bibfnamefont {A.}~\bibnamefont {Ceschini}}, \bibinfo
  {author} {\bibfnamefont {C.-H.}\ \bibnamefont {Chou}}, \bibinfo {author}
  {\bibfnamefont {R.-H.}\ \bibnamefont {Li}}, \bibinfo {author} {\bibfnamefont
  {K.}~\bibnamefont {Pandya}},\ and\ \bibinfo {author} {\bibfnamefont
  {A.}~\bibnamefont {Summer}},\ }\href
  {https://doi.org/https://doi.org/10.1016/j.physrep.2024.03.002} {\bibfield
  {journal} {\bibinfo  {journal} {Physics Reports}\ }\textbf {\bibinfo {volume}
  {1068}},\ \bibinfo {pages} {1} (\bibinfo {year} {2024})}\BibitemShut
  {NoStop}%
\bibitem [{\citenamefont {Misra-Spieldenner}\ \emph {et~al.}(2023)\citenamefont
  {Misra-Spieldenner}, \citenamefont {Bode}, \citenamefont {Schuhmacher},
  \citenamefont {Stollenwerk}, \citenamefont {Bagrets},\ and\ \citenamefont
  {Wilhelm}}]{Misra-Spieldenner2023}%
  \BibitemOpen
  \bibfield  {author} {\bibinfo {author} {\bibfnamefont {A.}~\bibnamefont
  {Misra-Spieldenner}}, \bibinfo {author} {\bibfnamefont {T.}~\bibnamefont
  {Bode}}, \bibinfo {author} {\bibfnamefont {P.~K.}\ \bibnamefont
  {Schuhmacher}}, \bibinfo {author} {\bibfnamefont {T.}~\bibnamefont
  {Stollenwerk}}, \bibinfo {author} {\bibfnamefont {D.}~\bibnamefont
  {Bagrets}},\ and\ \bibinfo {author} {\bibfnamefont {F.~K.}\ \bibnamefont
  {Wilhelm}},\ }\href {https://doi.org/10.1103/PRXQuantum.4.030335} {\bibfield
  {journal} {\bibinfo  {journal} {PRX Quantum}\ }\textbf {\bibinfo {volume}
  {4}},\ \bibinfo {pages} {030335} (\bibinfo {year} {2023})}\BibitemShut
  {NoStop}%
\bibitem [{\citenamefont {Chandarana}\ \emph {et~al.}(2022)\citenamefont
  {Chandarana}, \citenamefont {Hegade}, \citenamefont {Paul}, \citenamefont
  {Albarr\'an-Arriagada}, \citenamefont {Solano}, \citenamefont {del Campo},\
  and\ \citenamefont {Chen}}]{Chandarana:2022}%
  \BibitemOpen
  \bibfield  {author} {\bibinfo {author} {\bibfnamefont {P.}~\bibnamefont
  {Chandarana}}, \bibinfo {author} {\bibfnamefont {N.~N.}\ \bibnamefont
  {Hegade}}, \bibinfo {author} {\bibfnamefont {K.}~\bibnamefont {Paul}},
  \bibinfo {author} {\bibfnamefont {F.}~\bibnamefont {Albarr\'an-Arriagada}},
  \bibinfo {author} {\bibfnamefont {E.}~\bibnamefont {Solano}}, \bibinfo
  {author} {\bibfnamefont {A.}~\bibnamefont {del Campo}},\ and\ \bibinfo
  {author} {\bibfnamefont {X.}~\bibnamefont {Chen}},\ }\href
  {https://doi.org/10.1103/PhysRevResearch.4.013141} {\bibfield  {journal}
  {\bibinfo  {journal} {Phys. Rev. Research}\ }\textbf {\bibinfo {volume}
  {4}},\ \bibinfo {pages} {013141} (\bibinfo {year} {2022})}\BibitemShut
  {NoStop}%
\bibitem [{\citenamefont {Chai}\ \emph {et~al.}(2022)\citenamefont {Chai},
  \citenamefont {Han}, \citenamefont {Wu}, \citenamefont {Li}, \citenamefont
  {Dou},\ and\ \citenamefont {Guo}}]{Chai:2022}%
  \BibitemOpen
  \bibfield  {author} {\bibinfo {author} {\bibfnamefont {Y.}~\bibnamefont
  {Chai}}, \bibinfo {author} {\bibfnamefont {Y.-J.}\ \bibnamefont {Han}},
  \bibinfo {author} {\bibfnamefont {Y.-C.}\ \bibnamefont {Wu}}, \bibinfo
  {author} {\bibfnamefont {Y.}~\bibnamefont {Li}}, \bibinfo {author}
  {\bibfnamefont {M.}~\bibnamefont {Dou}},\ and\ \bibinfo {author}
  {\bibfnamefont {G.-P.}\ \bibnamefont {Guo}},\ }\href
  {https://doi.org/10.1103/PhysRevA.105.042415} {\bibfield  {journal} {\bibinfo
   {journal} {Phys. Rev. A}\ }\textbf {\bibinfo {volume} {105}},\ \bibinfo
  {pages} {042415} (\bibinfo {year} {2022})}\BibitemShut {NoStop}%
\bibitem [{\citenamefont {Wurtz}\ and\ \citenamefont
  {Love}(2022)}]{Wurtz2022counterdiabaticity}%
  \BibitemOpen
  \bibfield  {author} {\bibinfo {author} {\bibfnamefont {J.}~\bibnamefont
  {Wurtz}}\ and\ \bibinfo {author} {\bibfnamefont {P.~J.}\ \bibnamefont
  {Love}},\ }\href {https://doi.org/10.22331/q-2022-01-27-635} {\bibfield
  {journal} {\bibinfo  {journal} {{Quantum}}\ }\textbf {\bibinfo {volume}
  {6}},\ \bibinfo {pages} {635} (\bibinfo {year} {2022})}\BibitemShut {NoStop}%
\bibitem [{\citenamefont {Chandarana}\ \emph {et~al.}(2023)\citenamefont
  {Chandarana}, \citenamefont {Hegade}, \citenamefont {Montalban},
  \citenamefont {Solano},\ and\ \citenamefont
  {Chen}}]{chandarana2022digitizedcounterdiabatic}%
  \BibitemOpen
  \bibfield  {author} {\bibinfo {author} {\bibfnamefont {P.}~\bibnamefont
  {Chandarana}}, \bibinfo {author} {\bibfnamefont {N.~N.}\ \bibnamefont
  {Hegade}}, \bibinfo {author} {\bibfnamefont {I.}~\bibnamefont {Montalban}},
  \bibinfo {author} {\bibfnamefont {E.}~\bibnamefont {Solano}},\ and\ \bibinfo
  {author} {\bibfnamefont {X.}~\bibnamefont {Chen}},\ }\href
  {https://doi.org/10.1103/PhysRevApplied.20.014024} {\bibfield  {journal}
  {\bibinfo  {journal} {Phys. Rev. Appl.}\ }\textbf {\bibinfo {volume} {20}},\
  \bibinfo {pages} {014024} (\bibinfo {year} {2023})}\BibitemShut {NoStop}%
\bibitem [{\citenamefont {Chalupnik}\ \emph {et~al.}(2022)\citenamefont
  {Chalupnik}, \citenamefont {Melo}, \citenamefont {Alexeev},\ and\
  \citenamefont {Galda}}]{Chalupnik2022}%
  \BibitemOpen
  \bibfield  {author} {\bibinfo {author} {\bibfnamefont {M.}~\bibnamefont
  {Chalupnik}}, \bibinfo {author} {\bibfnamefont {H.}~\bibnamefont {Melo}},
  \bibinfo {author} {\bibfnamefont {Y.}~\bibnamefont {Alexeev}},\ and\ \bibinfo
  {author} {\bibfnamefont {A.}~\bibnamefont {Galda}},\ }in\ \href
  {https://doi.org/10.1109/QCE53715.2022.00028} {\emph {\bibinfo {booktitle}
  {2022 IEEE International Conference on Quantum Computing and Engineering
  (QCE)}}}\ (\bibinfo {year} {2022})\ pp.\ \bibinfo {pages}
  {97--103}\BibitemShut {NoStop}%
\bibitem [{\citenamefont {Chen}\ \emph {et~al.}(2010)\citenamefont {Chen},
  \citenamefont {Lizuain}, \citenamefont {Ruschhaupt}, \citenamefont
  {Gu\'ery-Odelin},\ and\ \citenamefont {Muga}}]{Chen2010}%
  \BibitemOpen
  \bibfield  {author} {\bibinfo {author} {\bibfnamefont {X.}~\bibnamefont
  {Chen}}, \bibinfo {author} {\bibfnamefont {I.}~\bibnamefont {Lizuain}},
  \bibinfo {author} {\bibfnamefont {A.}~\bibnamefont {Ruschhaupt}}, \bibinfo
  {author} {\bibfnamefont {D.}~\bibnamefont {Gu\'ery-Odelin}},\ and\ \bibinfo
  {author} {\bibfnamefont {J.~G.}\ \bibnamefont {Muga}},\ }\href
  {https://doi.org/10.1103/PhysRevLett.105.123003} {\bibfield  {journal}
  {\bibinfo  {journal} {Phys. Rev. Lett.}\ }\textbf {\bibinfo {volume} {105}},\
  \bibinfo {pages} {123003} (\bibinfo {year} {2010})}\BibitemShut {NoStop}%
\bibitem [{\citenamefont {Gu\'ery-Odelin}\ \emph {et~al.}(2019)\citenamefont
  {Gu\'ery-Odelin}, \citenamefont {Ruschhaupt}, \citenamefont {Kiely},
  \citenamefont {Torrontegui}, \citenamefont {Mart\'{\i}nez-Garaot},\ and\
  \citenamefont {Muga}}]{Guery-Odelin2019}%
  \BibitemOpen
  \bibfield  {author} {\bibinfo {author} {\bibfnamefont {D.}~\bibnamefont
  {Gu\'ery-Odelin}}, \bibinfo {author} {\bibfnamefont {A.}~\bibnamefont
  {Ruschhaupt}}, \bibinfo {author} {\bibfnamefont {A.}~\bibnamefont {Kiely}},
  \bibinfo {author} {\bibfnamefont {E.}~\bibnamefont {Torrontegui}}, \bibinfo
  {author} {\bibfnamefont {S.}~\bibnamefont {Mart\'{\i}nez-Garaot}},\ and\
  \bibinfo {author} {\bibfnamefont {J.~G.}\ \bibnamefont {Muga}},\ }\href
  {https://doi.org/10.1103/RevModPhys.91.045001} {\bibfield  {journal}
  {\bibinfo  {journal} {Rev. Mod. Phys.}\ }\textbf {\bibinfo {volume} {91}},\
  \bibinfo {pages} {045001} (\bibinfo {year} {2019})}\BibitemShut {NoStop}%
\bibitem [{\citenamefont {Vizzuso}\ \emph {et~al.}(2023)\citenamefont
  {Vizzuso}, \citenamefont {Passarelli}, \citenamefont {Cantele},\ and\
  \citenamefont {Lucignano}}]{vizzuso2023}%
  \BibitemOpen
  \bibfield  {author} {\bibinfo {author} {\bibfnamefont {M.}~\bibnamefont
  {Vizzuso}}, \bibinfo {author} {\bibfnamefont {G.}~\bibnamefont {Passarelli}},
  \bibinfo {author} {\bibfnamefont {G.}~\bibnamefont {Cantele}},\ and\ \bibinfo
  {author} {\bibfnamefont {P.}~\bibnamefont {Lucignano}},\ }\href@noop {}
  {\bibfield  {journal} {\bibinfo  {journal} {New Journal of Physics}\ }
  (\bibinfo {year} {2023})}\BibitemShut {NoStop}%
\bibitem [{\citenamefont {Albash}\ and\ \citenamefont
  {Lidar}(2018)}]{albash:2018}%
  \BibitemOpen
  \bibfield  {author} {\bibinfo {author} {\bibfnamefont {T.}~\bibnamefont
  {Albash}}\ and\ \bibinfo {author} {\bibfnamefont {D.~A.}\ \bibnamefont
  {Lidar}},\ }\href {https://doi.org/10.1103/RevModPhys.90.015002} {\bibfield
  {journal} {\bibinfo  {journal} {Rev. Mod. Phys.}\ }\textbf {\bibinfo {volume}
  {90}},\ \bibinfo {pages} {015002} (\bibinfo {year} {2018})}\BibitemShut
  {NoStop}%
\bibitem [{\citenamefont {{White}}\ \emph {et~al.}(2021)\citenamefont
  {{White}}, \citenamefont {{Gui}}, \citenamefont {{Saleem}},\ and\
  \citenamefont {{Suchara}}}]{White2021}%
  \BibitemOpen
  \bibfield  {author} {\bibinfo {author} {\bibfnamefont {N.}~\bibnamefont
  {{White}}}, \bibinfo {author} {\bibfnamefont {K.}~\bibnamefont {{Gui}}},
  \bibinfo {author} {\bibfnamefont {Z.}~\bibnamefont {{Saleem}}},\ and\
  \bibinfo {author} {\bibfnamefont {M.}~\bibnamefont {{Suchara}}},\ }in\
  \href@noop {} {\emph {\bibinfo {booktitle} {APS March Meeting Abstracts}}},\
  \bibinfo {series} {APS Meeting Abstracts}, Vol.\ \bibinfo {volume} {2021}\
  (\bibinfo {year} {2021})\ p.\ \bibinfo {pages} {H71.157}\BibitemShut
  {NoStop}%
\bibitem [{\citenamefont {Hoffman}\ and\ \citenamefont
  {Padberg}(2001)}]{Hoffman2001}%
  \BibitemOpen
  \bibfield  {author} {\bibinfo {author} {\bibfnamefont {K.~L.}\ \bibnamefont
  {Hoffman}}\ and\ \bibinfo {author} {\bibfnamefont {M.}~\bibnamefont
  {Padberg}},\ }\bibinfo {title} {Traveling salesman problem (tsp)traveling
  salesman problem},\ in\ \href {https://doi.org/10.1007/1-4020-0611-X_1068}
  {\emph {\bibinfo {booktitle} {Encyclopedia of Operations Research and
  Management Science}}},\ \bibinfo {editor} {edited by\ \bibinfo {editor}
  {\bibfnamefont {S.~I.}\ \bibnamefont {Gass}}\ and\ \bibinfo {editor}
  {\bibfnamefont {C.~M.}\ \bibnamefont {Harris}}}\ (\bibinfo  {publisher}
  {Springer US},\ \bibinfo {address} {New York, NY},\ \bibinfo {year} {2001})\
  pp.\ \bibinfo {pages} {849--853}\BibitemShut {NoStop}%
\bibitem [{\citenamefont {Graham}\ and\ \citenamefont
  {Hell}(1985)}]{Graham1985}%
  \BibitemOpen
  \bibfield  {author} {\bibinfo {author} {\bibfnamefont {R.}~\bibnamefont
  {Graham}}\ and\ \bibinfo {author} {\bibfnamefont {P.}~\bibnamefont {Hell}},\
  }\href {https://doi.org/10.1109/MAHC.1985.10011} {\bibfield  {journal}
  {\bibinfo  {journal} {Annals of the History of Computing}\ }\textbf {\bibinfo
  {volume} {7}},\ \bibinfo {pages} {43} (\bibinfo {year} {1985})}\BibitemShut
  {NoStop}%
\bibitem [{\citenamefont {Salkin}\ and\ \citenamefont
  {De~Kluyver}(1975)}]{Salkin1975}%
  \BibitemOpen
  \bibfield  {author} {\bibinfo {author} {\bibfnamefont {H.~M.}\ \bibnamefont
  {Salkin}}\ and\ \bibinfo {author} {\bibfnamefont {C.~A.}\ \bibnamefont
  {De~Kluyver}},\ }\href {https://doi.org/10.1002/nav.3800220110} {\bibfield
  {journal} {\bibinfo  {journal} {Naval Research Logistics Quarterly}\ }\textbf
  {\bibinfo {volume} {22}},\ \bibinfo {pages} {127–144} (\bibinfo {year}
  {1975})}\BibitemShut {NoStop}%
\bibitem [{\citenamefont {Papadimitriou}\ and\ \citenamefont
  {Steiglitz}(1998)}]{papadimitriou1998combinatorial}%
  \BibitemOpen
  \bibfield  {author} {\bibinfo {author} {\bibfnamefont {C.~H.}\ \bibnamefont
  {Papadimitriou}}\ and\ \bibinfo {author} {\bibfnamefont {K.}~\bibnamefont
  {Steiglitz}},\ }\href@noop {} {\emph {\bibinfo {title} {Combinatorial
  optimization: algorithms and complexity}}}\ (\bibinfo  {publisher} {Courier
  Corporation},\ \bibinfo {year} {1998})\BibitemShut {NoStop}%
\bibitem [{\citenamefont {Trevisan}(2012)}]{Trevisan2012}%
  \BibitemOpen
  \bibfield  {author} {\bibinfo {author} {\bibfnamefont {L.}~\bibnamefont
  {Trevisan}},\ }\href {https://doi.org/10.1137/090773714} {\bibfield
  {journal} {\bibinfo  {journal} {SIAM Journal on Computing}\ }\textbf
  {\bibinfo {volume} {41}},\ \bibinfo {pages} {1769} (\bibinfo {year}
  {2012})},\ \Eprint {https://arxiv.org/abs/https://doi.org/10.1137/090773714}
  {https://doi.org/10.1137/090773714} \BibitemShut {NoStop}%
\bibitem [{\citenamefont {Barahona}(1983)}]{BARAHONA1983107}%
  \BibitemOpen
  \bibfield  {author} {\bibinfo {author} {\bibfnamefont {F.}~\bibnamefont
  {Barahona}},\ }\href
  {https://doi.org/https://doi.org/10.1016/0167-6377(83)90016-0} {\bibfield
  {journal} {\bibinfo  {journal} {Operations Research Letters}\ }\textbf
  {\bibinfo {volume} {2}},\ \bibinfo {pages} {107} (\bibinfo {year}
  {1983})}\BibitemShut {NoStop}%
\bibitem [{\citenamefont {Poljak}\ and\ \citenamefont
  {Rendl}(1995)}]{POLJAK1995249}%
  \BibitemOpen
  \bibfield  {author} {\bibinfo {author} {\bibfnamefont {S.}~\bibnamefont
  {Poljak}}\ and\ \bibinfo {author} {\bibfnamefont {F.}~\bibnamefont {Rendl}},\
  }\href {https://doi.org/https://doi.org/10.1016/0166-218X(94)00155-7}
  {\bibfield  {journal} {\bibinfo  {journal} {Discrete Applied Mathematics}\
  }\textbf {\bibinfo {volume} {62}},\ \bibinfo {pages} {249} (\bibinfo {year}
  {1995})}\BibitemShut {NoStop}%
\bibitem [{\citenamefont {Cook}(2000)}]{cook2000p}%
  \BibitemOpen
  \bibfield  {author} {\bibinfo {author} {\bibfnamefont {S.}~\bibnamefont
  {Cook}},\ }\href@noop {} {\bibfield  {journal} {\bibinfo  {journal} {Clay
  Mathematics Institute}\ }\textbf {\bibinfo {volume} {2}},\ \bibinfo {pages}
  {6} (\bibinfo {year} {2000})}\BibitemShut {NoStop}%
\bibitem [{\citenamefont {Kadowaki}\ and\ \citenamefont
  {Nishimori}(1998)}]{kadowaki:1998}%
  \BibitemOpen
  \bibfield  {author} {\bibinfo {author} {\bibfnamefont {T.}~\bibnamefont
  {Kadowaki}}\ and\ \bibinfo {author} {\bibfnamefont {H.}~\bibnamefont
  {Nishimori}},\ }\href {https://doi.org/10.1103/PhysRevE.58.5355} {\bibfield
  {journal} {\bibinfo  {journal} {Phys. Rev. E}\ }\textbf {\bibinfo {volume}
  {58}},\ \bibinfo {pages} {5355} (\bibinfo {year} {1998})}\BibitemShut
  {NoStop}%
\bibitem [{\citenamefont {Serafini}(1994)}]{simulated-annealing-book}%
  \BibitemOpen
  \bibfield  {author} {\bibinfo {author} {\bibfnamefont {P.}~\bibnamefont
  {Serafini}},\ }in\ \href@noop {} {\emph {\bibinfo {booktitle} {Multiple
  Criteria Decision Making}}},\ \bibinfo {editor} {edited by\ \bibinfo {editor}
  {\bibfnamefont {G.~H.}\ \bibnamefont {Tzeng}}, \bibinfo {editor}
  {\bibfnamefont {H.~F.}\ \bibnamefont {Wang}}, \bibinfo {editor}
  {\bibfnamefont {U.~P.}\ \bibnamefont {Wen}},\ and\ \bibinfo {editor}
  {\bibfnamefont {P.~L.}\ \bibnamefont {Yu}}}\ (\bibinfo  {publisher} {Springer
  New York},\ \bibinfo {address} {New York, NY},\ \bibinfo {year} {1994})\ pp.\
  \bibinfo {pages} {283--292}\BibitemShut {NoStop}%
\bibitem [{\citenamefont {Hegde}\ \emph {et~al.}(2022)\citenamefont {Hegde},
  \citenamefont {Passarelli}, \citenamefont {Scocco},\ and\ \citenamefont
  {Lucignano}}]{phegde:ga}%
  \BibitemOpen
  \bibfield  {author} {\bibinfo {author} {\bibfnamefont {P.~R.}\ \bibnamefont
  {Hegde}}, \bibinfo {author} {\bibfnamefont {G.}~\bibnamefont {Passarelli}},
  \bibinfo {author} {\bibfnamefont {A.}~\bibnamefont {Scocco}},\ and\ \bibinfo
  {author} {\bibfnamefont {P.}~\bibnamefont {Lucignano}},\ }\href
  {https://doi.org/10.1103/PhysRevA.105.012612} {\bibfield  {journal} {\bibinfo
   {journal} {Phys. Rev. A}\ }\textbf {\bibinfo {volume} {105}},\ \bibinfo
  {pages} {012612} (\bibinfo {year} {2022})}\BibitemShut {NoStop}%
\bibitem [{\citenamefont {Passarelli}\ \emph {et~al.}(2019)\citenamefont
  {Passarelli}, \citenamefont {Cataudella},\ and\ \citenamefont
  {Lucignano}}]{gpassarelli:qa1}%
  \BibitemOpen
  \bibfield  {author} {\bibinfo {author} {\bibfnamefont {G.}~\bibnamefont
  {Passarelli}}, \bibinfo {author} {\bibfnamefont {V.}~\bibnamefont
  {Cataudella}},\ and\ \bibinfo {author} {\bibfnamefont {P.}~\bibnamefont
  {Lucignano}},\ }\href {https://doi.org/10.1103/PhysRevB.100.024302}
  {\bibfield  {journal} {\bibinfo  {journal} {Phys. Rev. B}\ }\textbf {\bibinfo
  {volume} {100}},\ \bibinfo {pages} {024302} (\bibinfo {year}
  {2019})}\BibitemShut {NoStop}%
\bibitem [{\citenamefont {Passarelli}\ \emph
  {et~al.}(2020{\natexlab{a}})\citenamefont {Passarelli}, \citenamefont {Yip},
  \citenamefont {Lidar}, \citenamefont {Nishimori},\ and\ \citenamefont
  {Lucignano}}]{gpassarelli:qa2}%
  \BibitemOpen
  \bibfield  {author} {\bibinfo {author} {\bibfnamefont {G.}~\bibnamefont
  {Passarelli}}, \bibinfo {author} {\bibfnamefont {K.-W.}\ \bibnamefont {Yip}},
  \bibinfo {author} {\bibfnamefont {D.~A.}\ \bibnamefont {Lidar}}, \bibinfo
  {author} {\bibfnamefont {H.}~\bibnamefont {Nishimori}},\ and\ \bibinfo
  {author} {\bibfnamefont {P.}~\bibnamefont {Lucignano}},\ }\href
  {https://doi.org/10.1103/PhysRevA.101.022331} {\bibfield  {journal} {\bibinfo
   {journal} {Phys. Rev. A}\ }\textbf {\bibinfo {volume} {101}},\ \bibinfo
  {pages} {022331} (\bibinfo {year} {2020}{\natexlab{a}})}\BibitemShut
  {NoStop}%
\bibitem [{\citenamefont {Passarelli}\ \emph
  {et~al.}(2020{\natexlab{b}})\citenamefont {Passarelli}, \citenamefont
  {Cataudella}, \citenamefont {Fazio},\ and\ \citenamefont
  {Lucignano}}]{gpassarelli:qa3}%
  \BibitemOpen
  \bibfield  {author} {\bibinfo {author} {\bibfnamefont {G.}~\bibnamefont
  {Passarelli}}, \bibinfo {author} {\bibfnamefont {V.}~\bibnamefont
  {Cataudella}}, \bibinfo {author} {\bibfnamefont {R.}~\bibnamefont {Fazio}},\
  and\ \bibinfo {author} {\bibfnamefont {P.}~\bibnamefont {Lucignano}},\ }\href
  {https://doi.org/10.1103/PhysRevResearch.2.013283} {\bibfield  {journal}
  {\bibinfo  {journal} {Phys. Rev. Research}\ }\textbf {\bibinfo {volume}
  {2}},\ \bibinfo {pages} {013283} (\bibinfo {year}
  {2020}{\natexlab{b}})}\BibitemShut {NoStop}%
\bibitem [{\citenamefont {Passarelli}\ \emph {et~al.}(2018)\citenamefont
  {Passarelli}, \citenamefont {De~Filippis}, \citenamefont {Cataudella},\ and\
  \citenamefont {Lucignano}}]{gpassarelli:qa4}%
  \BibitemOpen
  \bibfield  {author} {\bibinfo {author} {\bibfnamefont {G.}~\bibnamefont
  {Passarelli}}, \bibinfo {author} {\bibfnamefont {G.}~\bibnamefont
  {De~Filippis}}, \bibinfo {author} {\bibfnamefont {V.}~\bibnamefont
  {Cataudella}},\ and\ \bibinfo {author} {\bibfnamefont {P.}~\bibnamefont
  {Lucignano}},\ }\href {https://doi.org/10.1103/PhysRevA.97.022319} {\bibfield
   {journal} {\bibinfo  {journal} {Phys. Rev. A}\ }\textbf {\bibinfo {volume}
  {97}},\ \bibinfo {pages} {022319} (\bibinfo {year} {2018})}\BibitemShut
  {NoStop}%
\bibitem [{\citenamefont {Passarelli}\ \emph
  {et~al.}(2022{\natexlab{a}})\citenamefont {Passarelli}, \citenamefont
  {Fazio},\ and\ \citenamefont {Lucignano}}]{gpassarelli:qa5}%
  \BibitemOpen
  \bibfield  {author} {\bibinfo {author} {\bibfnamefont {G.}~\bibnamefont
  {Passarelli}}, \bibinfo {author} {\bibfnamefont {R.}~\bibnamefont {Fazio}},\
  and\ \bibinfo {author} {\bibfnamefont {P.}~\bibnamefont {Lucignano}},\ }\href
  {https://doi.org/10.1103/PhysRevA.105.022618} {\bibfield  {journal} {\bibinfo
   {journal} {Phys. Rev. A}\ }\textbf {\bibinfo {volume} {105}},\ \bibinfo
  {pages} {022618} (\bibinfo {year} {2022}{\natexlab{a}})}\BibitemShut
  {NoStop}%
\bibitem [{\citenamefont {Passarelli}\ \emph
  {et~al.}(2022{\natexlab{b}})\citenamefont {Passarelli}, \citenamefont {Yip},
  \citenamefont {Lidar},\ and\ \citenamefont {Lucignano}}]{gpassarelli:qa6}%
  \BibitemOpen
  \bibfield  {author} {\bibinfo {author} {\bibfnamefont {G.}~\bibnamefont
  {Passarelli}}, \bibinfo {author} {\bibfnamefont {K.-W.}\ \bibnamefont {Yip}},
  \bibinfo {author} {\bibfnamefont {D.~A.}\ \bibnamefont {Lidar}},\ and\
  \bibinfo {author} {\bibfnamefont {P.}~\bibnamefont {Lucignano}},\ }\href
  {https://doi.org/10.1103/PhysRevA.105.032431} {\bibfield  {journal} {\bibinfo
   {journal} {Phys. Rev. A}\ }\textbf {\bibinfo {volume} {105}},\ \bibinfo
  {pages} {032431} (\bibinfo {year} {2022}{\natexlab{b}})}\BibitemShut
  {NoStop}%
\bibitem [{\citenamefont {Passarelli}\ and\ \citenamefont
  {Lucignano}(2023)}]{gpassarelli:qa7}%
  \BibitemOpen
  \bibfield  {author} {\bibinfo {author} {\bibfnamefont {G.}~\bibnamefont
  {Passarelli}}\ and\ \bibinfo {author} {\bibfnamefont {P.}~\bibnamefont
  {Lucignano}},\ }\href {https://doi.org/10.1103/PhysRevA.107.022607}
  {\bibfield  {journal} {\bibinfo  {journal} {Phys. Rev. A}\ }\textbf {\bibinfo
  {volume} {107}},\ \bibinfo {pages} {022607} (\bibinfo {year}
  {2023})}\BibitemShut {NoStop}%
\bibitem [{\citenamefont {Hegde}\ \emph {et~al.}(2023)\citenamefont {Hegde},
  \citenamefont {Passarelli}, \citenamefont {Cantele},\ and\ \citenamefont
  {Lucignano}}]{10.1088/1367-2630/ace547}%
  \BibitemOpen
  \bibfield  {author} {\bibinfo {author} {\bibfnamefont {P.~R.}\ \bibnamefont
  {Hegde}}, \bibinfo {author} {\bibfnamefont {G.}~\bibnamefont {Passarelli}},
  \bibinfo {author} {\bibfnamefont {G.}~\bibnamefont {Cantele}},\ and\ \bibinfo
  {author} {\bibfnamefont {P.}~\bibnamefont {Lucignano}},\ }\bibfield
  {journal} {\bibinfo  {journal} {New Journal of Physics}\ }\href
  {https://doi.org/10.1088/1367-2630/ace547} {10.1088/1367-2630/ace547}
  (\bibinfo {year} {2023})\BibitemShut {NoStop}%
\bibitem [{\citenamefont {{Farhi}}\ \emph {et~al.}(2000)\citenamefont
  {{Farhi}}, \citenamefont {{Goldstone}}, \citenamefont {{Gutmann}},\ and\
  \citenamefont {{Sipser}}}]{farhi:2000}%
  \BibitemOpen
  \bibfield  {author} {\bibinfo {author} {\bibfnamefont {E.}~\bibnamefont
  {{Farhi}}}, \bibinfo {author} {\bibfnamefont {J.}~\bibnamefont
  {{Goldstone}}}, \bibinfo {author} {\bibfnamefont {S.}~\bibnamefont
  {{Gutmann}}},\ and\ \bibinfo {author} {\bibfnamefont {M.}~\bibnamefont
  {{Sipser}}},\ }\href@noop {} {\bibfield  {journal} {\bibinfo  {journal}
  {arXiv e-prints}\ } (\bibinfo {year} {2000})},\ \Eprint
  {https://arxiv.org/abs/quant-ph/0001106} {arXiv:quant-ph/0001106 [quant-ph]}
  \BibitemShut {NoStop}%
\bibitem [{\citenamefont {del Campo}(2013)}]{delCampo2013Shortcuts}%
  \BibitemOpen
  \bibfield  {author} {\bibinfo {author} {\bibfnamefont {A.}~\bibnamefont {del
  Campo}},\ }\href {https://doi.org/10.1103/PhysRevLett.111.100502} {\bibfield
  {journal} {\bibinfo  {journal} {Phys. Rev. Lett.}\ }\textbf {\bibinfo
  {volume} {111}},\ \bibinfo {pages} {100502} (\bibinfo {year}
  {2013})}\BibitemShut {NoStop}%
\bibitem [{\citenamefont {Casas}\ \emph {et~al.}(2012)\citenamefont {Casas},
  \citenamefont {Murua},\ and\ \citenamefont {Nadinic}}]{casas:zassenhaus}%
  \BibitemOpen
  \bibfield  {author} {\bibinfo {author} {\bibfnamefont {F.}~\bibnamefont
  {Casas}}, \bibinfo {author} {\bibfnamefont {A.}~\bibnamefont {Murua}},\ and\
  \bibinfo {author} {\bibfnamefont {M.}~\bibnamefont {Nadinic}},\ }\href
  {https://doi.org/https://doi.org/10.1016/j.cpc.2012.06.006} {\bibfield
  {journal} {\bibinfo  {journal} {Computer Physics Communications}\ }\textbf
  {\bibinfo {volume} {183}},\ \bibinfo {pages} {2386} (\bibinfo {year}
  {2012})}\BibitemShut {NoStop}%
\bibitem [{\citenamefont {Byrd}\ \emph {et~al.}(1995)\citenamefont {Byrd},
  \citenamefont {Lu}, \citenamefont {Nocedal},\ and\ \citenamefont
  {Zhu}}]{Byrd1995}%
  \BibitemOpen
  \bibfield  {author} {\bibinfo {author} {\bibfnamefont {R.~H.}\ \bibnamefont
  {Byrd}}, \bibinfo {author} {\bibfnamefont {P.}~\bibnamefont {Lu}}, \bibinfo
  {author} {\bibfnamefont {J.}~\bibnamefont {Nocedal}},\ and\ \bibinfo {author}
  {\bibfnamefont {C.}~\bibnamefont {Zhu}},\ }\href
  {https://doi.org/10.1137/0916069} {\bibfield  {journal} {\bibinfo  {journal}
  {SIAM Journal on Scientific Computing}\ }\textbf {\bibinfo {volume} {16}},\
  \bibinfo {pages} {1190} (\bibinfo {year} {1995})},\ \Eprint
  {https://arxiv.org/abs/https://doi.org/10.1137/0916069}
  {https://doi.org/10.1137/0916069} \BibitemShut {NoStop}%
\bibitem [{\citenamefont {Santoro}\ \emph {et~al.}(2002)\citenamefont
  {Santoro}, \citenamefont {Marto{\v n}{\'a}k}, \citenamefont {Tosatti},\ and\
  \citenamefont {Car}}]{santoro-martonak}%
  \BibitemOpen
  \bibfield  {author} {\bibinfo {author} {\bibfnamefont {G.~E.}\ \bibnamefont
  {Santoro}}, \bibinfo {author} {\bibfnamefont {R.}~\bibnamefont {Marto{\v
  n}{\'a}k}}, \bibinfo {author} {\bibfnamefont {E.}~\bibnamefont {Tosatti}},\
  and\ \bibinfo {author} {\bibfnamefont {R.}~\bibnamefont {Car}},\ }\href
  {https://doi.org/10.1126/science.1068774} {\bibfield  {journal} {\bibinfo
  {journal} {Science}\ }\textbf {\bibinfo {volume} {295}},\ \bibinfo {pages}
  {2427} (\bibinfo {year} {2002})}\BibitemShut {NoStop}%
\bibitem [{\citenamefont {Zhou}\ \emph {et~al.}(2020)\citenamefont {Zhou},
  \citenamefont {Wang}, \citenamefont {Choi}, \citenamefont {Pichler},\ and\
  \citenamefont {Lukin}}]{LeoZhou2020}%
  \BibitemOpen
  \bibfield  {author} {\bibinfo {author} {\bibfnamefont {L.}~\bibnamefont
  {Zhou}}, \bibinfo {author} {\bibfnamefont {S.-T.}\ \bibnamefont {Wang}},
  \bibinfo {author} {\bibfnamefont {S.}~\bibnamefont {Choi}}, \bibinfo {author}
  {\bibfnamefont {H.}~\bibnamefont {Pichler}},\ and\ \bibinfo {author}
  {\bibfnamefont {M.~D.}\ \bibnamefont {Lukin}},\ }\href
  {https://doi.org/10.1103/PhysRevX.10.021067} {\bibfield  {journal} {\bibinfo
  {journal} {Phys. Rev. X}\ }\textbf {\bibinfo {volume} {10}},\ \bibinfo
  {pages} {021067} (\bibinfo {year} {2020})}\BibitemShut {NoStop}%
\bibitem [{\citenamefont {Bishop}\ \emph {et~al.}(2023)\citenamefont {Bishop},
  \citenamefont {Montangero},\ and\ \citenamefont {Wilhelm}}]{bishop2023set}%
  \BibitemOpen
  \bibfield  {author} {\bibinfo {author} {\bibfnamefont {G.}~\bibnamefont
  {Bishop}}, \bibinfo {author} {\bibfnamefont {S.}~\bibnamefont {Montangero}},\
  and\ \bibinfo {author} {\bibfnamefont {F.~K.}\ \bibnamefont {Wilhelm}},\
  }\href@noop {} {\bibinfo {title} {A set of annealing protocols for optimized
  system dynamics and classification of fully connected spin glass problems}}
  (\bibinfo {year} {2023}),\ \Eprint {https://arxiv.org/abs/2310.10442}
  {arXiv:2310.10442 [quant-ph]} \BibitemShut {NoStop}%
\bibitem [{\citenamefont {Kato}(1950)}]{kato1950}%
  \BibitemOpen
  \bibfield  {author} {\bibinfo {author} {\bibfnamefont {T.}~\bibnamefont
  {Kato}},\ }\href {https://doi.org/10.1143/JPSJ.5.435} {\bibfield  {journal}
  {\bibinfo  {journal} {Journal of the Physical Society of Japan}\ }\textbf
  {\bibinfo {volume} {5}},\ \bibinfo {pages} {435} (\bibinfo {year} {1950})},\
  \Eprint {https://arxiv.org/abs/https://doi.org/10.1143/JPSJ.5.435}
  {https://doi.org/10.1143/JPSJ.5.435} \BibitemShut {NoStop}%
\end{thebibliography}
\end{document}